%% file: GBprc.tex
\documentstyle[12pt,epsf]{report}
\makeatletter
\@addtoreset{equation}{section}
\makeatother

\def\nq{n_\ssq}
\def\ns{n_\ssS}

\def\leff{\Scl^{\rm eff}}

\def\Tc{$T_c$}
\def\GB{{\sss GB}}
\def\pGB{{\sss pGB}}
\def\SC{{\sss SC}}
\def\AF{{\sss AF}}
\def\inv{{\rm inv}}
\def\sb{{\rm sb}}
\def\opt{{\rm opt}}

\begin{document}
%
%
\title{Goldstone and Pseudo-Goldstone Bosons in Nuclear,
Particle and Condensed-Matter Physics }
\author{C.P. Burgess}
\date{``Effective Theories in Matter''\\
Nuclear Physics Summer School and Symposium\\
Seoul National University, Korea \\
McGill-98/25\\
$ $ \\
\copyright June 1998\\ 
$ $  \\
$ $  \\
$ $  \\
{\sl Dedicated to:} \\
Caroline, Andrew, Matthew, Ian and Michael}
\maketitle
\pagenumbering{roman}
\tableofcontents
\listoffigures
\listoftables
\setlength{\baselineskip}{5ex}

\input notemacros

\pagenumbering{arabic}

\chapter{Goldstone Bosons}

Goldstone bosons are  weakly-coupled states which appear in
the low-energy spectrum of any system for which a rigid (or
global) symmetry is spontaneously broken (that is, the
symmetry is not preserved by the system's ground state).  A
great deal is known about the properties of these bosons,
since at low energies their properties are largely governed
by the nature of the symmetries which are spontaneously
broken, depending only weakly on the details of the system
itself.

This review is devoted to explaining the modern {\it
effective lagrangian} method for identifying Goldstone
boson properties. These methods are much more efficient
than are the older current-algebra techniques of yore. 

\section{Introduction}

It is a common feature of many physical systems that their
behaviour is relatively simple when examined only at low
energies (or temperatures) compared to the system's own
characteristic scales. It often happens that there are
relatively few states which can participate in low-energy
processes, and their interactions can sometimes become less
and less important the lower the energies that are examined.
Very general theoretical tools exist to exploit this
simplicity, when it arises.

One such tool is the technique of effective lagrangians.
The guiding idea for this method is the belief, first
clearly enunciated by Weinberg, that there is no loss of
generality in using a field theory to capture the
low-energy behaviour of any system. This is because field
theory in itself contains very little content beyond
ensuring the validity of general `motherhood' properties
like unitarity, cluster decomposition, and so on. 
According to this point of view, if a field theory is
identified which is the most general consistent with the
low-energy degrees of freedom and symmetries of any
particular system (together with a few `motherhood'
properties) then this field theory {\it must} provide a
good description of the system's low-energy limit.

This is a particularly useful observation when the
low-energy degrees of freedom are weakly interacting ({\it
regardless} of how strongly interacting their higher-energy
counterparts might be), because then the resulting field
theory may be simple enough to be used to predict
explicitly the system's low-energy properties. This
simplicity is somewhat paradoxical since, as we shall see,
the low-energy effective lagrangians are typically very
complicated, involving all possible powers of the various
fields and their derivatives. Simplicity is achieved in
spite of the complicated effective lagrangian because, for
weakly-coupled theories, general power-counting arguments
exist which permit an efficient identification of the
comparatively few interactions which appear at any given
order in a low-energy expansion.

Remarkably, there turns out to be a very important
situation for which very general results are known
concerning the existence of very light degrees of freedom
whose low-energy interactions are weak. This occurs
whenever a continuous global symmetry is spontaneously
broken (\ie\ which is a symmetry of the hamiltonian but
{\em not} a symmetry of the ground state), since when this
happens Goldstone's theorem guarantees the existence of
low-energy Goldstone bosons, as well as determining a great
deal about their interactions. These theorems, and their
description in terms of an effective lagrangian
formulation, are the subject of this review.

\subsection{A Road Map}

This section outlines how the material covered in this
review is organized.

\begin{enumerate}
\item
{\bf General Formalism:}
All of the general results may be found in Chapter 1,
starting with a statement of the key theorems --- those of
Noether and Goldstone --- which underlie everything else.
This is followed by a motivational discussion of the
simplest example to which Goldstone's theorem applies.
Although the properties of the Goldstone bosons are
guaranteed by general theorems, the moral of the example is
that these properties are generally not manifest in a
low-energy effective lagrangian unless a special choice of
variables is used. These variables are identified and
exploited first for spontaneously-broken abelian internal
symmetries, and then the process is repeated for nonabelian
internal symmetries. Both lorentz-invariant and
nonrelativistic systems are considered. For the
nonrelativistic case, special attention given to the
breaking of time reversal, since this qualitatively affects
the nature of the low-energy effective lagrangian. The
spontaneous breaking of spacetime symmetries, like
rotations, translations and lorentz transformations, is not
discussed in this review.

\item
{\bf Applications:}
Chapters 2 through 4 are devoted to specific applications
of the methods of Chapter 1 to examples in
high-energy/nuclear and condensed-matter physics. Chapter 2
starts with the classic relativistic example of pions as
pseudo-Goldstone bosons, whose study gave birth to many of
the techniques described in Chapter 1. (A pseudo-Goldstone
boson is the Goldstone boson for an {\it approximate}
symmetry, as opposed to an exact symmetry.) This is
followed in Chapter 3 by a study of spin waves (magnons) in
both ferromagnets and antiferromagnets. Chapter 4 then
closes with a recent, more speculative, application of
these ideas to the $SO(5)$ proposal for the
high-temperature superconductors.

\item
{\bf Bibliography:}
Finally, Chapter 5 contains a brief bibiography. It is not
meant to be exhaustive, as a great many articles have
emerged over the past decades of applications of these
methods.  I therefore restrict myself to listing those
papers and reviews of which I am most familiar. I apologize
in advance to the authors of the many excellent  articles I
have omitted.
\end{enumerate}

The review is aimed at upper-year graduate students, or
practicing researchers, since it presupposes a familiarity
with quantum field theory. It was written with an
audience of high-energy and nuclear phycisists in mind, and
so for the most part units are used for which $\hbar = c =
1$. However, I hope it will prove useful to
condensed-matter physicists as well. Enjoy!

\section{Noether's Theorem}

We start with a statement of Noether's theorem, since this
plays a role in the statement of Goldstone's theorem, which
is the main topic of this chapter.

For a field theory Noether's theorem guarantees the
existence of a conserved current, $j^\mu$, for every global
continuous symmetry of the action. To low orders in the
derivative expansion it is usually enough to work with
actions which depend only on the fields and their first
derivatives, so we restrict our statement of the theorem to
this case.

Consider therefore a system governed by an action $S = \int
d^4x \; \Scl(\phi, \partial_\mu \phi)$, where $\phi(x)$
generically denotes the fields relevant to the problem. We
imagine that $S$ is invariant under a set of
transformations of these fields, $\delta \phi = \xi_a(\phi)
\, \omega^a$, where $\omega^a$ denote a collection of
independent, spatially constant symmetry parameters.
Invariance of $S$ implies that the lagrangian density,
$\Scl$, must vary at most into a total derivative:
\eq
\label{varoflingeneral}
\delta \Scl \equiv \partial_\mu \; \Bigl( \omega^a \,
V_a^\mu \Bigr) ,
\eeq
for some quantities $V^\mu_a(\phi)$. This equation is meant
to hold as an identity, for arbitrary field configurations,
$\phi$, and for arbitrary constant parameters, $\omega^a$.
Rewriting the variation of $\Scl$ directly in terms of the
variations of the fields, and equating to zero the
coefficient of the arbitrary constant 
$\omega^a$ in the result then  gives:
\bg
\label{varoflagain}
\partial_\mu \; V_a^\mu  &=& { \partial \Scl \over \partial
\phi} \; \xi_a + {
\partial \Scl \over \partial (\partial_\mu \phi)} \;
\partial_\mu  \xi_a \nn\\
&=& \left[ { \partial \Scl \over \partial \phi} -
\partial_\mu \left( {
\partial
\Scl \over \partial (\partial_\mu \phi)} \right) \right] 
\; \xi_a  +
\partial_\mu \left( { \partial \Scl \over \partial
(\partial_\mu
\phi)} \; \xi_a \right).
\nd

The statement of the theorem follows from this last
equation. It states that the quantities
\eq
\label{defofj}
j^\mu_a \equiv - \; {\partial \Scl \over \partial
(\partial_\mu \phi)} \; \xi_a
+ V_a^\mu,
\eeq
satisfy the special property
\eq
\label{conscondition}
\partial_\mu j^\mu_a = 0,
\eeq
when they are evaluated at a solution to the equations of
motion for $S$ --- \ie\ on field configurations for which
the square bracket on the right-hand-side of
eq.~\pref{varoflagain} vanishes.

Even though we have used relativistic notation in this
argument, the  conclusion, eq.~\pref{conscondition}, is
equally valid for nonrelativistic systems. For these
systems, if we write $\rho_a = 
j^0_a$ for the temporal component of $j^\mu_a$, and denote
its spatial components by the three-vector $\bfj_a$, then
current conservation (eq.~\pref{conscondition}) is
equivalent to the familiar continuity equation
\eq
\label{noethersthm}
{ \partial \rho_a \over \partial t} + \nabla \cdot \bfj_a =
0 .
\eeq

Eq.~\pref{conscondition} or eq.~\pref{noethersthm}, are
called conservation laws because they guarantee that the
{\em charges}, $Q_a$, defined by
\eq
\label{chargint}
Q_a(t) = \int_{{\rm fixed} ~ t} d^3\bfr \; \rho_a(\bfr,t) =
\int d^3\bfr \;
j^0_a(x),
\eeq
are conserved in the sense that they are independent of
$t$. These charges are sometimes called the generators of
the symmetry because their commutator with the fields give
the symmetry transformations themselves  
\eq
\label{gencondition}
i\omega^a \; [Q_a, \phi(x) ] = \omega^a \, \xi_a = \delta
\phi.
\eeq

The existence of such a conserved current carries special
information if the symmetry involved should be
spontaneously broken, as we now describe.

\section{Goldstone's Theorem}

Whenever the ground state of a system does not respect one
of the system's global continuous symmetries, there are
very general implications for the low-energy theory. This
is the content of Goldstone's theorem, which we now
present. This theorem is central to the purpose of this
chapter, which is devoted to making its implications
manifest in a low-energy effective theory.

Goldstone's theorem states that any system for which a
continuous, global symmetry is spontaneously broken, must
contain in its spectrum a state, $\ket{G} $ --- called a
{\em Goldstone mode}, or {\em Goldstone boson} since it
must be a boson\footnote{Supersymmetry is an exception to
this statement, since spontaneously broken global
supersymmetry ensures the existence of a Goldstone fermion,
the goldstino.} --- which has the defining property that it
is created from the ground state by performing a
spacetime-dependent symmetry transformation. In equations,
$\ket{G}$ is defined by the condition that the following
matrix element cannot vanish:\footnote{We use
nonrelativistic notation here to emphasize that the
conclusions are not specific to relativistic systems. This
will prove useful when nonrelativistic applications are
considered in later sections.}
\eq
\label{gbdefn}
\bra{G} \rho(\bfr,t) \ket{\Omega}  \neq 0. \eeq
Here, $\ket{\Omega}$ represents the ground state of the
system, and $\rho = j^0$ is the density for the conserved
charge --- guaranteed to exist by Noether's theorem --- for
the spontaneously broken symmetry. 

Before turning to its implications, we outline the proof of
this result. The starting point is the assumption of the
existence of a {\em local order parameter}. 
This can be defined to be a field, 
$\phi(x)$, in the problem which satisfies two defining
conditions. Firstly, it transforms nontrivially under the
symmetry in question: 
\ie\ there is another field, $\psi(x)$, for which:
\eq
\label{gbassone}
\delta \psi \equiv i [ Q, \psi(x)] = \phi(x). \eeq
$Q$ is the conserved charge defined by integrating the
density 
$\rho(\bfr,t)$ throughout all of space. Secondly, the field
$\phi$ must have a nonzero expectation in the ground state:
\eq
\label{gbasstwo}
\Avg{\phi} \equiv \bra{\Omega} \phi(x) \ket{\Omega}  \equiv
v \neq 0.
\eeq
This last condition would be inconsistent with 
eq.~\pref{gbassone}\  if the ground state were invariant
under the symmetry of interest, since this would mean $Q
\ket{\Omega} = 0$, implying the right-hand-side of
eq.~\pref{gbasstwo} must vanish.

The proof of the theorem now proceeds from the following
steps. ($i$) Substitute eq.~\pref{gbassone} into
eq.~\pref{gbasstwo}; ($ii$) Use  $Q = \int \rho\; d^3\bfr$
in the result, as is guaranteed to be possible by Noether's
theorem; ($iii$) Insert a partition of unity as a sum over
energy eigenstates, $1 = \sum_n |
n \rangle \bra{n}$, on either side of the operator $\rho$.
The resulting expression shows that if no energy eigenstate exists
which satisfies the defining condition, eq.~\pref{gbdefn},
then the right-hand-side of eq.~\pref{gbasstwo} must
vanish, in contradiction with the starting assumptions.
This proves the theorem. We next elaborate on its
consequences.

The defining matrix element, eq.~\pref{gbdefn}, and the
conservation law, eq.~\pref{noethersthm}, together imply
that Goldstone bosons must have a number of important
properties. Besides determining their spin and their
statistics, it implies two properties which are of
particular importance:

\begin{enumerate}

\item
The Goldstone boson must be {\em gapless}, in that its
energy must vanish in the limit that its (three-) momentum
vanishes. That is:
\eq
\label{gapless}
\lim_{p \to 0} E(p) = 0.
\eeq
To see why this follows from eq.~\pref{gbdefn}, it
is helpful to make the dependence on position and time
in this equation explicit by using the identities $\rho_a(\bfr,t)
= e^{-iHt} \, \rho_a(\bfr,0) \, e^{iHt}$ and $\bfj_a(\bfr,t)
= e^{i{\bf P}\cdot \bfr} \, \bfj_a(0,t) \, e^{-i{\bf P}
\cdot \bfr}$, together with the energy- and 
momentum-eigenstate conditions: 
$H \ket{\Omega} = {\bf P} \ket{\Omega}
= 0$, ${\bf P} \ket{G(p)} = \bfp \ket{G(p)}$ and 
$H \ket{G(p)} = E_p \ket{G(p)}$. Then, differentiation
of  eq.~\pref{gbdefn} with respect to $t$, and use of
the continuity equation, eq.~\pref{noethersthm}, gives:
\bg
\label{whymassless}
-i E_p \, e^{-i E_p \, t} \; \bra{G} \rho_a(\bfr,0)
\ket{\Omega} &=& \bra{G} {\partial \rho_a
\over \partial t}\,(\bfr,t) \ket{\Omega} \nn\\
&=& - \; \bra{G} \nabla \cdot \bfj_a(\bfr,t) \ket{\Omega} \\
&=& - i \, \bfp \cdot \bra{G} \bfj_a(\bfr,t) \ket{\Omega} . \nn
\nd
Eq.~\pref{gapless} follows from this last equality in the
limit $\bfp \to 0$, given that the matrix element, 
$\bra{G} \rho_a(\bfr,0) \ket{\Omega}$, cannot vanish {\it
by definition} for a Goldstone boson. 

In relativistic systems, for which $E(p) = \sqrt{p^2 +
m^2}$ where $m$ is the particle mass, the gapless condition,
eq.~\pref{gapless}, is equivalent to the masslessness of the 
Goldstone particle.

\item
More generally, the argument just made can be 
extended to more complicated matrix elements. One finds
in this way that the Goldstone boson for any exact symmetry
must completely decouple from all of its interactions in 
the limit that its momentum vanishes. Physically,
this is because eq.~\pref{gbdefn} states
that in the zero-momentum limit the Goldstone state
literally is a symmetry transformation of the ground state.
As a result it is {\em completely indistinguishable} from
the vacuum in this limit. 

\end{enumerate}

These properties have a lot of implications for the
low-energy behaviour of any system which satisfies the
assumptions of the theorem. The first guarantees that the
Goldstone boson must itself be one of the light states of
the theory, and so it must be included in any effective
lagrangian analysis of this low energy behaviour. The
second property ensures that the Goldstone mode must be
weakly coupled in the low-energy limit, and strongly limits
the possible form its interactions can take.

The properties of gaplessness and low-energy decoupling 
can also be useful even if the spontaneously broken
`symmetry' in question is really not an exact symmetry.
To the extent that the symmetry-breaking terms, $H_{\rm sb}$, 
of the system's Hamiltonian are small, the symmetry may
be regarded as being approximate. In this case the violation of
the gapless and decoupling properties can usefully be 
treated perturbatively in $H_{\rm sb}$. The Goldstone
particles for any such approximate symmetry --- called
{\it pseudo}-Goldstone bosons --- are then systematically light
and weakly coupled at low energies, instead of being
strictly massless or exactly decoupled. 

The purpose of the remainder of this chapter is to show in
detail how these properties are encoded into the low-energy
effective lagrangian. By considering simple examples we find
that although these properties are always true, they need
not be manifest in the lagrangian in an arbitrary theory.
They can be made manifest, however, by performing an
appropriate field redefinition to a standard set of field
variables. We first identify these variables, and use them
to extract the implications of Goldstone's theorem for the
low-energy effective theory in the simplest case, for which
the symmetry group of interest is abelian. The results are
then generalized in subsequent sections to the nonabelian
case.

\section{Abelian Internal Symmetries}

In order to see the issues which are involved, it is
instructive to consider a simple field theory for which a
symmetry is spontaneously broken. We therefore first
consider a simple model involving a single complex scalar
field, $\phi$.

\subsection{A Toy Example}

The lagrangian density:
\bg
\label{abeltoymodel}
\Scl &=& - \; \partial_\mu \phi^* \partial^\mu \phi - V(\phi^*
\phi), \nn\\
\hbox{with} \qquad V &=& {\lambda \over 4} \; \left( \phi^*
\phi - {\mu^2
\over \lambda} \right)^2,
\nd
is invariant with respect to a $U(1)$ group of symmetries: 
$\phi \to e^{i \alpha} \; \phi$. This is a global symmetry
because the term involving derivatives of $\phi$ is only
invariant if the symmetry parameter, $\alpha$, is a
constant throughout spacetime. It is called an internal
symmetry since the symmetry acts only on fields and does
not act at all on the spacetime coordinate, $x^\mu$. For
later reference, the Noether current for this symmetry is:
\eq
\label{toync}
j_\mu = -i \left( \phi^* \partial_\mu \phi - \phi \;
\partial_\mu \phi^* \right).
\eeq

For small $\lambda$ this system is well approximated by a
semiclassical expansion, provided that the field $\phi$ is
$O \left( 
\lambda^{-\hf} \right)$ in size. This may be seen by
redefining 
$\phi = \tw\phi / \sqrt{\lambda}$, and noticing that all of
the 
$\lambda$-dependence then scales out of the lagrangian: 
$\Scl(\phi,\mu,\lambda) = {1\over \lambda} \Scl(\tw\phi,
\mu, 1)$ --- for which the limit $\lambda \to 0$ is seen to
be equivalent to $\hbar \to 0$ in the semiclassical limit.

The vacuum of the theory is therefore well described, for
small 
$\lambda$, by the classical configuration of minimum
energy. Since the classical energy density is a sum of
positive terms,  
$\Sch =  \dot{\phi}^* \dot{\phi} +  \del\phi^* \cdot
\del\phi + 
V(\phi^*\phi)$, it is simple to minimize. The vacuum
configuration is a constant throughout spacetime, $\dot\phi
= \del\phi = 0$, and its constant value, $\phi = v$, must
minimize the classical potential: $V(v^*v) = 0$. We may use
the $U(1)$ symmetry to choose $v$ to be real, and if $\mu^2$
is positive then the solution becomes $v = \mu
/\sqrt{\lambda}$. Happily this configuration lies within
the conditions of validity of this semiclassical analysis.

Since the vacuum configuration, $\phi = v \ne 0$, is not
invariant under the $U(1)$ transformations, $\phi \to e^{i
\alpha} \phi$, the $U(1)$ symmetry is seen to be
spontaneously broken. Goldstone's theorem should apply, and
so we now identify the Goldstone degree of freedom.

The spectrum may be identified by changing variables to the
real and imaginary parts of the deviation of the field
$\phi$ from its vacuum configuration. Defining $\Scr \equiv
\sqrt{2} \; \Re (\phi - v)$ and $\Sci \equiv \sqrt{2} \; \Im
\phi$ diagonalizes the kinetic and mass terms, and the
scalar potential in terms of these variables becomes:
\eq
\label{newabelpotl}
V = {m_\ssr^2 \over 2} \, \Scr^2 + {g_{30} \over 3!} \,
\Scr^3 + {g_{12} \over
2} \, \Scr \Sci^2 + {g_{40} \over 4!} \, \Scr^4 +
{g_{22} \over 4} \, \Scr^2 \Sci^2 + {g_{04} \over 4!} \,
\Sci^4 ,
\eeq
where the couplings and masses in this potential are given
in terms of the original parameters, $\lambda$ and $\mu$,
by: 
\eq
\label{potcouplings}
m_\ssr^2 = \lambda \mu^2, \qquad {g_{30} \over 3!} =
{g_{12} \over 2} =
{\lambda v \over 2 \sqrt{2}}, \qquad {g_{40} \over 4!} =
{g_{04} \over 4!} =
{g_{22} \over 4} = {\lambda \over 16}.
\eeq

Notice the existence of a massless field, $\Sci$, as is
required by Goldstone's theorem. We can verify that $\Sci$
really is the Goldstone boson by writing the Noether
current, 
eq.~\pref{toync}, in terms of the mass eigenstates, $\Scr$
and 
$\Sci$:
\eq
\label{newtoync}
j_\mu = v \sqrt{2} \; \partial_\mu \Sci + \left( \Scr \;
\partial_\mu \Sci - \Sci \;
\partial_\mu \Scr \right).
\eeq
Clearly the matrix element:
\eq
\label{testme}
\bra{ \Sci(p) } j^\mu(x) \ket{0} \propto v \sqrt{2} \;
p^\mu \; e^{-ipx}
\eeq
does not vanish (unless $v=0$), as is required of a
Goldstone boson.

\begin{figure}
\epsfbox[-70 670 650 670]{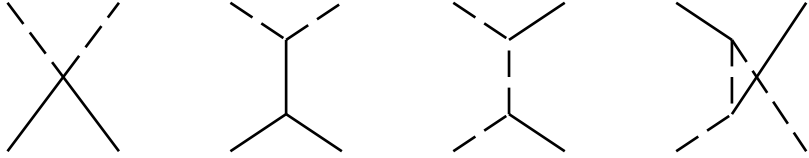}
\vspace{1.5in}
\caption{The Feynman graphs which describe $\Scr-\Sci$
scattering at tree level. Solid lines denote $\Scr$ and
dashed lines represent $\Sci$.}
\end{figure}

A puzzle with the potential of eqs.~\pref{newabelpotl} and 
\pref{potcouplings} is that the Goldstone boson, $\Sci$,
appears in the scalar potential, and so its couplings do
not appear to vanish in the limit of vanishing momentum.
This is only an appearance, however, and $\Sci$ really does
decouple at low energies, as can be tested by computing
Goldstone boson scattering in this limit. For example, the
$S$-matrix at tree level for $\Sci - \Scr$ scattering may
be computed by evaluating the Feynman graphs of Fig.~(1.1).
The result is:
\eq
\label{smatrixdef}
S[\Scr(r) + \Sci(s) \to \Scr(r') + \Sci(s')] = {i \Sca \;
\delta^4(r + s - r' - s') \over (2\pi)^2 \sqrt{16 s^0 r^0
s'^0 r'^0}} \; ,
\eeq
with
\eq
\label{smatrixresult}
\Sca = - \, g_{22} + { g_{12} \; g_{30} \over 
(s+s')^2 + m_\ssr^2 - i\eps} +
g_{12}^2 \left[ {1 \over (s+r)^2 -i \eps} + {1 \over (s -
r')^2 - i\eps}
\right].
\eeq
In the limit $s^\mu , s'^\mu \to 0$ this becomes (using the
condition $r^2 = r'^2 = - m_\ssr^2$):
\bg
\label{zeromomlim}
\Sca & \to & - \, g_{22} + { g_{12} \; g_{30} \over 
m_\ssr^2} - \, {2 g_{12}^2 \over m_\ssr^2} , \nn\\
&=& \lambda \; \left( - \, \hf + {3 \over 2} \;  - 1
\right) = 0.
\nd

The scattering amplitude indeed vanishes in the
zero-momentum limit, as it must according to Goldstone's
theorem. This vanishing is not manifest in the lagrangian,
however, and is only accomplished through a nontrivial
cancellation of terms in the $S$-matrix. For many purposes,
not least when constructing an effective theory to describe
the low energy interactions of the Goldstone bosons, it
would be preferable to have this decoupling be manifest in
the lagrangian. We will now do so, by making a field
redefinition to a new set of variables for which decoupling
becomes explicit.

\subsection{A Better Choice of Variables}

In order to identify which variables would make the
decoupling of Goldstone bosons more explicit in the
lagrangian, it is useful to recall the definition of what
the Goldstone mode physically is. Its defining condition,
eq.~\pref{gbdefn}, can be interpreted to mean that the
Goldstone modes are obtained from the ground state by
performing a symmetry transformation, but with a spacetime
dependent transformation parameter.

In the example considered in the previous section the
ground state configuration is $\phi = v$, and so a local
symmetry transformation of this ground-state would be $\phi
= v e^{i 
\theta(x)}$. If this is substituted into the lagrangian of 
eq.~\pref{abeltoymodel}, we find $\Scl \left( \phi = v e^{i
\theta(x)} \right) = - v^2 \partial_\mu \theta \partial^\mu
\theta$. 
$\theta$ does not drop out of the problem because, although
the lagrangian vanishes when it is evaluated at $\phi =v$,
the configuration $v e^{i\theta(x)}$ is only related to
$\phi =v$ by a symmetry when $\theta$ is a constant. This
fact that $\theta$ parameterizes a symmetry direction when
it is restricted to constant field configurations
guarantees that any $\theta$-dependence of $\Scl$ {\em
must} involve at least one derivative of $\theta$, thereby
dropping out of the problem in the limit of small
derivatives --- \ie\ small momenta, or long wavelengths.

All of this suggests that $\theta$ would make a good
representation for the Goldstone mode, since this is
precisely what a Goldstone mode is supposed to do: decouple
from the problem in the limit of small momenta. We are led
to the suggestion of using polar coordinates in field
space, 
\eq
\label{polcoords}
\phi(x) = \chi(x) \; e^{i \theta(x)},
\eeq
in order to better exhibit the Goldstone boson properties.
In this expression both $\theta$ and $\chi$ are defined to
be real. Substituting this into the lagrangian gives:
\eq
\label{linpolcoords}
\Scl = - \partial_\mu \chi \partial^\mu \chi - \chi^2
\partial_\mu \theta
\partial^\mu \theta - V(\chi^2).
\eeq
It is clear that these variables do the trick, since the
fact that $\theta$ appears in the definition,
eq.~\pref{polcoords}, in the same way as does a symmetry
parameter guarantees that it completely drops out of the
scalar potential, as must a Goldstone boson if its
low-energy decoupling is to be made explicit.

A price has been paid in exchange for making the low-energy
decoupling of the Goldstone boson explicit, however. This
price is most easily seen once the fields are canonically
normalized, which is acheived by writing $\chi = v +
\nth{\sqrt{2}} \; \chi'$ and 
$\theta = \nth{v \sqrt{2}} \; \varphi$. With these
variables the lagrangian is seen to have acquired nominally
nonrenormalizable interactions:
\eq
\label{nrints}
\Scl_{\rm nr} = - \left[ {\chi' \over \sqrt{2} \; v^2} +
{\chi'^2 \over 4 v^2}
\right] \; \partial_\mu \varphi \partial^\mu \varphi. \eeq
Of course, the $S$-matrix for the theory in these variables
is identical to that derived from the manifestly
renormalizable lagrangian expressed in terms of the
variables $\Scr$ and $\Sci$. So the $S$-matrix remains
renormalizable even when computed using the variables
$\chi'$ and $\varphi$. (The same is not true for {\em
off-shell} quantities like Green's functions, however,
since the renormalizability of these quantities need not
survive a nonlinear field redefinition.)

In this toy model there is therefore a choice to be made
between making the lagrangian manifestly display either the
renormalizability of the theory, or the Goldstone boson
nature of the massless particle. Which is best to keep
explicit will depend on which is more convenient for the
calculation that is of interest. Since, as we shall see,
renormalizability is in any case given up when dealing with
effective low-energy field theories, it is clear that the
variables which keep the Goldstone boson properties
explicit are the ones of choice in this case.

\subsection{The General Formulation}

The reason why the above redefinition works may be seen by
asking how the $U(1)$ symmetry acts on the new variables.
The key observation is that the symmetry  transformation
becomes 
{\em inhomogeneous}: $\theta \to \theta + \alpha$, where 
$\alpha$ is a constant. In terms of the canonically
normalized field, $\varphi$, this transformation law
becomes:
\eq
\label{gbabeltransf}
\varphi \to \varphi + \sqrt{2} \; v \; \alpha. \eeq
This kind of transformation rule is the hallmark of a
Goldstone boson, since it enforces the explicit nature of
all of the Goldstone boson properties in the lagrangian. In
fact --- as can be expected from the generality of
Goldstone's theorem --- they can all be derived purely on
the grounds of this symmetry transformation, and do not
rely at all on the details of the underlying model which
motivated its consideration.

To show that this is true, imagine writing an arbitrary
effective theory for a real scalar field, $\varphi$,
subject only to the symmetry of eq.~\pref{gbabeltransf}
(and, for simplicity, to Poincar\'e invariance).  The most
general lagrangian which is invariant under this
transformation is an arbitrary function of the derivatives,
$\partial_\mu\varphi$, of the field. An expansion in
interactions of successively higher dimension then gives:
\eq
\label{abelgbaction}
\leff(\varphi) = - \hf \; \partial_\mu \varphi \partial^\mu
\varphi - {a \over
4\, v^4} \;  \partial_\mu \varphi \partial^\mu \varphi \;
\partial_\nu \varphi
\partial^\nu \varphi + \cdots,
\eeq
where we have inserted a power of $v$ as appropriate to
ensure that the parameter $a$ is dimensionless. This
accords with the expectation that it is the
symmetry-breaking scale, $v$, which sets the natural scale
relative to which the low energy limit is to be taken. In
the toy model just considered, integrating out the heavy
field, $\chi'$ produces these powers of $v$ through the
appearance of the inverse of the heavy mass, $m_\ssr$. The
result must be an effective lagrangian of the form of 
eq.~\pref{abelgbaction}, but with a specific, calculable
coefficient for the parameter $a$.

This, most general, lagrangian automatically ensures that
$\varphi$ has all of the Goldstone boson properties. For
instance, since the symmetry implies that $\leff$ can only
depend on derivatives of 
$\varphi$, it ensures that $\varphi$ cannot appear at all
in the scalar potential, and so in particular ensures that
$\varphi$ is massless. Similarly, applying Noether's
theorem to the kinetic term for $\varphi$ implies that
there is a contribution to the Noether current, $j^\mu$,
which is linear in $\varphi$: 
\eq
\label{abelgbncex}
j^\mu = \sqrt{2} \; v \; \left[ \partial^\mu \varphi + {a
\over v}
(\partial^\nu \varphi \, \partial_\nu \varphi) \,
\partial^\mu \varphi
+ \cdots \right] \, . \eeq
The ellipses in this expression represent contributions to
$j^\mu$ which come from other terms in the lagrangian
besides the 
$\varphi$ kinetic term. Clearly this ensures that the
matrix element $\bra{G} j^\mu \ket{0} \neq 0$ so long as $v
\neq 0$.

Such an understanding of the Goldstone nature of a field,
like 
$\varphi$, as an automatic consequence of a symmetry is
clearly invaluable when constructing effective lagrangians
for systems subject to spontaneous symmetry breaking. We
next turn to the generalization of these results to the
more general case of nonabelian internal symmetries.

\section{Nonabelian Internal Symmetries}

The lesson learned from the abelian example is that half of
the art of constructing effective lagrangians for Goldstone
bosons lies in the choice of a convenient set of variables
in terms of which their properties are explicitly displayed
in the lagrangian. In this section the above construction is
generalized to the case of nonabelian, global, internal
symmetries.

\subsection{A Second Toy Model}

As guidance towards an appropriate choice of field
variables we once more start with a simple toy model for
which the underlying theory is explicitly known. Consider,
therefore, a system of $N$ real scalar fields, $\phi^i, i =
1,\dots,N$, which for simplicity of notation we arrange into
an $N$-component column vector, denoted by $\phi$ with no
superscript. Notice that there is no loss of generality in
working with real fields, since any complex fields could
always be decomposed into  real and imaginary parts. We
take as lagrangian density
\eq
\label{nonabeltm}
\Scl = - \hf \; \partial_\mu \phi^\sst \partial^\mu \phi -
V(\phi),
\eeq
where the superscript `T' denotes the transpose, and where
$V(\phi)$ is a potential whose detailed form is not
required. The kinetic term of this lagrangian is manifestly
invariant under the Lie group, $O(N)$, of orthogonal
rotations amongst the $N$ real fields: $\phi \to O \phi$,
where the $O$'s are independent of spacetime position,
$\partial_\mu O  =0$, and $O^\sst O = 1$. In general, the
potential $V(\phi)$ need not be also invariant under these
$O(N)$ transformations, but may only preserve some subgroup
of these, $G \subset O(N)$. That is, if $g \in G$, then $V(g
\phi) = V(\phi)$ for all fields $\phi$.

Suppose now that for some regime of parameters this model
is well described by the semiclassical approximation, and
further that the potential, $V$, is minimized for some
nonzero value for the fields: $\phi = v \neq 0$. If this is
the case, then the symmetry group $G$ may be spontaneously
broken to some subgroup, $H 
\subset G$, which is defined by: $h v = v$, for all $h \in
H$.

\subsection{A Group-Theoretic Aside}

It is important to notice that the current whose existence
is guaranteed by Noether's theorem --- and so which plays
the central role in Goldstone's theorem --- arises only
if the symmetry of interest is {\it continuous}. Continuous
here means that the group elements may be parameterized 
by a continuous parameter (like a rotation angle), as opposed
to a discrete label. Groups with continuous labels are
called Lie groups provided their labels are sufficiently
smooth. Before proceeding it is useful to pause to 
record some mathematical properties of such 
Lie groups, and their associated Lie algebras. 

\begin{enumerate}

\item
Typically the continuous symmetry groups which arise in
physical applications do so as explicit finite-dimensional
unitary matrices. As a result a special role is played by
compact groups, for which the parameter space of the group
is a compact set. Compact groups are of such special
interest since it is only for compact groups that
finite-dimensional, unitary and faithful matrix
representations exist.\footnote{A representation is
faithful if there is a one-to-one correspondence between
the group elements and the matrices which represent them.
Since the groups of interest are usually {\em defined} by a
finite-dimensional and unitary representation, this
representation is, by definition, faithful.} We assume
compact groups throughout what follows, and we work
explicitly with representations involving
finite-dimensional and unitary matrices, $g^\dagger =
g^{-1}$.

\item
There is also no loss of generality in assuming our
representation matrices, $g$, to be real: $g = g^*$. This
is because any complex representation may always be
decomposed into its real and imaginary parts. This
convention is ensured in the scalar-field example we are
considering by choosing to employ only real fields. We do
{\em not} assume these matrices to be irreducible. 
Recall that if the matrices are reducible, then 
there is a basis in which they can be written in a 
block-diagonal form:
\eq
\label{reducibledef}
g = \pmatrix{ g_{(1)} & & \cr & \ddots & \cr && g_{(n)}
\cr}. 
\eeq

\item
It is useful to phrase much of what follows in terms of the
Lie algebra of $G$ and $H$ rather than in terms of the Lie
groups themselves. That is, we take advantage of the fact
that any group element which is connected to the identity
element, $g = 1$, may be written as a matrix exponential:
$g = \exp \left[ i \alpha^a T_a 
\right]$, of a linear combination of a collection of basis
matrices, or generators, $T_a$, $a = 1,...,d$ where $d$ is
called the dimension of the group. The $T_a$'s lie inside
what is called the Lie algebra of $G$. The unitarity and
reality of the group elements, $g$, imply the matrices
$T_a$ to be hermitian and imaginary:
\eq
\label{whatthetssatisfy}
T_a = T^\dagger_a = - T^*_a = - T^\sst_a. \eeq

\item
Since the generators, $T_a$, are finite dimensional and
hermitian, it follows that the matrix $N_{ab} = \Tr(T_a
T_b)$ is positive definite. As a result we are free to
redefine the generators to ensure that $N_{ab} =
\delta_{ab}$. With this choice there is no distinction to
be made between indices $a$ and $b$ which are superscripts
and subscripts. We assume this convenient choice to have
been made in what follows.

\item
Closure of the group multiplication law --- \ie\ the statement
that $g_1, g_2 \in G$ implies $g_1 g_2 \in G$ ---implies
commutation relations for the $T_a$'s: $T_a \,T_b - 
T_b \, T_a = i \, c_{abd} 
\, T_d$ where the $c_{abd}$'s are a set of constant
coefficients which are characteristic of the group
involved. From its definition it is clear that $c_{abd}$ is
antisymmetric under the interchange of the indices $a$ and
$b$. Whenever the generators are chosen so that $N_{ab} =
\delta_{ab}$ it also turns out that $c_{abd}$ is completely
antisymmetric under the interchange of {\em any} two indices.

\item
For the present purposes it is convenient to choose a basis
of generators which includes the generators of the subgroup
$H$ as a subset. That is, choose $\{ T_a \} = \{ t_i,
X_\alpha \}$, where the $t_i$'s generate the Lie algebra of
$H$, and the $X_\alpha$'s constitute the rest. Since $H$ is
defined as the group which preserves the vacuum
configuration, its generators must satisfy $t_i v = 0$. The
closure of the subgroup, $H$, under multiplication ensures
that $t_i \, t_j - t_j \, t_i = i \, c_{ijk} \; t_k$ (with
no $X_\alpha$'s on the right-hand-side). Schematically this
can be written $c_{ij\alpha} = 0$.

\item
The $X_\alpha$'s do not lie within the Lie algebra of $H$,
and so satisfy $X_\alpha v \neq 0$. They are said to
generate the space, $G/H$, of {\em cosets}. A coset is an
equivalence class which is defined to contain all of the
elements of $G$ that are related by the multiplication by
an element of $H$. Physically, the $X_\alpha$'s represent
those generators of the symmetry group, $G$, which are
spontaneously broken.

\item
The group property of $H$ described above, together with
the complete antisymmetry of the $c_{abd}$'s implies a
further condition: $c_{i\alpha j} = 0$. This states that
$t_i \, X_\alpha - X_\alpha \, t_i = i \, c_{i \alpha
\beta} X_\beta$ (with no $t_j$'s on the right-hand-side).
This states that the $X_\alpha$'s fall into a (possibly
reducible) representation of $H$. Once exponentiated into a
statement about group multiplication, the condition $tX - Xt
\propto X$ implies,  for any $h \in H$, that $h X_\alpha
h^{-1} = {L_\alpha}^\beta X_\beta$ for some coefficients,
${L_\alpha
}^\beta$.

By contrast, $X_\alpha \,X_\beta - X_\beta \, X_\alpha$
need not have a particularly simple form, and can be
proportional to both $X_\gamma$'s and $t_i$'s.\footnote{If
the right-hand-side of $X_\alpha \,X_\beta - X_\beta \,
X_\alpha$ were assumed not to contain any  $X_\gamma$'s,
then the coset $G/H$ would be called a {\em symmetric}
space, but we do not make this assumption here.}

\end{enumerate}

\subsection{The Toy Model Revisited}

Returning to the toy model defined by the lagrangian,
eq.~\pref{nonabeltm}, we know Goldstone's theorem 
implies that the assumed symmetry-breaking pattern
must give rise to a collection of massless Goldstone
bosons, whose interactions we wish to exhibit explicitly.
The Goldstone modes are, intuitively, obtained by
performing symmetry transformations on the ground state.
Since an infinitesimal symmetry transformation on the
ground state corresponds to the directions $X_\alpha v$ in
field space, we expect the components of $\phi$ in this
direction, $v^\sst X_\alpha \phi$, to be the Goldstone
bosons. It is indeed straightforward to verify that the
$G$-invariance of the lagrangian ensures the masslessness
of these modes. There is consequently precisely one
Goldstone degree of freedom for each generator of $G/H$.

More generally, in order to make low-energy decoupling of
these Goldstone bosons manifest we require that they do not
appear at all in the scalar potential. Following the example
taken for the case of the abelian symmetry, we therefore
change variables from 
$\phi(x) = \{ \phi^i\}$ to $\chi(x) = \{ \chi^n \}$ and
$\theta(x) = 
\{ \theta^\alpha \}$, where   
\eq
\label{nonabeldef}
\phi = U(\theta) \; \chi ,
\eeq
and $U(\theta) = \exp[ i \theta^\alpha(x) X_\alpha]$ is a
spacetime-dependent symmetry transformation in the
direction of the broken generators, $X_\alpha$. 

In order for eq.~\pref{nonabeldef} to provide a well
defined change of variables, $\chi$ must satisfy some kind
of constraint. We therefore require that $\chi$ be
perpendicular (in field space) to the Goldstone directions,
$X_\alpha v$. That is:
\eq
\label{chicond}
v^\sst X_\alpha \chi = 0, \qquad \hbox{for all} \; x^\mu \;
\hbox{and} \;
X_\alpha.
\eeq
Notice that this constraint --- together with the identity
$v^\sst 
X_\alpha v = 0$, which follows from the antisymmetry of the
$X_\alpha$'s --- is precisely what is required to ensure the
vanishing of the cross terms, proportional to $\partial_\mu 
\theta^\alpha \partial^\mu \chi'^n$, in the quadratic part
of the expansion of the kinetic terms about the ground
state configuration: $\chi = v + \chi'$.

Since $U(\theta)$ is an element of $G$, the variable 
$\theta$ is guaranteed to drop out of the scalar potential. 
Of course, this is the point of this change of variables,
and it happens because $G$-invariance requires the 
potential to satisfy $V(U \chi) = V(\chi)$. As a result, all 
of the terms in $\Scl$ which involve the Goldstone bosons, 
$\theta$, vanish when $\partial_\mu \theta^\alpha =0$, 
and eqs.~\pref{nonabeldef} and 
\pref{chicond} define the change of variables which makes
explicit the low-energy Goldstone-boson decoupling.

We pause now to briefly argue that it is always possible to
satisfy eq.~\pref{chicond} starting from an arbitary smooth
field configuration, $\phi(x)$. That is, we argue that it
is always possible to find a spacetime dependent group
element, $U(\theta) \in G$, for which 
$\chi = U^{-1} \phi$ satisfies eq.~\pref{chicond}. 
To this end consider the following
function, $\Scf[g(\alpha)] \equiv v^\sst g(\alpha) \phi$,
where $g(\alpha)$ is an arbitrary, spacetime-dependent
element of $G$. Focus, for a moment, on $\Scf$ as a function
of the parameters, $\alpha^a$, of the group for a fixed
spacetime position, $x^\mu$. Since all of the variables,
$\phi$, $v$ and $g$ have been chosen to be real, and since
the group, $G$, is compact, $\Scf(\alpha)$ defines a
real-valued function having a compact range. It is a
theorem that any such function must have a maximum and a
minimum, and so there exist group elements, $\ol g =
g(\ol\alpha)$, for which $\left. 
(\partial \Scf / \partial\alpha^a) \right|_{\alpha =
\ol\alpha}$ vanishes. Repeating this condition for each
point in spacetime defines functions $\alpha^a(x)$ whose
smoothness follows from the assumed smoothness of $\phi(x)$.

The final step in the argument is to show that the
existence of these stationary points of $\Scf$ also give
solutions to the problem of finding a $U$ for which $\chi =
U^{-1} \phi$ satisfies 
eq.~\pref{chicond}. This last step follows by explicitly
taking the derivative of $g$ with respect to $\alpha^a$:
and using parameters $\alpha$ such that $(\partial g/
\partial \alpha^a) g^{-1} = T_a$. In this case the
vanishing of $\partial \Scf/ \partial\alpha^a$, when
evaluated at $g = \ol{g} = g(\ol\alpha)$, implies $v^\sst
T_a \ol{g} \phi = 0$. We see that the choice $\chi = U^{-1}
\phi$, with $U = \ol{g}^{-1}$, therefore satisfies
eq.~\pref{chicond}, and the existence of such a solution
follows from the existence of a maximum and minimum of
$\Scf$.

This concludes the argument. We next explore the properties
of the new variables.

\subsection{The Nonlinear Realization}

Having motivated this choice of variables, we now determine
how they transform under the group $G$ of symmetry
transformations. The transformation rules which we obtain
--- and which we show to be unique, up to field
redefinitions, below --- carry all of the information
concerning Goldstone boson properties, and so requiring
low-energy lagrangians to be invariant under these
transformations automatically encodes these properties into
the low-energy theory.

The starting point is the transformation rule for $\phi$:
$\phi \to 
\tw\phi = g \phi$, where $g = \exp[ i \alpha^a T_a ]$. The
transformation rule for the new variables is then $\theta
\to 
\tw\theta$ and $\chi \to \tw\chi$, where $\phi = U(\theta)
\chi$ and $\tw\phi = U(\tw\theta) \tw\chi$. That is, under
any transformation, $g \in G$, $\theta, \chi, \tw\theta$
and $\tw\chi$ are related by:
\eq
\label{prelimreln}
g U(\theta) \chi = U(\tw\theta) \tw\chi.
\eeq

This last equation states that the matrix $\gamma \equiv
\tw U^{-1} g U$  (where $\tw U$ denotes $U(\tw\theta)$) has
the property that $\gamma \chi = \tw\chi$. The central
result to be now proven is that this condition implies that
$\gamma$ must lie within the subgroup, $H$ of unbroken
transformations, and so may be written $\gamma = \exp(i u^i
t_i)$, for some function $u^i(\theta,g)$. Once this has been
demonstrated, the transformation law therefore becomes:
\eq
\label{gbtransrule}
\theta^\alpha \to \tw\theta^\alpha(\theta, g) \qquad
\hbox{and} \qquad \chi
\to \tw\chi(\theta,g,\chi),
\eeq
where
\bg
\label{gbtransruletwo}
g \; e^{i \theta^\alpha X_\alpha} &=& e^{i \tw\theta^\alpha
X_\alpha} \; e^{i
u^i t_i}, \nn\\
\tw\chi &=& e^{i u^i t_i} \; \chi.
\nd

The first of the eqs.~\pref{gbtransruletwo} should be read
as defining the nonlinear functions
$\tw\theta^\alpha(\theta,g)$ and $u^i(\theta,g)$. They are
defined by finding the element, $g 
e^{i \theta \cdot X} \in G$, and then decomposing this
matrix into the product of a factor, $e^{\tw\theta \cdot
X}$, lying in $G/H$ times an element, $e^{iu \cdot t}$, in
$H$. The second line of eqs.~\pref{gbtransruletwo} then
defines the transformation rule for the non-Goldstone
fields, $\chi$.

These rules generally define transformation laws which are
nonlinear in the Goldstone fields, $\theta^\alpha$. They
furnish, nonetheless, a faithful realization of the
symmetry group $G$, in that $\tw\theta(\theta,g_1 g_2) =
\twi{\theta}
(\tw\theta(\theta,g_2),g_1)$ \etc. This may either be
directly verified using the definitions of
eqs.~\pref{gbtransruletwo}, or by noticing that this
property is inherited from the faithfulness of the original
linear representation of $G$ on $\phi$.

There is a particularly interesting special case for which 
eqs.~\pref{gbtransruletwo} can be explicitly solved for
$\gamma = e^{i u \cdot t}$ and $\tw U = e^{i \tw \theta
\cdot X}$. This is when $g = h$ lies in $H$. In this case,
the solution is easily seen to be: $\gamma = h$ and $\tw U
= h U h^{-1}$. Both $\chi$ and $\theta$ therefore transform
{\em linearly} under the unbroken symmetry transformations,
$H$. That is: 
\bg 
\label{spcaseofh} 
\theta^\alpha X_\alpha &\to& \tw \theta^\alpha X_\alpha = h
\; \theta^\alpha X_\alpha \; h^{-1}, \nn\\ 
\chi &\to& \tw\chi = h \chi,
\nd
for all $h \in H$.

For the broken symmetries, $g \in G/H$, it is useful to
specialize to an infinitesimal transformation, $g = 1 +
i\omega^\alpha 
X_\alpha + \cdots$. In this case we have $\gamma = 1 + i 
u^i(\theta,\omega) t_i + \cdots$, and $U(\tw\theta) =
U(\theta) [ 1 + i \Delta^\alpha(\theta,\omega) X_\alpha +
\cdots]$, where 
$u^i(\theta,\omega)$ and $\Delta(\theta,\omega)$ must both
also be infinitesimal quantities. Eq.~\pref{gbtransruletwo}
gives them explicitly to be:
\bg
\label{explicittransfnone}
\Delta_\alpha &=& \Tr \Bigl[ X_\alpha e^{-i \theta \cdot X}
(\omega \cdot X)
e^{i \theta \cdot X} \Bigr], \nn\\
&\approx& \omega_\alpha - c_{\alpha \beta \gamma}
\omega^\beta \theta^\gamma +
O(\theta^2) ; \\
\label{explicittransfntwo}
u_i &=& \Tr \Bigl[ t_i e^{-i \theta \cdot X} (\omega \cdot
X)  e^{i
\theta \cdot X} \Bigr], \nn\\
&\approx& - c_{i \alpha \beta} \omega^\alpha \theta^\beta +
O(\theta^2)
\nd
where we used: $\Tr(X_\alpha X_\beta) =
\delta_{\alpha\beta}$, $\Tr(t_i t_j) = \delta_{ij}$ and
$\Tr(t_i X_\alpha) = 0$.

This last expression for $\Delta_\alpha(\theta,\omega)$ can
be re-expressed in terms of the change,
$\delta \theta^\alpha \equiv \xi^\alpha(\theta,\omega) \equiv
\tw\theta^\alpha - \theta^\alpha$, of the Goldstone-boson
fields. The relation between $\Delta_\alpha$ and
$\xi^\alpha$ is linear: 
$\Delta_\alpha = M_{\alpha\beta}(\theta) \, \xi^\beta$,
where the matrix, $M_{\alpha\beta}$, of coefficients may be
computed using the following useful identity, which holds
for any two square matrices, $A$ and $B$:    
\bg
\label{identity}
e^{-iA} \; e^{i(A+B)} &=& 1 + i \int_0^1 ds \; e^{-isA} B
e^{is(A+B)} , \nn\\
&=& 1 + i \int_0^1 ds \; e^{-isA} B e^{isA} + O(B^2). \nd
Using this with $A = \theta \cdot X$ and $B = \xi \cdot X$
gives:
\eq
\label{deltaxireln}
M_{\alpha\beta} = \int_0^1 ds \; \Tr \Bigl[ X_\alpha e^{-i 
s\theta \cdot X}
X_\beta e^{i s \theta \cdot X} \Bigr].
\eeq

The transformation rules for the $\theta^\alpha$ with
respect to the broken symmetries in $G/H$ have two
important properties. The first crucial property is  that
the transformation law is {\em inhomogeneous} in the broken
symmetry parameters, since
\eq
\label{nonabelinhomo}
\delta\theta^\alpha = \omega^\alpha -
{c^\alpha}_{\beta\gamma} \omega^\beta
\theta^\gamma + O(\theta^2).
\eeq
As was observed earlier for the abelian example, it is this
property which enforces the decoupling of the Goldstone
bosons at low energies.

The second important property is that, for a nonabelian
group, the symmetries in $G/H$ act {\em nonlinearly} on the
fields 
$\theta^\alpha$. This property is also significant since it
ruins many of the consequences which would otherwise hold
true for symmetries which are linearly realized. For
example, the masses of those particles whose fields lie in
a linear representation of a symmetry group necessarily
have equal masses, \etc. The same is not true for particles
whose fields are related by nonlinear transformations.

There is a corollary which follows from the nonlinearity of
the realization of the symmetries in $G/H$. The fact that
the transformation of $\theta^\alpha$ and $\chi^n$ are both
field dependent implies that the action of these symmetries
are spacetime dependent. For example, even though the
transformation parameters themselves, $\omega^\alpha$, are
constants --- since $G$ is a global symmetry --- the
transformation matrix $\gamma = e^{i u \cdot t}$ which
appears in the $\chi$ transformation law is not a constant,
$\partial_\mu \gamma \neq 0$. This fact complicates the
construction of lagrangians which are invariant with
respect to these symmetries.

\section{Invariant Lagrangians}

With the transformation rules for the Goldstone boson
fields in hand we may now turn to the construction of
invariant Lagrangians which can describe their low-energy
interactions. The main complication here is in the
construction of the kinetic terms, since the transformation
rules for the fields are spacetime dependent due to their
complicated dependence on the fields.

A clue as to how to proceed can be found by reconsidering
the toy model of scalar fields $\phi$. In this case the
kinetic term, proportional to $\partial_\mu \phi^\sst
\partial^\mu \phi$ is manifestly $G$ invariant. It must
therefore remain so after performing the change of
variables to $\theta$ and $\chi$. To see how this is comes
about, we notice that after the replacement 
$\phi = U \chi$ we have: $\partial_\mu \phi = U
(\partial_\mu \chi + U^{-1} \partial_\mu U \chi )$. In
terms of the new variables, the kinetic term is invariant
because the combination $\Scd_\mu 
\chi = \partial_\mu \chi + U^{-1} \partial_\mu U \chi$
transforms covariantly: $\Scd_\mu \chi \to \gamma \Scd_\mu
\chi$.  It does so because $U^{-1} \partial_\mu U$
transforms like a gauge-potential:
\bg
\label{gaugepot}
U^{-1} \partial_\mu U &\to& \tw U^{-1} \partial_\mu \tw U,
\\
&=& \gamma \; ( U^{-1} \partial_\mu U ) \; \gamma^{-1} -
\partial_\mu
\gamma \; \gamma^{-1} . \nn
\nd

More information emerges if we separate out the parts of
$U^{-1} \partial_\mu U$ which are proportional to
$X_\alpha$ from those which are proportional to $t_i$ since
the inhomogeneous term, 
$\partial_\mu\gamma \; \gamma^{-1}$, is purely proportional
to $t_i$. That is, if we define: 
\eq
\label{decompn}
U^{-1} \partial_\mu U = -i \Sca^i_\mu t_i  + i e^\alpha_\mu
X_\alpha,
\eeq
then each of these terms transforms separately under $G$
transformations: 
\bg
\label{decomptrans}
-i \Sca^i_\mu(\theta) t_i &\to& -i \Sca^i_\mu(\tw\theta)
t_i =
\gamma \; [ -i \Sca^i_\mu(\theta) t_i ] \; \gamma^{-1} - 
\partial_\mu
\gamma \; \gamma^{-1} , \nn\\
i e^\alpha_\mu(\theta) X_\alpha &\to&  i
e^\alpha_\mu(\tw\theta) X_\alpha =
\gamma \; [ i e^\alpha_\mu(\theta) X_\alpha ] \;
\gamma^{-1}. 
\nd

We see that the quantity $\Sca^i_\mu$ transforms like a
gauge potential. For infinitesimal transformations, $g
\approx 1 + 
i \omega^\alpha X_\alpha$ and $\gamma(\theta,g) \approx 1 + 
i u^i(\theta,\omega) t_i$, we have: 
\eq
\label{infatransf}
\delta  \Sca^i_\mu(\theta) = \partial_\mu
u^i(\theta,\omega) - {c^i}_{jk}
u^j(\theta, \omega) \Sca^k_\mu(\theta).
\eeq
Similarly, $e^\alpha_\mu(\theta)$ transforms covariantly,
with
\eq
\label{infetransf}
\delta  e^\alpha_\mu(\theta) = - {c^\alpha}_{i \beta}
u^i(\theta,\omega) \;
e^\beta_\mu(\theta).
\eeq
In this last expression, the structure constants define
representation matrices, $(\Sct_i)_{\alpha\beta} = 
c_{\alpha i \beta}$, of the Lie algebra of $H$. These are
the same matrices which define the representation of $H$
that the generators $X_\alpha$ form, and it is important
for later purposes to recall that these representation
matrices need {\it not} be irreducible. If this
representation {\em is} reducible then it is possible to
define more $G$-invariant quantities with which to build
the low energy lagrangian than would otherwise be possible.

If we extract the overall factor of $\partial_\mu
\theta^\alpha$, so that $\Sca^i_\mu = \Sca^i_\alpha(\theta)
\; \partial_\mu 
\theta^\alpha$ and $e^\alpha_\mu = {e^\alpha}_\beta(\theta) 
\; \partial_\mu \theta^\beta$, then the identity,
eq.~\pref{identity}, gives the following expressions for
the coefficients:
\bg
\label{gaugeexp}
\Sca^i_\alpha(\theta) &=& - \int_0^1 ds \; \Tr \Bigl[ t_i
e^{-is \theta \cdot
X} X_\alpha e^{is \theta \cdot X} \Bigr], \nn\\ &\approx&
\hf \, c_{i\alpha
\beta} \theta^\beta + O(\theta^2) ,
\nd
and
\bg
\label{nbeinexp}
{e^\alpha}_\beta(\theta) &=& \int_0^1 ds \; \Tr \Bigl[
X_\alpha e^{-is \theta
\cdot X} X_\beta e^{is \theta \cdot X} \Bigr], \nn\\
&\approx&
\delta_{\alpha\beta} - \hf \, c_{\alpha \beta \gamma}
\theta^\gamma
+ O(\theta^2) .
\nd

With these tools it is now clear how to build $G$-invariant
couplings among the $\theta^\alpha$, and between the 
$\theta^\alpha$'s and other fields, such as $\chi$ from the
scalar-field example.

It is simplest to build self-interactions for the Goldstone
bosons. An invariant Lagrangian density may be built by
combining the covariant quantity, $e^\alpha_\mu =
{e^\alpha}_\beta \partial_\mu \theta^\beta$ in all possible
$H$-invariant ways. This is simple to do since this quantity
transforms very simply under $G$: $e_\mu 
\cdot X \to \gamma e_\mu \cdot X \gamma^{-1}$.

Derivatives of $e^\alpha_\mu$ can also be included by
differentiating using the covariant derivative constructed
from $\Sca^i_\mu t_i$:  
\eq
\label{derivofe}
(D_\mu e_\nu)^\alpha = \partial_\mu e^\alpha_\nu  +
{c^\alpha}_{i\beta}
\Sca^i_\mu \, e^\beta_\nu  , \eeq
which transforms in the same way as does $e^\alpha_\mu$: 
$\delta (D_\mu e_\nu)^\alpha = - {c^\alpha}_{i\beta} u^i
(D_\mu e_\nu)^\beta$.

The lagrangian is $\Scl(e_\mu, D_\mu e_\nu, \dots)$, where
the ellipses denote terms involving higher covariant
derivatives. Provided only that this lagrangian is
constrained to be globally $H$ invariant:  
\eq
\label{hinvcondn}
\Scl(h e_\mu h^{-1}, h D_\mu e_\nu h^{-1}, \dots) \equiv
\Scl(e_\mu, D_\mu
e_\nu, \dots),
\eeq
the result is guaranteed to be {\em automatically} globally
$G$ invariant, as required. For a Poincar\'e invariant
system, the term involving the fewest derivatives therefore
becomes:
\eq
\label{fewestdsinv}
\Scl_{\sss GB} = - \, \hf \, f_{\alpha \beta} \,
\eta^{\mu\nu} \;
e^\alpha_\mu \, e^\beta_\nu  + \hbox{(higher-derivative
terms)}.
\eeq

In this expression, positivity of the kinetic energy
implies that the matrix $f_{\alpha\beta}$ must be positive
definite. $G$-invariance dictates that it must also satisfy
$f_{\lambda\beta} {c^\lambda}_{i \alpha} + 
f_{\alpha\lambda} {c^\lambda}_{i \beta} = 0$. It
was remarked earlier that on general grounds the matrices,
$X_\alpha$, fill out a representation, $\Scr$, of the
unbroken subgroup $H$ with representation matrices given by
$(\Sct_i)_{\alpha\beta} = c_{\alpha i \beta}$, and in terms
of these matrices $G$-invariance requires the vanishing of
all of the commutator, $[\Sct_i , f]$, for all of the
generators, $\Sct_i$. If this representation, $\Scr$, of
$H$ is irreducible then, by Schur's lemma,
$f_{\alpha\beta}$ must be proportional to the unit matrix,
with positive coefficient: $f_{\alpha\beta} = F^2
\delta_{\alpha\beta}$. Otherwise, if $\Scr$ is reducible
into $n$ irreducible diagonal blocks, then $f_{\alpha
\beta}$ need only be block diagonal, with each diagonal
element being proportional to a unit matrix:
\eq
\label{redcaseforf}
f_{\alpha \beta} = \pmatrix{ F_1^2 \delta_{\alpha_1
\beta_1} & & \cr & \ddots &
\cr & & F_n^2 \delta_{\alpha_n \beta_n} \cr}, \eeq 
for $n$ independent positive constants, $F_n^2$. We see
that the lowest-dimension terms in the most general
low-energy Goldstone-boson self-coupling lagrangian is
parameterizable in terms of these $n$ constants.

If other fields --- denoted here collectively by $\chi$ ---
also appear in the low-energy theory then, since the
symmetry $H$ is not broken by the ground state, the fields
$\chi$ must also transform linearly under $H$: $\chi \to h
\chi$, where the matrices $\{ h \}$ form a (possibly
reducible) representation of $H$. In this case the starting
point for inferring the coupling of the Goldstone bosons is
an arbitrary, globally $H$-invariant lagrangian: 
$\Scl(\chi, \partial_\mu \chi, \dots)$ with $ \Scl(h \chi,
h 
\partial_\mu \chi, \dots) \equiv \Scl(\chi, \partial_\mu
\chi, \dots)$, for $\partial_\mu h = 0$. This lagrangian
will be automatically promoted to become $G$-invariant by
appropriately coupling the Goldstone bosons.

The promotion to $G$ invariance proceeds by assigning to
$\chi$ the nonlinear $G$-transformation rule: $\chi \to
\gamma \chi$, where $\gamma = \gamma(\theta,g) \in H$ is
the field-dependent $H$ matrix which is defined by the
nonlinear realization, 
eq.~\pref{gbtransruletwo}. An arbitrary globally
$H$-invariant 
$\chi$-lagrangian then becomes $G$ invariant if all
derivatives are replaced by the $\theta$-dependent
covariant derivative: 
$\partial_\mu \chi \to D_\mu \chi = \partial_\mu \chi - i
\Sca^i 
t_i \chi$, for which $D_\mu \chi \to \gamma D_\mu \chi$.

The general lagrangian therefore becomes: $\Scl(e_\mu,
\chi, 
D_\mu e_\nu, D_\mu \chi, \dots)$, where $G$-invariance is
ensured provided only that $\Scl$ is constrained by global
$H$ invariance:
\eq
\label{genlagrinvc}
\Scl(h e_\mu h^{-1}, h \, \chi, h D_\mu e_\nu h^{-1}, h \,
D_\mu \chi, \dots)
\equiv \Scl(e_\mu, \chi, D_\mu e_\nu, D_\mu \chi, \dots).
\eeq

\section{Uniqueness}

The previous construction certainly defines a $G$-invariant
lagrangian for the interactions of the Goldstone bosons
which arise from the symmetry-breaking pattern $G \to H$,
given the transformation rules which were derived in
earlier sections. Our goal in the present section is to
show that this construction gives the most general such
invariant lagrangian. That is, we wish to show that the
most general lagrangian density which is invariant under
the transformation rules of eqs.~\pref{gbtransruletwo} may
be constructed using only the quantities
$e^\alpha_\mu(\theta)$ and $\Sca^i_\mu(\theta)$ in addition
to any other fields, $\chi$.

We start with a general lagrangian density,
$\Scl(\theta,\partial_\mu \theta, \chi,\partial_\mu \chi)$,
involving the fields $\theta^\alpha$, $\chi^n$ and their
derivatives. We do not include a dependence on second and
higher derivatives of these fields, but this extension is
straightforward to make along the lines that are described
in this section. It is more convenient in what follows to
trade the assumed dependence of $\Scl$ on
$\partial_\mu\theta$ for a dependence on the combinations
$e^\alpha_\mu = 
{e^\alpha}_\beta(\theta) \; \partial_\mu \theta^\beta$ and 
$\Sca^i_\mu = \Sca^i_\alpha(\theta) \; \partial_\mu
\theta^\alpha$. There is no loss of generality in doing so,
since any function of 
$\theta$ and $\partial_\mu \theta$ can always be written as
a function of $\theta$, $e^\alpha_\mu$ and $\Sca^i_\mu$.
This equivalence is most easily seen in terms of the matrix
variable $U(\theta) = e^{ i \theta \cdot X}$. Any function
of $\theta$ and $\partial_\mu \theta$ can equally well be
written as a function of $U$ and $\partial_\mu U$, or
equivalently as a function of $U$ and $U^{-1} \partial_\mu
U$. But an arbitrary function of 
$U^{-1} \partial_\mu U$ is equivalent to a general function
of 
$e^\alpha_\mu$ and $\Sca^i_\mu$, as may be seen from
expression \pref{decompn}.

The condition that a general function, $\Scl(\theta^\alpha, 
e^\alpha_\mu, \Sca^i_\mu, \chi, \partial_\mu \chi)$, be
invariant with respect to $G$ transformations is:
\eq
\label{ginvconditionforl}
\delta \Scl = {\partial \Scl \over \partial \theta^\alpha}
\; \delta
\theta^\alpha + {\partial \Scl \over \partial e^\alpha_\mu}
\; \delta
e^\alpha_\mu + {\partial \Scl \over \partial \Sca^i_\mu} \;
\delta
\Sca^i_\alpha + {\partial \Scl \over \partial \chi^n} \;
\delta
\chi^n + {\partial \Scl \over \partial ( \partial_\mu
\chi^n)} \;
\delta \partial_\mu \chi^n = 0.
\eeq
We first specialize to the special case where the symmetry
transformation lies in $H$: $g = e^{i \eta \cdot t} \in H$.
In this case we must use, in eq.~\pref{ginvconditionforl},
the transformations: 
\bg
\label{htransfns}
&& \delta \theta^\alpha = - {c^\alpha}_{i\beta} \eta^i
\theta^\beta, \qquad
\delta e^\alpha_\mu = - {c^\alpha}_{i\beta} \eta^i
e^\beta_\mu, \qquad
\delta \Sca^i_\mu = - {c^i}_{jk} \eta^i \Sca^k_\mu, \nn\\
&& \quad \hbox{and}
\qquad \delta \chi^n = i \eta^i (t_i \chi)^n , \qquad
\delta \partial_\mu \chi^n = i \eta^i (t_i \partial_\mu
\chi)^n.
\nd
Requiring $\delta\Scl = 0$ for all possible transformation
parameters, $\eta^i$, then implies the following identity: 
\eq
\label{hinvcondnforl}
{\partial \Scl \over \partial \theta^\alpha} \;
{c^\alpha}_{i\beta}
\theta^\beta + {\partial \Scl \over \partial e^\alpha_\mu}
\;
{c^\alpha}_{i\beta} e^\beta_\mu + {\partial \Scl \over
\partial
\Sca^j_\mu} \;{c^j}_{ik}\Sca^k_\mu - {\partial \Scl \over
\partial
\chi^n} \; i (t_i \chi)^n - {\partial \Scl \over \partial
(\partial_\mu
\chi^n)} \; i (t_i \partial_\mu \chi)^n = 0. \eeq
This identity simply states that $\Scl$ must be constructed
to be an $H$-invariant function of its arguments, all of
which transform linearly with respect to $H$
transformations.

For the remaining symmetry transformations which do not lie
in $H$, $g = e^{i \omega \cdot X} \in G/H$, we instead
evaluate eq.~\pref{ginvconditionforl} using the following
transformations:
\bg
\label{gmodhtransfns}
&& \delta \theta^\alpha = {\xi^\alpha}_{\beta}
\omega^\beta, \qquad
\delta e^\alpha_\mu = - {c^\alpha}_{i\beta} u^i
e^\beta_\mu, \qquad
\delta \Sca^i_\mu = \partial_\mu u^i - {c^i}_{jk} u^j
\Sca^k_\mu, \nn\\
&& \quad \hbox{and} \qquad \delta \chi^n = i u^i (t_i
\chi)^n \qquad
\delta \partial_\mu \chi^n = i u^i (t_i \partial_\mu
\chi)^n ,
\nd
where $\xi^\alpha = {\xi^\alpha}_\beta(\theta) \,
\omega^\beta$ and $u^i = u^i_\alpha(\theta) \,
\omega^\alpha$ are the nonlinear functions of $\theta$ that
are defined by 
eq.~\pref{gbtransruletwo}. Using these in 
eq.~\pref{ginvconditionforl}, and simplifying the resulting
expression using eq.~\pref{hinvcondnforl}, leads to the
remaining condition for $G$ invariance:
\eq
\label{gmodhinvcond}
{\partial \Scl \over \partial \theta^\alpha} \; 
\Bigl({\xi^\alpha}_\beta +
{c^\alpha}_{i\gamma} u^i_\beta \theta^\gamma \Bigr) +
{\partial
\Scl \over \partial \Sca^j_\mu} \; \partial_\mu u^i_\beta +
{\partial \Scl
\over
\partial (\partial_\mu \chi^n)} \; i \partial_\mu u^i_\beta
(t_i \chi)^n = 0.
\eeq

This last identity contains two separate pieces of
information. The first piece can be extracted by
specializing to $\theta^\alpha = 0$. In this case, since
$\partial_\mu u^i_\beta = \partial_\alpha 
u^i_\beta \; \partial_\mu \theta^\alpha$ vanishes when 
$\theta^\alpha = 0$, and since eq.~\pref{nonabelinhomo}
implies ${\xi^\alpha}_\beta(\theta = 0) =
\delta^\alpha_\beta$, we find:
\eq
\label{indofthetacond}
\left. { \partial \Scl \over \partial \theta^\alpha}
\right|_{\theta = 0} = 0.
\eeq
But, since the group transformation law for $\theta^\alpha$
is inhomogeneous, we may always perform a symmetry
transformation to ensure that $\theta^\alpha = 0$ for {\em
any} point $p \in G/H$. As a result,
eq.~\pref{indofthetacond} implies the more general
statement:
\eq
\label{indofthetacondgen}
{ \partial \Scl \over \partial \theta^\alpha}\equiv 0
\qquad 
\hbox{throughout}
\qquad G/H.
\eeq

The rest of the information contained in 
eq.~\pref{gmodhinvcond} may now be extracted by using 
$\partial \Scl / \partial \theta^\alpha = 0$ to eliminate
the first term. One finds the remaining condition:  
\eq
\label{covderivcond}
\left( {\partial \Scl \over \partial \Sca^j_\mu} + 
{\partial \Scl \over \partial
(\partial_\mu \chi^n)} \; i(t_j \chi)^n \right)  
\partial_\mu u^j_\beta = 0.
\eeq
This equation has a very simple meaning. It states that
$\Scl$ can depend on the two variables, $\Sca^j_\mu$ 
and $\partial_\mu \chi^n$, only through the one 
combination: $(D_\mu \chi)^n \equiv \partial_\mu \chi^n 
- i \Sca^j_\mu (t_j \chi)^n$. That is, 
$\chi$ can appear differentiated in $\Scl$ only through its
covariant derivative, $D_\mu \chi$.

We see from these arguments that the $G$-invariance of
$\Scl$ is equivalent to the statement that $\Scl$ must be
an $H$-invariant function constructed from the
covariantly-transforming variables $e^\alpha_\mu$, $\chi$
and $D_\mu \chi$. If higher derivatives of $\theta$ had
been considered, then the vanishing of the terms in 
$\delta \Scl$ that are proportional to more than one
derivative of $u^i$ would similarly imply that derivatives
of $e^\alpha_\mu$ must also only appear through its
covariant derivative, $(D_\mu e_\nu)^\alpha$, defined by
eq.~\pref{derivofe}. This proves the uniqueness of the
construction of invariant lagrangians using these covariant
quantities.

\section{The Geometric Picture}

There is an appealing geometric description of the
resulting effective lagrangian, which makes available many
powerful techniques from differential geometry to effective
lagrangian methods. We pause here to briefly outline this
connection.

Consider, for simplicity, only the self-interactions of the
Goldstone bosons: $\Scl_{\sss GB}(\theta)$. This can be
expanded into terms having increasing numbers of
derivatives acting on $\theta$. The first few terms that
are consistent with Poincar\'e invariance are:  
\eq
\label{gblagronly}
\Scl_{\sss GB}(\theta) = - V(\theta) - \hf \, 
g_{\alpha\beta}(\theta) \;
\partial_\mu \theta^\alpha \partial^\mu \theta^\beta +
\cdots .
\eeq
Positivity of the kinetic energy for fluctuations about any
configuration, $\theta^\alpha$, requires the symmetric
matrix $g_{\alpha\beta}$ to be positive definite for all
$\theta$.

The geometrical interpretation arises once we recall that
the fields $\theta$ take values in the coset space $G/H$,
and so each Goldstone boson field configuration can be
considered to be a map from spacetime into $G/H$. The
function, $V$, can then be considered to be a real-valued
function which is defined on the space $G/H$. Similarly,
the positive symmetric matrix, 
$g_{\alpha\beta}$, defines a metric tensor on $G/H$. These
identifications of $V$ and $g_{\alpha\beta}$ with
geometrical objects on $G/H$ are consistent with their
transformation properties under field redefinitions,
$\delta \theta^\alpha = 
\xi^\alpha(\theta)$, which are the analogues of coordinate
transformations on $G/H$. To see this, perform this
transformation in the lagrangian of eq.~\pref{gblagronly}.
The result is to replace $V$ and $g_{\alpha\beta}$ by the
quantities $V + \Lie{\xi} V$ and $g_{\alpha\beta} +
\Lie{\xi} g_{\alpha
\beta}$, where the linear operator, $\Lie{\xi}$, is known
as the Lie derivative in the direction specified by
$\xi^\alpha$, and is given explicitly for a scalar and a
covariant tensor by:
\bg
\label{coordtransfrules}
\Lie{\xi} V &=& \xi^\alpha \partial_\alpha V, \nn\\ 
\hbox{and} \qquad \Lie{\xi}
g_{\alpha\beta} &=& \xi^\lambda \partial_\lambda
g_{\alpha\beta} + g_{\lambda \beta} \partial_\alpha 
\xi^\lambda + g_{\alpha
\lambda} \partial_\beta \xi^\lambda.
\nd
In these expressions derivatives, like $\partial_\alpha V$, 
$\partial_\lambda g_{\alpha\beta}$ or $\partial_\alpha 
\xi^\lambda$, all represent differentiation with respect to 
$\theta^\alpha$, and not with respect to spacetime
position, 
$x^\mu$.

Clearly, the $G$ invariance of the first few terms of
$\Scl_{\sss GB}$ therefore becomes equivalent to the
problem of finding a scalar and a metric for which
$\Lie{\xi} V = \Lie{\xi} g_{\alpha
\beta} = 0$ for each of the $\xi^\alpha(\theta)$'s which
describe the action of $G$ on $G/H$. Since every point on
$G/H$ can be reached from any other by performing such a
$G$ transformation --- \ie\ $G/H$ is a {\em homogeneous
space} --- it follows that the only possible invariant
function, $V$, is a constant which is independent of
$\theta$. This expresses the low-energy decoupling of the
Goldstone bosons since it shows that $G$ invariance
completely forbids their appearance in the scalar potential.

Similarly, the condition for the $G$ invariance of the
kinetic terms is that the metric $g_{\alpha\beta}$ must
also be invariant under the action of all of the
$\xi^\alpha(\theta)$'s which generate $G$ on $G/H$. That
is, all of these $\xi$'s must generate {\em isometries} of
the metric $g_{\alpha\beta}$. The problem of finding the
most general invariant kinetic term is therefore equivalent
to constructing the most general $G$-invariant metric on
$G/H$. A comparison of the lagrangian of 
eq.~\pref{gblagronly}, with our earlier result, 
eq.~\pref{fewestdsinv}, gives a representation of the
metric 
$g_{\alpha\beta}$ in terms of the quantities
${e^\alpha}_\beta
(\theta)$. We have
\bg
\label{itsannbein}
g_{\alpha\beta} &=& f_{\gamma\lambda} \; 
{e^\gamma}_\alpha {e^\lambda}_\beta ,
\nn\\
&=& \sum_{r=1}^n F_r^2 \, \delta_{\gamma_r 
\lambda_r} \; {e^{\gamma_r}}_\alpha
{e^{\lambda_r}}_\beta , \\
&\approx& \sum_{r=1}^n F_r^2 \, \Bigl[ 
\delta_{\alpha_r \beta_r} -
c_{\alpha_r \beta_r \gamma} \theta^\gamma + 
O(\theta^2) \Bigr] . \nn
\nd

Eq.~\pref{itsannbein} also has
a geometric interpretation. It shows that the object
${e^\alpha}_\beta$ can be interpreted as a $G$-covariant
{\em vielbein} for the space $G/H$. In Reimannian
geometry a vielbein is the name given to a set of $N$ 
linearly independent vectors, ${e^a}_\beta$, $a=1,
\dots,N$, which are tangent to an $N$-dimensional space. 
(The name is German for `many legs', with {\em viel} 
meaning `many', and so such vectors are also called
{\em zweibein} in two dimensions --- with {\em zwei} = 
`two' --- or {\em vierbein} in four dimensions --- with 
{\em vier} = `four'.) Part of the utility of identifying
such a set of vectors is that it is always possible to 
reconstruct from them the space's metric, using: 
$g_{\alpha\beta} = \delta_{ab} \, {e^a}_\alpha
\, {e^b}_\beta$. 

The geometrical interpretation of ${e^\alpha}_\beta$ as a
vielbein, and the uniqueness of the construction of
invariant lagrangians proven in the previous section, gives
the general solution to the geometrical problem of constructing
$G$-invariant metrics on the space 
$G/H$. We see that there is an $n$-parameter family of such
metrics, where $n$ counts the number of irreducible
representations of $H$ which are formed by the generators, 
$X_\alpha$, of $G/H$. The $n$ parameters are given
explicitly by the constants $F_r^2, r=1,\dots,n$.

For many physical applications the representation of $H$
that is furnished by the $X_\alpha$ is irreducible, and in
this case the $G$-invariant metric is uniquely determined
up to its overall normalization: $g_{\alpha\beta} = F^2
\hat{g}_{\alpha\beta}$, with $\hat{g}_{\alpha\beta} =
\delta_{\alpha\beta} + O(\theta)$. For such systems there
is precisely one constant in the effective lagrangian for
Goldstone bosons which is undetermined by the symmetries of
the problem, if we include only the fewest possible (two) 
derivatives:
\eq
\label{genlowdterm}
\Scl_{\sss GB} = - \, {F^2 \over 2} \;
\hat{g}_{\alpha\beta}(\theta) \;
\partial_\mu \theta^\alpha \partial^\mu \theta^\beta 
+ \hbox{(higher derivative terms)} .
\eeq
Once the one constant, $F^2$, is determined, either by
calculation from an underlying theory, or by appeal to
experiment, the lowest-order form for all of the Goldstone
boson interactions are completely determined by the
symmetry breaking pattern. The resulting model-independent
predictive power has wide applications throughout physics,
as we shall see when we consider examples in subsequent
chapters.

\section{Nonrelativistic Lagrangians}

Before turning to examples, we pause to outline the changes
in the above analysis which become necessary when it is
applied to nonrelativistic systems, for which the spacetime
symmetry is not Poincar\'e invariance. This is what is
appropriate, for example, in condensed-matter applications
for which there is a preferred frame, defined by the
centre-of-mass of the medium of interest.

So long as our attention is restricted to internal
symmetries, most of the considerations of this chapter
apply just as well to nonrelativistic problems as they do
to relativistic ones. In particular, the expressions
obtained for the nonlinear realization of broken symmetries
on the Goldstone boson fields does not depend at all on the
spacetime symmetries which are assumed.

We assume for simplicity here that the system remains
invariant with respect to translations and rotations,
although the generalization to different spacetime groups
is straightforward in principle. In practice, the results
we obtain also apply to some lattice systems, for which
translations and rotations are not symmetries. This is
because, at least for the first few derivatives in a
derivative expansion of the lagrangian, the lattice group
for some lattices, such as a cubic lattice for example,
implies the same restrictions as do translation and
rotation invariance.

There are two cases, which we consider separately below,
depending on whether or not time reversal is a good 
symmetry of the system. 

\subsection{When Time-Reversal is a Good Symmetry}

For unbroken time reversal, the main difference from the
relativistic situation lies in the fact that the time- and
space-derivatives become independent since they are
unrelated by any symmetries. For instance, assuming
unbroken translation and rotation invariance, the most
general self-couplings amongst the Goldstone bosons for the
symmetry-breaking pattern $G \to H$ are:
\bg
\label{genlowdtermnr}
\Scl_{\sss GB} &=& - \hf \sum_{r=1}^n \Bigl[ F_{(r),t}^2 \;
\hat{g}^{(r)}_{\alpha\beta}(\theta) \; \dot \theta^\alpha 
\dot \theta^\beta +
\,
F_{(r),s}^2 \; \hat{g}^{(r)}_{\alpha\beta}(\theta) \; \del  
\theta^\alpha \cdot
\del \theta^\beta \Bigr] \nn\\ && \qquad \qquad 
\qquad + \hbox{(higher
derivative terms)} .
\nd
Compared to the relativistic case, twice as many --- that
is, $2n$ --- constants, $F_{(r),t}^2$ and $F_{(r),s}^2$,
are required to parameterize the terms containing the
fewest number of derivatives. Once these two constants are
determined, all other interactions at this order of the
derivative expansion are clearly completely determined.

\subsection{When Time-Reversal is Broken}

New possibilities arise for the lagrangian when
time-reversal symmetry is broken, as is the case for a
ferromagnet, for example. In this case, it is possible to
write down terms in $\Scl_{\sss GB}$ which involve an odd
number of time derivatives. In particular, the dominant,
lowest-dimension term involving time derivatives involves
only one:
\eq
\label{tvterm}
\Delta \Scl_{\sss GB} = - A_\alpha(\theta) \; \dot
\theta^\alpha.
\eeq
The coefficient function, $A_\alpha(\theta)$, can be
considered to define a vector field on the coset space
$G/H$. It is to be chosen to ensure the $G$ invariance of
the low-energy theory.

If $\Delta \Scl_{\sss GB}$ is required to be invariant
under $G$ transformations, then it must be built using 
the quantity ${e^\alpha}_\mu(\theta) = {e^\alpha}_\beta(\theta)
\, \partial_\mu\theta^\beta$ and its covariant derivatives. If it is to
involve only a single time derivative, then it must be
proportional only to $e_0(\theta)$. But the only such
$G$-invariant quantity is: 
$k_\alpha {e^\alpha}_\beta \dot \theta^\beta$, where the
constants, $k_\alpha$, must satisfy $k_\alpha
{c^\alpha}_{i\beta} = 0$ for all indices $i$ and $\beta$.
Such a $k_\alpha$ exists only if the corresponding
generator, $X_\alpha$, commutes with all of the generators
of the unbroken subgroup, $H$. This quite restrictive
condition is not ever satisfied in many situations of
physical interest, and for these systems it appears --- at
least superficially --- that no terms involving only a
single time derivative are consistent with $G$-invariance.

This conclusion would be too strong, however, because it is
too restrictive a condition to demand the $G$-invariance of
the lagrangian density, $\Scl$. We are only required by $G$
symmetry to demand the invariance of the action. The
lagrangian density need not be invariant, provided that its
variation is a total derivative. It is therefore worth
re-examining the time-reversal-violating term of
eq.~\pref{tvterm} in this light.

Once we drop total derivatives in $\Delta \Scl_{\sss GB}$,
it is clear that the coefficient $A_\alpha(\theta)$ is only
defined up to the addition of a gradient. That is, any two
choices $A_\alpha$ and $\tw A_\alpha \equiv A_\alpha +
\partial_\alpha \Omega(\theta)$, differ in their
contribution to $\Delta \Scl_{\sss GB}$ only by the total
derivative:  
\eq
\label{totderivdiff}
\Delta \Scl_{\sss GB}(A+\partial\Omega) - 
\Delta \Scl_{\sss GB}(A) =
- \partial_\alpha \Omega(\theta) \; \dot \theta = - 
\, \dot \Omega(\theta).
\eeq
In geometrical terms we may therefore regard the
coefficient function, $A_\alpha(\theta)$, as defining a
gauge potential on the coset space $G/H$.

The condition that $\Delta \Scl_{\sss GB}$ contribute a
$G$-invariant term to the action therefore only requires
the coefficient $A_\alpha(\theta)$ to be $G$-invariant {\em
up to a gauge transformation}. In equations, $G$-invariance
of the action only requires:
\eq
\label{ginvcondfora}
\Lie{\xi} A_\alpha \equiv \xi^\beta 
\partial_\beta A_\alpha + A_\beta
\partial_\alpha \xi^\beta = \partial_\alpha 
\Omega_\xi, \eeq
for each generator $\delta\theta^\alpha = \xi^\alpha$ of
the action of $G$ on $G/H$, and for some scalar functions, 
$\Omega_\xi(\theta)$, on $G/H$. This last condition is
equivalent to demanding the invariance of the
gauge-invariant quantity: 
$F_{\alpha\beta} = \partial_\alpha A_\beta - \partial_\beta 
A_\alpha$. That is,
\eq
\label{ginvcondforf}
\Lie{\xi} F_{\alpha\beta} \equiv \xi^\lambda 
\partial_\lambda F_{\alpha\beta}
+ F_{\lambda \beta} \partial_\alpha \xi^\lambda 
+ F_{\alpha \lambda}
\partial_\beta \xi^\lambda = 0.
\eeq

We shall find that this condition {\em does} admit
solutions in many cases of interest --- most notably for
the example of a ferromagnet.

\section{Power Counting}

Before proceeding to real-life applications, a final
important issue must be addressed.  Since the lagrangians
expressed using Goldstone boson variables is typically
nonrenormalizable, it is necessary to know how to use
nonrenormalizable lagrangians  when making quantitative
calculations. 

The key to doing so is to consider the Goldstone Boson
lagrangians to which we have been led in previous sections
to be `effective theories' which describe only the
low-energy behaviour of the system of interest. For
instance, in our toy models the Goldstone bosons
($\theta^\alpha$) are massless while the other degrees of
freedom ($\chi$) are not. (Although the pseudo-Goldstone
bosons for an approximate symmetry are not exactly
massless, they may nonetheless appear in the low-energy
theory so long as their mass, $m$, is  sufficiently small.)
A lagrangian involving only Goldstone bosons or
pseudo-Goldstone bosons can only hope to describe physics
at energies, $q$, below the mass threshhold, $M$, for
producing the heavier particles. (It is often the case that
$M$  is proportional to the symmetry-breaking scale(s),
$v$). The predictions of such a lagrangian are to be
regarded as reproducing, in powers of $q/M$, whatever the
`underlying' (or `microscopic') theory --- \ie\ the theory
involving the heavy $\chi$ states --- might be.

In order to make this concrete, consider one such a
Lagrangian, having the form:
\eq
\label{leffpc}
\leff = f^4 \sum_n {c_n \over M^{d_n}} \; 
\Sco_n \left( {\phi \over v}
\right) ,
\eeq
where $\phi$ denotes a generic boson field, $c_n$ are a set
of dimensionless coupling constants which we imagine to be
at most $O(1)$, and $f, M$ and $v$ are mass scales of the
underlying problem. (For example, in the application to
pions which follows we will have $f = \sqrt{\fpi \L_\chi}$,
$M = \L_\chi$ and $v = 
\fpi$, where $\fpi$ and $\L_\chi$ are scales which
characterize the strength of the appropriate symmetry
breaking in the strong interactions.) $d_n$ is the
dimension of the operator $\Sco_n$, in powers of mass, as
computed by counting only the dimensions of the  field,
$\phi$, and derivatives, $\partial\phi$.

Imagine using this lagrangian to compute a scattering
amplitude, 
$\Sca_\sse(q)$, involving the scattering of $E$ particles
whose four-momenta are collectively denoted by $q$. We wish
to focus on the contribution to $\Sca$ due to a Feynman
graph having $I$ internal lines and $V_{ik}$ vertices. The
labels $i$ and $k$ here indicate two characteristics of the
vertices: $i$ counts the number of lines which converge at
the vertex, and $k$ counts the power of momentum which
appears in the vertex. Equivalently, $i$ counts the number
of powers of the fields, $\phi$, which appear in the
corresponding interaction term in the lagrangian, and $k$
counts the number of derivatives of these fields which
appear there.

\subsection{Some Useful Identities}

The positive integers, $I$, $E$ and $V_{ik}$, which
characterize the Feynman graph in question are not all
independent since they are related by the rules for
constructing graphs from lines and vertices. This relation
can be obtained by equating two equivalent methods of
counting the number of ways that internal and external
lines can end in a graph. On the one hand, since all lines
end at a vertex, the number of ends is given by summing
over all of the ends which appear in all of the vertices:
$\sum_{ik} i 
\, V_{ik}$. On the other hand, there are two ends for each
internal line, and one end for each external line in the
graph: $2 I + E$. Equating these gives the identity which
expresses the `conservation of ends':
\eq
\label{consofends}
2 I + E = \sum_{ik} i \,  V_{ik}, \qquad
\hbox{(Conservation of Ends)}.
\eeq

A second useful identity {\em defines} of the number of
loops, $L$, for each (connected) graph:
\eq
\label{loopdef}
L = 1 + I - \sum_{ik} V_{ik}, 
\qquad \hbox{(Definition of $L$)}.
\eeq
For simple planar graphs, this definition agrees with the
intuitive notion what the number of loops in a graph means.

\subsection{Dimensional Estimates}

We now collect the dependence of $\Sca_\sse(a)$ on the 
parameters in $\leff$.

Reading the Feynman rules from the lagrangian of
eq.~\pref{leffpc} shows that the vertices in the Feynman
graph contribute the following factor:
\eq
\label{vertexcont}
\hbox{(Vertex)} =  \prod_{ik} \left[ 
i (2 \pi)^4 \delta^4(p) \; \left( {p
\over M} \right)^k \; \left( {f^4 \over v^i} 
\right) \right]^{V_{ik}},
\eeq
where $p$ generically denotes the various momenta running
through the vertex.

Similarly, each internal line in the graph contributes the
additional factors:
\eq
\label{internallinecont}
\hbox{(Internal Line)} = \left[ -i \int {d^4 p 
\over (2 \pi)^4} \;
\left( {M^2 v^2 \over f^4} \right) \; {1 
\over p^2 + m^2} \right]^I,
\eeq
where, again, $p$ denotes the generic momentum flowing
through the line. $m$ denotes the mass of the light
particles which appear in the effective theory, and it is
assumed that the kinetic terms which define their
propagation are those terms in $\leff$ involving two
derivatives and two powers of the fields, $\phi$.

As usual for a connected graph, all but one of the
momentum-conserving delta functions in
eq.~\pref{vertexcont} can be used to perform one of the
momentum integrals in 
eq.~\pref{internallinecont}. The one remaining delta
function which is left after doing so depends only on the
external momenta, $\delta^4(q)$, and expresses the overall
conservation of four-momentum for the process. Future
formulae are less cluttered if this factor is extracted
once and for all, by defining the reduced amplitude,
$\tilde \Sca$, by
\eq
\label{redampdef}
\Sca_\sse(q) = i (2 \pi)^4 \delta(q) \; \tilde\Sca_\sse(q).
\eeq

The number of four-momentum integrations which are left
after having used all of the momentum-conserving delta
functions is then $I - \sum_{ik} V_{ik} + 1 = L$. This last
equality uses the definition, eq.~\pref{loopdef}, of the
number of loops, $L$.

We now wish to estimate the result of performing the
integration over the internal momenta. In general these
are complicated integrals for which a simple result is not
always possible to give. Considerable simplifications arise,
however, if all of the masses and energies of the
particles in the low-energy theory are of the same order
of magnitude, since in this case much can be said about
the order of magnitude of the momentum integrals purely on 
dimensional grounds. (Although this is often the situation
of interest when employing effective theories, it must be
borne in mind that it does not always apply. For instance
it excludes a situation of considerable practical interest, 
where the low-energy theory includes very massive but 
slowly-moving, nonrelativistic particles. Power counting
for such systems is beyond the scope of this review.)

In order to proceed with a dimensional argument
it is most convenient to regulate the 
ultraviolet divergences which arise in the momentum
integrals using dimensional regularization. For
dimensionally-regularized integrals, the key observation
is that the size of the result  is set on dimensional
grounds by the light masses or external momenta of the
theory. That is, if all external energies, $q$, are
comparable to (or larger than) the masses, $m$, of the
light particles whose scattering is being calculated, then
$q$ is the light scale controlling the size of the
momentum integrations, so dimensional analysis implies that
an estimate of the size of the momentum integrations is:
\eq
\label{newdimgrounds}
\int \cdots \int \left( {d^n p\over (2 \pi)^n} \right)^A 
\; {p^B \over (p^2 +
q^2)^C }  \sim \left( {1 \over 4 \pi} \right)^{2A}
q^{nA + B - 2C} ,
\eeq
with a dimensionless prefactor which carries all of the
complicated dependence on dimensionless ratios
like $q/m$. The prefactor also depends on the dimension, $n$, 
of spacetime, and may be singular in the limit that 
$n \to 4$.\footnote{We ignore here any
logarithmic infrared mass singularities which may arise in
this limit.} 

With this estimate for the size of the momentum
integrations, we find the following contribution to the
amplitude $\tilde 
\Sca_\sse(q)$:
\eq
\label{intcontribution}
\int \cdots \int \left( {d^4 p\over (2 \pi)^4} 
\right)^L \; {p^{\sum_{ik} k
V_{ik}} \over (p^2 + q^2)^I }  \sim \left( {1 
\over 4 \pi}
\right)^{2L} q^{4L - 2I + \sum_{ik} k V_{ik}} , \eeq
which, with liberal use of the identities 
\pref{consofends} and \pref{loopdef}, 
gives as estimate for $\tilde\Sca_\sse(q)$:
\eq
\label{aedwdimreg}
\tilde\Sca_\sse(q) \sim f^4 \; \left( {1 \over v} 
\right)^E \; \left( {M^2
\over
4 \pi f^2} \right)^{2L} \; \left( {q \over M} 
\right)^{2 + 2L + \sum_{ik} (k -
2) V_{ik}} .
\eeq
This expression is the principal result of this section. 
Its utility lies in the fact that it links the
contributions of the various effective interactions in the
effective lagrangian, \pref{leffpc}, with the dependence of
observables on small mass ratios such as $q/M$. As a result
it permits the determination of which interactions in the
effective lagrangian are required to reproduce any given
order in $q/M$ in physical observables.

Most importantly, eq.~\pref{aedwdimreg}  shows how  to
calculate using nonrenormalizable theories. It implies that
even though the lagrangian can contain arbitrarily many
terms, and so potentially arbitrarily many  coupling
constants, it is nonetheless predictive {\it so long as 
its predictions are only made for low-energy processes, for
which $q/M \ll 1$}.  (Notice also that the factor
$(M/f)^{4L}$ in eq.~\pref{aedwdimreg} implies,  all other
things being equal, the scale $f$ cannot be taken to be
systematically smaller than $M$ without ruining the
validity of the loop expansion in the effective low-energy
theory.)

Before stating more explicitly the effective-lagrangian
logic, which eq.~\pref{aedwdimreg} summarizes, we pause to
generalize it to include fermions in the low-energy
effective theory.

\subsection{Including Fermions}

It is straightforward to extend these results to include
light fermions in the effective theory, although once again
subject to the important assumption that all masses and
energies are small in the effective theory. To this end, first
generalize the starting form assumed for the lagrangian to
include fermion fields, $\psi$, in addition to boson
fields, $\phi$:
\eq
\label{leffpcwfermions}
\leff = f^4 \sum_n {c_n \over M^{d_n}} \; 
\Sco_n \left( {\phi \over v_\ssb},
{\psi \over v_\ssf^{3/2}} \right).
\eeq

An important difference between fermion and boson
propagators lies in the way each falls off for large
momenta. Whereas a boson propagator varies like $1/p^2$ for
large $p$, a fermion propagator goes only like $1/p$. This
leads to a difference in their contribution to the power
counting of a Feynman graph. It is therefore important to
keep separate track of the number of fermion and boson
lines, and we therefore choose to now label vertices using
{\em three} indices: $k$, $i_\ssb$ and $i_\ssf$. As before,
$k$ labels the numbers of derivatives in the corresponding
interaction, but now $i_\ssb$ and $i_\ssf$ separately count
the number of bose and fermi lines which terminate at the
vertex of interest. The number of vertices in a graph which
carry a given value for $k$, $i_\ssb$ and $i_\ssf$ we now
label by $V_{i_\ssb i_\ssf k}$.

Consider now computing an amplitude which has $E_\ssb$
external bosonic lines, $E_\ssf$ external fermion lines,
and 
$I_\ssb$ and $I_\ssf$ internal bose and fermi lines.
Repeating, with the lagrangian of
eq.~\pref{leffpcwfermions}, the power counting argument
which led (using dimensional regularization) to
eq.~\pref{aedwdimreg} now gives instead the following
result:
\eq
\label{pcresultwfermions}
\tilde\Sca_{E_\ssb,E_\ssf}(q) \sim f^4 \; 
\left( {1 \over v_\ssb}
\right)^{E_\ssb}
\; \left( {1 \over v_\ssf} \right)^{3E_\ssf/2}  \; 
\left( {M^2 \over 4 \pi f^2}
\right)^{2L} \; \left( {q \over M} \right)^P, \eeq
where the power $P$ can be written in either of the
following two equivalent ways:
\bg
\label{equivwaysforp}
P &=& 4 - E_\ssb - {3\over 2} \, E_\ssf + 
\sum_{i_\ssb,i_\ssf,k} \left( k +
i_\ssb + {3 \over 2} \, i_\ssf - 4 \right) 
V_{i_\ssb i_\ssf k} , \nn\\
&=& 2 + 2L - \hf \, E_\ssf + \sum_{i_\ssb,i_\ssf,k} 
\left( k +\hf \, i_\ssf - 2 \right) V_{i_\ssb i_\ssf k} .
\nd

\section{The Effective-Lagrangian Logic}

The powercounting estimates just performed show how to
organize calculations using nominally nonrenormalizable
theories, considering them as effective field theories. 
They suggest the following general logic concerning 
their use.

\begin{description}

\item[Step I]
Choose the accuracy (\eg\ one part per mille) with which
observables, such as $\Sca_\sse(q)$, are to be computed.

\item[Step II]
Determine the order in the small mass ratios $q/M$ or $m/M$
that must be required in order to acheive the desired
accuracy.

\item[Step III]
Use the power counting result, eq.~\pref{aedwdimreg}, to
find which terms in the effective lagrangian are needed in
order to compute to the desired order in $q/M$.
Eq.~\pref{aedwdimreg} also determines which order in the
loop expansion is required for each effective interaction
of interest.

\item[Step IVa]
Compute the couplings of the required effective
interactions using the full underlying theory. If this step
should prove to be impossible, due either to ignorance of
the underlying theory or to the intractability of the
required calculation, then it may be replaced by the
following alternative:

\item[Step IVb]
If the coefficients of the required terms in the effective
lagrangian cannot be computed then they may instead be
regarded as unknown parameters which are to be taken from
experiment. Once a sufficient number of observables are
used to determine these parameters, all other observables
may be unambiguously predicted using the effective theory.

\end{description}

A number of points bear emphasizing at this point.

\begin{enumerate}

\item
The possibility of treating the effective lagrangian
phenomenologically, as in Step IVb above, immeasurably
broadens the utility of effective lagrangian techniques,
since they need not be restricted to situations for which
the underlying theory is both known and calculationally
simple. Implicit in such a program is the underlying
assumption that there is no loss of generality in working
with a local field theory. This assumption has been borne
out in all known examples of physical systems. It is based
on the conviction that the restrictions which are implicit
in working with local field theories are simply those that
follow from general physical principles, such as unitarity
and cluster decomposition.

\item
Since eq.~\pref{aedwdimreg} --- or 
eqs.~\pref{pcresultwfermions} and 
\pref{equivwaysforp} --- states that only a finite
number of terms in $\leff$ contribute to any fixed order in
$q/M$, and these terms need appear in only a finite number
of loops, it follows that only a finite amount of labour is
required to obtain a fixed accuracy in observables.

Renormalizable theories represent the special case for
which it suffices to work only to zeroeth order in the
ratio $q/M$. This can be expected to eventually
dominate at sufficiently low energies compared to
$M$, which is the reason why renormalizable 
theories play such an important role throughout 
physics.

\item
An interesting corollary of the above observations is the
fact that only a finite number of renormalizations are
required in the low-energy theory in order to make finite
the predictions for observables to any fixed order in
$q/M$. Thus, although an effective lagrangian is not
renormalizable in the traditional sense, it nevertheless
{\em is} predictive in the same way that a renormalizable
theory is.

\end{enumerate}

\chapter{Pions: A Relativistic Application}

We now present a relativistic application of these
techniques to the low-energy interactions of pions and
nucleons. This example provides a very useful, and
experimentally successful, description of the low-energy
limit of the strong interactions, and so illustrates how
effective lagrangians can remain predictive even if it is
impossible to predict their effective couplings from an
underlying theory. This example is also of historical
interest, since the study of low-energy pion scattering
comprises the context within which the above
Goldstone-boson formalism initially arose.

\section{The Chiral Symmetries of QCD}

The modern understanding of the strong interactions is
based on the theory of mutually interacting spin-half
quarks and spin-one gluons that is called Quantum
Chromodynamics (QCD). This theory is described by the
lagrangian density 
\eq
\label{qcdlagr}
\Scl_{\sss QCD} = - \, \nth{4} \, G^a_{\mu\nu} 
G_a^{\mu\nu} - \sum_n \qbr_n
(\Dslsh + m_{q_n}) q_n,
\eeq
where $G^a_{\mu\nu} = \partial_\mu G^a_\nu - \partial_\nu 
G^a_\mu + g {f^a}_{bc} \, G^b_\mu g^c_\nu$ is the field
strength tensor for the gluon fields, $G^a_\mu$. Here 
$a=1,\dots,8$ labels the generators of the gauge symmetry
group of the theory, $SU_c(3)$, for which the ${f^a}_{bc}$
are the structure constants. The subscript `$c$' of
$SU_c(3)$ stands for `colour', which is the name given to
the strong charge.

The quarks are represented by Dirac spinors, $q_n$, where
$n=1,\cdots,6$ counts the six kinds of quarks. In order of
increasing mass, these are: $u,d,s,c,b$ and $t$. For the
purposes of later comparison we list here the quark masses,
$m_{q_n}$, in GeV: $m_u = 0.0015 - 0.005$, 
$m_d = 0.003 - 0.009$, $m_s = 0.06 - 0.17$,
$m_c = 1.1 - 1.4$, $m_b = 4.1 - 4.4$ and 
$m_t = 173.8 \pm 5.2$. All of these
quarks are assumed to transform in the three-dimensional
representation of the colour symmetry group, $SU_c(3)$, and
so their covariant derivative (which appears in the
combination $\Dslsh = \gamma^\mu 
D_\mu$ in the lagrangian) is: $D_\mu q_n = \partial_\mu q_n
-\, {i\over 2}\, g\, G^a_\mu \lambda_a q_n$. The eight matrices,
$\lambda_a$, denote the three-by-three Gell-Mann matrices,
which act on the (unwritten) colour index of each of the
quarks. The explicit form for these matrices is not
required in what follows. In all of these expressions $g$
represents the coupling constant whose value controls the
strength of the quark-gluon and gluon-gluon couplings.

The strong interactions as given by the above lagrangian
density are believed to bind the quarks and gluons into
bound states, which correspond to the observed strongly
interacting particles (or, {\em hadrons}), such as protons
($p$), neutrons ($n$), pions 
($\pi$), kaons ($K$), \etc. Table 2.1 lists the masses and
some of the quantum numbers for all of the hadrons whose
masses are less than 1 GeV. Considerably more states have
masses above 1 GeV.

\begin{table}
\begin{center}
\begin{tabular}{ccccc}
Particle & Quark Content & Mass (GeV) 
& Spin & Isospin \\ &&&& \\
$\pi^- (\pi^+) [\pi^0]$ & $d \ubr (u \dbr) 
[u\ubr,d\dbr]$ & 0.140 [0.135]& 0 & 1 \\
$\pi^0$ & $u\ubr, d \dbr$ & 0.135 & 0 & 1 \\ 
$K^+ (K^0)$ & $u \sbr (d \sbr)$ &
0.494 (0.498) & 0 & $\hf$ \\
$K^- (\ol{K}^0)$ & $s \ubr (s \dbr)$ & 
0.494 (0.498) & 0 & $\hf$ \\
$\eta$ & $u \ubr, d \dbr, s\sbr$ & 0.547 & 0 & 0 \\ 
$\rho^- (\rho^+) [\rho^0]$
& $d \ubr (u \dbr) [u\ubr, d\dbr]$ & 0.770 & 1 &
1 \\
$\omega$ & $u \ubr, d \dbr, s\sbr$ & 0.782 & 1 & 0 \\  
$K^{*+} (K^{*0})$ & $u
\sbr (d \sbr)$ & 0.892 (0.896) & 1 & $\hf$ \\
$K^{*-} (\ol{K}^{*0})$ & $s \ubr (s \dbr)$ 
& 0.892 (0.896) & 1 & $\hf$ \\
$\eta'$ & $u \ubr, d \dbr, s\sbr$ & 0.958 & 0 & 0 \\  
$f_0$ & $u \ubr, d \dbr,
s\sbr$ & 0.980 & 0 & 0 \\  $a_0$ & $u \ubr, 
d \dbr, s\sbr$ & 0.980 & 0 & 1 \\
&&&& \\
$p (n)$ & $uud (ddu)$ & 0.938 (0.940) & $\hf$ & $\hf$ \\ 
\end{tabular}
\caption{Masses and Quantum Numbers of the Lightest
Hadrons} \end{center}
\end{table}

For the present purposes the most significant feature about
this particle spectrum is that the lightest two quarks, $u$
and $d$, have masses which are much smaller than all of the
masses of the states which make up the particle spectrum.
This suggests that the QCD dynamics may be well
approximated by taking $m_u , 
m_d \approx 0$, and working perturbatively in these masses
divided by a scale, $\Lambda$, which is typical of the
strong interactions. {}From the observed bound-state
spectrum we expect $\Lambda$ to be roughly 1 GeV.

The approximation for which $m_u$ and $m_d$ vanish turns
out to be a very useful one. This is because the QCD
lagrangian acquires the very useful symmetries
\eq
\label{chiralsym}
\pmatrix{ u \cr d \cr} \to \Bigl( U_\ssl \; \Pl + 
U_\ssr \; \Pr \Bigr) \;
\pmatrix{u \cr d \cr},
\eeq
in this limit, where $U_\ssl$ and $U_\ssr$ are arbitrary
two-by-two unitary matrices having unit determinant. The
Dirac matrix $\Pl = \hf(1+ \gamma_5)$ projects onto the 
left-handed part of each of the quarks, $u$ and $d$, 
while $\Pr = \hf(1
- \gamma_5)$ projects onto their right-handed part. The
group of symmetries which is obtained in this way is $G =
SU_\ssl(2) \times 
SU_\ssr(2)$, with the subscripts `$L$' and `$R$' indicating
the handedness of the quarks on which the corresponding
factor of the group acts. A symmetry such as this which
treats left- and right-handed fermions differently is
called a {\em chiral} symmetry. These transformations are
exact symmetries of QCD in the limit of vanishing $m_u$ and
$m_d$, but are only approximate symmetries when these masses
take their real values. Because the approximate symmetry
involved is chiral, the technique of expanding quantities
in powers of the light-quark masses is called {\em Chiral
Perturbation Theory}.

If this symmetry, $G$, were not spontaneously broken by the
QCD ground state, $\ket{\Omega}$, then we would expect all
of the observed hadrons to fall into representations of $G$
consisting of particles having approximately equal masses.
This is not seen in the spectrum of observed hadrons,
although the known particles 
{\em do} organize themselves into roughly degenerate
representations of the approximate symmetry of {\em
isospin}: $SU_\ssi(2)$. The isospin quantum number, $I$,
for the observed $SU_\ssi(2)$ representations of the
lightest hadrons are listed in Table 2.1. (The dimension of
the corresponding representation is $2I+1$.) Isospin
symmetry can be understood at the quark level to consist of
the diagonal subgroup of $G$, for which $U_\ssl = U_\ssr$.
That is, the approximate symmetry group which is seen to
act on the particle states is that for which the left- and
right-handed components of the quarks $u$ and $d$ rotate
equally.

This suggests that if QCD is to describe the experimentally
observed hadron spectrum, then its ground state must
spontaneously break the approximate symmetry group $G$ down
to the subgroup $H = SU_\ssi(2)$, for which:
\eq
\label{isospintrans}
\pmatrix{u \cr d \cr} \to U \; \pmatrix{u \cr d \cr}.  \eeq
There is indeed good theoretical evidence, such as from
numerical calculations, that the ground state of QCD really
does behave in this way.

Given this symmetry-breaking pattern, we know that the
low-energy spectrum of the theory must include the
corresponding Goldstone bosons. If $G$ had been an exact
symmetry, then the corresponding Goldstone bosons would be
exactly massless. But since $G$ is only a real symmetry in
the limit that $m_u$ and $m_d$ both vanish, it follows that
the Goldstone bosons for spontaneous $G$ breaking need only
vanish with these quark masses. So long as the $u$ and $d$
quarks are much lighter than the natural scale --- $\L
\approx 1$ GeV --- of the strong interactions, so must be
these Goldstone bosons. Indeed, the lightest hadrons in the
spectrum, $\pi^\pm$ and $\pi^0$, have precisely the quantum
numbers which are required for them to be the Goldstone
bosons for the symmetry-breaking pattern 
$SU_\ssl(2) \times SU_\ssr(2) \to SU_\ssi(2)$.\footnote{In
fact, the next-lightest particles, $K$ and $\eta$, together
with the pions have the quantum numbers to be the Goldstone
bosons for the pattern $SU_\ssl(3) \times SU_\ssr(3) \to
SU_\ssv(3)$, which would be appropriate in the limit that
the lightest three quarks, $u,d$ and $s$, were {\em all}
massless. The unbroken subgroup here, $SU_\ssv(3)$, again
is the diagonal, handedness-independent, subgroup.}
Particles such as these which are light, but not massless,
because they are the Goldstone bosons only of an
approximate symmetry of a problem are called {\em
pseudo-Goldstone bosons}.

Since the low-energy interactions of Golstone bosons are
strongly restricted by the symmetry-breaking pattern which
guarantees their existence, it is possible to
experimentally test this picture of pions as
pseudo-Goldstone bosons. The remainder of this chapter is
devoted to extracting some of the simplest predictions for
pion interactions which can be obtained in this way. The
fact that these predictions successfully describe the
low-energy interactions of real pions gives support to the
assumed symmetry-breaking pattern for the ground state of
the strong interactions.

\section{The Low-Energy Variables}

In order to proceed, we must first construct the nonlinear
realization for the case $G = SU_\ssl(2) \times SU_\ssr(2)$
and $H = SU_\ssi(2)$. To do so, we first write out the
representation we shall use for the elements of each of
these groups. Denoting the Pauli matrices, $\vec\tau =
\{\tau_n\}$, by $\tau_1 = 
\pmatrix{0 & 1 \cr 1 & 0 \cr}$, $\tau_2 = \pmatrix{0 & -i
\cr i & 0 \cr}$ and $\tau_3 = \pmatrix{1 & 0 \cr 0 & -1
\cr}$, we write:  %
\eq
\label{gelementex}
g = \pmatrix{ e^{{i \over 2} \;  
\vec\omega_\ssl \cdot \vec\tau} & 0 \cr 0 &
e^{{i \over 2} \; \vec\omega_\ssr 
\cdot \vec \tau} \cr} \in SU_\ssl(2) \times
SU_\ssr(2),
\eeq
and
\eq
\label{helementex}
h = \pmatrix{ e^{{i \over 2}\; 
\vec\omega_\ssi \cdot \vec\tau} & 0 \cr 0 &
e^{{i \over 2} \; \vec\omega_\ssi 
\cdot \vec\tau} \cr} \in SU_\ssi(2).
\eeq
We adopt here, and throughout the remainder of the chapter,
an obvious vector notation for the three-component
quantities $\omega^n$, $\theta^n$, $u^n$, \etc.

In this representation, the Goldtone boson field becomes: 
\eq
\label{gbfielddefex}
U(\vec\theta) = \pmatrix{ e^{{i \over 2} 
\; \vec\theta \cdot \vec\tau} & 0 \cr
0 & e^{-\, {i \over 2}\; \vec\theta \cdot 
\vec\tau} \cr}, \eeq
and the nonlinear $H$ transformation, $\gamma$, is: 
\eq
\label{gammafielddefex}
\gamma(\vec\theta,g) = \pmatrix{ 
e^{{i \over 2}\; \vec u \cdot \vec\tau} 
& 0  \cr
0 & e^{{i \over 2} \; \vec u \cdot \vec\tau} \cr}. \eeq

\subsection{A Notational Aside}

Before passing to the nonlinear realization, we briefly
pause to make contact between the variables as defined
here, and those that are often used in the literature. We
have defined the elements, $g \in G$, the matrices
$U(\vec\theta)$, and 
$\gamma(\vec\theta,g)$ in a block-diagonal form which
emphasizes the left- and right-handed parts of the
transformations:
\eq
\label{newway}
g = \pmatrix{ g_\ssl & 0 \cr 0 & g_\ssr \cr}, 
\quad U(\vec\theta) = \pmatrix{
u_\ssl(\vec\theta) & 0 \cr  0 & u_\ssr(\vec\theta)
\cr}, \quad \gamma(\vec\theta,g) = \pmatrix{ 
h(\vec\theta,g) & 0 \cr 0 &
h(\vec\theta,g) \cr}.
\eeq

In terms of these quantities, the transformation law $U \to
\tw 
U = g U \gamma^\dagger$ becomes $u_\ssl \to \tw{u}_\ssl = 
g_\ssl u_\ssl h^\dagger$ and $u_\ssr \to \tw{u}_\ssr =
g_\ssr 
u_\ssr h^\dagger$. It is common practice to work with the
composite quantity, $\Xi$, for which the transformation
rule does not depend on the implicitly-defined matrix $h$.
That is, if 
$\Xi \equiv u_\ssl u_\ssr^\dagger$, then $\Xi \to \tw\Xi = 
g_\ssl \Xi g_\ssr^\dagger$. This transformation law has the
advantage of involving only explicit, constant matrices. In
terms of the Goldstone boson fields, $\vec\theta$, we have
$u_\ssl = 
e^{{i \over 2}\; \vec\theta \cdot \vec\tau} =
u_\ssr^\dagger$, 
so $\Xi = u_\ssl u_\ssr^\dagger = e^{i \vec\theta \cdot
\vec\tau}$.

It is possible, and often convenient, to reformulate all of
the Goldstone boson self-couplings that are obtained
elsewhere in this chapter in terms of this field $\Xi$. It
is {\em not} possible to express the Goldstone-boson
couplings to other fields, $\chi$, in this way since the
matrix $\gamma$ cannot be removed from the transformation
law for these other fields.

\subsection{The Nonlinear Realization}

The nonlinear realization is now obtained by constructing
both $\vec{\tw\theta}(\vec\theta,g)$ and $\vec{u} =
\vec{u}(\vec\theta,g)$, using the condition $g \; U(\vec\theta)
= U(\vec{\tw\theta}) \; 
\gamma$. For the groups under consideration this
construction may be performed in closed form by using the
identity:
\eq
\label{pauliidentity}
\exp\Bigl[ i \vec\alpha \cdot \vec\tau \Bigr] 
= \cos\alpha  + i \hat\alpha
\cdot \vec\tau \; \sin \alpha ,
\eeq
where $\alpha = \sqrt{\vec\alpha \cdot \vec\alpha}$, and 
$\hat\alpha =  \vec\alpha/\alpha$.

Using this identity to multiply out both sides of the
defining equation $g \; U(\vec\theta) = U(\vec{\tw\theta})
\; \gamma$, and equating the coefficients of 1 and
$\vec\tau$, separately for the left- and right-handed parts
of the matrices, gives explicit expressions for
$\delta\vec\theta = \vec\xi$ and $\vec u$. If 
$g_{\ssl,\ssr} = \exp\left[ {i \over 2} \,
\vec\omega_{\ssl,\ssr} 
\cdot \vec\tau \right]$, and defining $\vec \omega_\ssi
\equiv 
\hf (\vec\omega_\ssl + \vec\omega_\ssr)$ and
$\vec\omega_\ssa  
\equiv \hf (\vec\omega_\ssl - \vec\omega_\ssr)$, then:     
\bg
\label{chpttransfnsone}
\vec \xi &=& \vec\theta \times \vec\omega_\ssv 
+ {\theta \over 2} \, \left(
\tan {\theta \over 2} + \cot {\theta \over 2} \right) 
[ \vec \omega_\ssa -
\hat\theta (\hat\theta \cdot \vec \omega_\ssa) ] 
+ \hat\theta (\hat\theta \cdot
\vec \omega_\ssa), \nn\\
&=& \vec \omega_\ssa + \vec \theta \times 
\omega_\ssv + O(\theta^2); \\
\label{chpttransfnstwo}
\vec u &=& \vec \omega_\ssv + 
(\hat\theta \times \vec\omega_\ssa) \; \tan
{\theta \over 2} \nn\\
&=& \vec \omega_\ssv +{ \vec \theta 
\times \vec \omega_\ssa \over 2} +
O(\theta^2).
\nd

For future reference we notice that the transformation law for 
$\vec \theta$ implies that the three broken generators of 
$G = SU_\ssl(2) \times SU_\ssr(2)$ form an irreducible,
three-dimensional representation of the unbroken subgroup, 
$H = SU_\ssi(2)$.

Similarly evaluating the combination 
\eq
\label{aedefsinu}
U^\dagger \partial_\mu U = {i \over 2} \; \vec e_\mu \cdot
\pmatrix{\vec\tau & \cr & - \vec\tau \cr} + {i \over 2} 
\; \vec \Sca_\mu \cdot \pmatrix{\vec\tau & \cr & \vec\tau \cr},
\eeq
gives the quantities with which the
invariant lagrangian is built: 
\bg
\label{chptcovquantitiesone}
\vec e_\mu &=& \left( {\sin \theta \over \theta} 
\right) \; \partial_\mu
\vec\theta - \left( {\sin \theta - 
\theta \over \theta^3} \right) \;
(\vec\theta \cdot \partial_\mu \vec \theta ) 
\; \vec \theta, \nn\\
&=& \partial_\mu \vec \theta \left(1 - \nth{6} 
\; \theta^2 \right) + \nth{6}
\; (\vec\theta \cdot \partial_\mu \vec\theta) \; 
\vec \theta + O(\theta^5); \\
\label{chptcovquantitiestwo}
\vec \Sca_\mu &=& - 2 \; \left( {\sin^2 
{\theta\over 2} \over \theta^2}
\right) \; ( \vec \theta \times \partial_\mu 
\vec \theta) \nn\\
&=& - \, \hf \; \vec\theta \times \partial_\mu 
\vec \theta + O(\theta^4).
\nd
Notice that $\vec e_\mu$ is odd, and $\vec \Sca_\mu$ is
even, under the interchange $\vec\theta \to -  \vec\theta$.
The low-energy Goldstone boson lagrangian will be required
to be invariant under such an inversion of $\vec\theta$,
since this is a consequence of the parity invariance of the
underlying QCD theory.

It is useful to also record here the $G$-transformation
rules for the other fields which can appear in the
low-energy theory. Of particular interest are the nucleons
--- neutrons and protons --- since low-energy pion-nucleon
interactions are amenable to experimental study. The
nucleon transformation rules under $G = SU_\ssl(2) \times
SU_\ssr(2)$ are completely dictated by their
transformations under the unbroken isospin subgroup, $H = 
SU_\ssi(2)$. Since the nucleons form an isodoublet, $N = 
\pmatrix{p \cr n\cr}$, they transform under isospin
according to 
$\delta N = {i \over 2} \; \vec\omega_\ssi \cdot \vec\tau
\; N$. The rule for the complete $G$ tranformations is
therefore simply
\eq
\label{nucleontransf}
\delta N = {i \over 2} \; \vec u \cdot \vec\tau \; N. \eeq
We therefore see that the appropriate covariant derivative
for nucleons is:
\eq
\label{ncovderiv}
D_\mu N = \partial_\mu N - \, {i \over 2}\;
\vec\Sca_\mu(\theta)  \cdot \vec\tau \; N.
\eeq

\section{Invariant Lagrangians}

We may now turn to the construction of the invariant
lagrangian which governs the low-energy form for pion
interactions. The lagrangian describing pion
self-interactions involving the fewest derivatives is
uniquely determined up to an overall normalizing constant.
As was discussed in detail in the previous chapter, this is
because of the irreducibility of the transformation rules of
the broken generators, $\vec X = \hf \; \vec\tau \;
\gamma_5$, under the unbroken isospin transformations. The
most general $G$-invariant lagrangian density involving
only two derivatives is  
\eq
\label{selfintforpions}
\Scl_{\pi \pi} = - \; {F^2 \over 2} \;
\hat{g}_{mn}(\vec\theta) \; \partial_\mu
\theta^m \partial^\mu \theta^n + 
\hbox{(higher-derivative terms)},
\eeq
where the metric, $\hat{g}_{mn}$, on $G/H$ is: 
\bg
\label{chptmetric}
g_{mn}(\theta) &=& \delta_{rs} \; 
{e^r}_m \, {e^s}_n = \delta_{mn} \; \left(
{\sin^2 \theta \over \theta^2} \right) + 
\theta_m \theta_n \; \left(
{\theta^2 - \sin^2 \theta \over \theta^4} \right) , \nn\\ 
&=& \delta_{mn} \:
\left(1 - \nth{3} \; \theta^2 \right) + \nth{3} \;
\theta_m \theta_n + O(\theta^4).
\nd

For applications to pion scattering it is useful to
canonically normalize the pion fields, that is, to ensure
that their kinetic terms take the form: $- \, \hf \;
\partial_\mu \vec\pi \cdot \partial^\mu 
\vec\pi$. This requires the rescaling: $\vec\theta = \vec
\pi / F$. With this choice we have:
\bg
\label{pionderivterms}
\Scl_{\pi\pi} &=& - \, \hf \; \left[ 
{ F^2 \sin^2 \left( {\pi \over F} \right)
\over \pi^2 } \right] \partial_\mu \vec 
\pi \cdot \partial^\mu \vec \pi +
\left[
{ \pi^2 - F^2 \sin^2 \left( { \pi \over F} 
\right) \over \pi^4 } \right]  (
\vec \pi \cdot \partial_\mu \vec \pi) 
( \vec \pi \cdot \partial^\mu \vec \pi)
\nn\\
&& \qquad \qquad + \hbox{(higher-derivative terms)}. \\ 
&=& - \, \hf \;
\partial_\mu \vec \pi \cdot \partial^\mu \vec \pi -  
\, {1 \over 2 F^2}\;  (\vec \pi \cdot \partial_\mu 
\vec \pi) \; (\vec\pi \cdot \partial^\mu
\vec \pi) + O(\pi^6) + \cdots.  \nn
\nd
An integration by parts has been performed in writing the
$\pi^4$ term of the expansion of the lagrangian.

The couplings between nucleons and pions to lowest order in
the derivative expansion involve only one derivative. The
most general form for these that is consistent with the
nonlinearly-realized $G$-invariance, and with parity
invariance, is:
\bg
\label{pionnucleonderiv}
\Scl_{\pi \ssn} &=& - \, \Nbr \left( \dslsh 
- {i \over 2}\; \vec {\Scaslsh}
( \vec
\theta) \cdot \vec \tau + m_\ssn \right) 
N - \, {ig \over 2} \, \left( \Nbr
\gamma^\mu \gamma_5 \vec \tau N \right) 
\; \vec e_\mu( \vec\theta ), \nn\\
&=& - \, \Nbr \left( \dslsh + m_\ssn \right) 
N - \; {ig \over 2F} \, \left(
\Nbr \gamma^\mu \gamma_5 \vec \tau 
N \right) \cdot \partial_\mu \vec \pi \\
&& \qquad \qquad - \, {i \over 2F^2} \, 
\left( \Nbr \gamma^\mu \vec \tau N
\right) \cdot (\vec \pi \times \partial_\mu 
\vec \pi) + \cdots. \nn
\nd
The ellipses here represent terms which involve either
three or more powers of the pion field, more than two
powers of the nucleon field, or involve more than one
derivative.

Clearly, only the one constant $F$ need be determined in
order to completely fix the dominant low-energy pion
self-interactions, and a second constant, $g$, is also
required to determine the lowest-order pion-nucleon
couplings.

\subsection{Conserved Currents}

For future reference it is instructive to compute the
Noether currents for the symmetry group $G= SU_\ssl(2)
\times 
SU_\ssr(2)$ in both the underlying theory (\ie\ QCD), and
in the effective low-energy pion-nucleon theory.

In QCD, the symmetry transformation under $G$ is given by 
$\delta q = {i \over 2} \; ( \vec \omega_\ssl \Pl + \vec 
\omega_\ssr \Pr) \cdot \vec \tau \; q$, where $q =
\pmatrix{u \cr d \cr}$ denotes the two-component quantity
containing the two lightest quarks. The corresponding
Noether currents that are obtained from the QCD lagrangian,
eq.~\pref{qcdlagr} are:   
\eq
\label{noethercforqcd}
\vec\jmath_\ssl^\mu = {i \over 2} \; \qbr 
\gamma^\mu \Pl \vec \tau \; q, \qquad
\hbox{and} \qquad \vec\jmath_\ssr^\mu = {i \over 2} \; 
\qbr \gamma^\mu \Pr \vec
\tau \; q .
\eeq
The current, $\vec\jmath^\mu_\ssi$, which corresponds to
the unbroken $SU_\ssi(2)$ isospin symmetry is therefore: 
$\vec\jmath^\mu_\ssi = \vec\jmath^\mu_\ssl + 
\vec\jmath^\mu_\ssr = {i \over 2} \; \qbr \gamma^\mu \vec 
\tau \; q$. The current for the broken symmetry is
similarly: 
$\vec\jmath^\mu_\ssa = \vec\jmath^\mu_\ssl - 
\vec\jmath^\mu_\ssr = {i \over 2} \; \qbr \gamma^\mu 
\gamma_5 \vec \tau \; q$.

In the effective pion-nucleon theory the corresponding
current may also be constructed using the known action of
$G$ on 
$\vec \pi$ and $N$, and using the lagrangian, whose
lowest-derivative terms are given by
eqs.~\pref{pionderivterms} and 
\pref{pionnucleonderiv}. Keeping only the terms involving a
single pion or only two nucleons, at the lowest order in the
derivative expansion, then gives:
\bg
\label{effcurrents}
\vec\jmath_\ssi^\mu &=& - \Bigl( \vec\pi \times 
\partial^\mu \vec \pi \Bigr) +
{i \over 2} \; \Nbr \gamma^\mu \vec \tau \; 
N + \cdots , \nn\\
\vec\jmath_\ssa^\mu &=& F \, \partial^\mu 
\vec \pi + {ig \over 2} \; \Nbr
\gamma^\mu \gamma_5 \vec \tau \; N + \cdots.  \nd
There are an infinite number of higher-order terms in these
currents corresponding to the infinite number of
interactions in the effective pion-nucleon lagrangian. All
of the terms not written explicitly above involve
additional factors of the fields $\vec\pi$ or $N$, or
involve more derivatives of these fields than do the terms
displayed.

\subsection{Determining $F$ and $g$}

These expressions for the Noether currents for $G$ turn out
to furnish a handle for experimentally determining the
constants $F$ and $g$. This is because, as is made explicit
in the following section, experimental information exists
concerning the value of some of the matrix elements of the
broken current 
$\vec\jmath^\mu_\ssa$.

This experimental information exists because it is
precisely the current $\vec\jmath^\mu_\ssa$ which appears
in that part of the weak-interaction lagrangian which
describes transitions from $d$ quarks to $u$ quarks. Since
these transitions are responsible for many reactions,
including all nuclear $\beta$-decays, free-neutron decay,
and $\pi^\pm$ decay, the corresponding matrix elements of
this current can be measured.

The terms in the underlying lagrangian which describe these
decays are obtained by supplementing the QCD interactions of
eq.~\pref{qcdlagr} with the weak-interaction term:
\eq
\label{weakintunderlying}
\Scl_{\rm weak} = {\GF \cos\theta_\ssc 
\over \sqrt{2}} \; \ubr
\gamma^\nu (1 + \gamma_5) d \; \ol\nu_\ell 
\gamma_\nu (1 + \gamma_5) \ell +
\hc.
\eeq
Here the Dirac spinor field $\ell$ and the Majorana field
$\nu_\ell$ respectively represent a charged lepton --- in
practice, the electron and muon --- and the corresponding
neutrino. $\GF$ is the {\em Fermi coupling constant}, which
is determined from the muon decay rate to be $\GF =
1.16649(2) \times 10^{-5}$ 
GeV${}^{-2}$. The angle $\theta_\ssc$ is called the {\em
Cabbibo angle}, and it parameterizes the fact that the size
of the coupling constant, $\GF \cos\theta_\ssc$, as seen in
superallowed nuclear $\beta$-decays is smaller than $\GF$
as is measured in muon decay. Numerically, $\cos\theta_\ssc
= 0.9753(6)$.

The main feature to be noticed from 
eq.~\pref{weakintunderlying} is that the quark combination
which appears is a linear combination of the conserved
$SU_\ssl(2) 
\times SU_\ssr(2)$ currents:
\bg
\label{wkcurrentreln}
i \ubr \gamma^\mu (1 + \gamma_5) d &=&
 i \qbr \gamma^\mu \Pl (\tau_1 + i \tau_2)
\; q \nn\\
&=& \Bigl[  (j_\ssi)^\mu_1 + i (j_\ssi)^\mu_2 \Bigr] 
+  \Bigl[ (j_\ssa)^\mu_1 +
i (j_\ssa)^\mu_2 \Bigr].
\nd
In preparation for using eqs.~\pref{effcurrents} we have
re-expressed the left-handed currents which appear in the
weak interactions in favour of the axial and vector
currents using: 
$\vec \jmath^\mu_\ssl = \hf \;( \vec \jmath^\mu_\ssi + \vec 
\jmath^\mu_\ssa)$.

To compute the decay rate for the reaction $\pi^+ \to \mu^+ 
\nu_\mu$ we require the following matrix element:
$\bra{\mu^+, 
\nu_\mu} \Scl_{\rm weak} \ket{\pi^+}$. The part of this
matrix element which involves strongly-interacting
particles is 
$\bra{\Omega} \vec \jmath^\mu_\ssa \ket{\pi^+}$, where 
$\ket{\Omega}$ is the QCD ground state. The isospin
current, 
$\vec \jmath^\mu_\ssi$, does not appear in $\pi^+$ decay
because its matrix element vanishes due to the parity
invariance of the strong interactions. The most general
form for this matrix element which is consistent with
Poincar\'e and isospin invariance is given by:
\eq
\label{pionmatrixelement}
\bra{\Omega} (j_\ssa)_n^\mu(x) \ket{\pi_m(q)}
 = {i \fpi \, q^\mu \; e^{iqx}
\over \sqrt{ (2\pi)^3 \, 2 q^0}} \; \delta_{mn} ,  \eeq
where it is conventional to extract the numerical factor
$1/\sqrt{(2 \pi)^3 2}$, and the pion states are labelled
here as members, 
$\ket{\pi_n}$ ($n=1,2,3$), of an isotriplet. These are
related to the physical states, having definite electric
charge, by: 
$\ket{\pi^\pm} = \nth{\sqrt{2}} \; ( \ket{\pi_1} \mp i 
\ket{\pi_2})$, and $\ket{\pi^0} = \ket{\pi_3}$.

The only unknown quantity in this matrix element is the
constant $\fpi$, which is inferred to be $\fpi = 92$ MeV by
comparing the prediction, $1/\tau_{\rm th} = (\GF^2
\cos^2\theta_\ssc \fpi^2 m_\mu^2 m_\pi/4 \pi) \; 
(1 - m_\mu^2 / m_\pi^2)^2$, with the observed mean 
lifetime, $\tau_{\rm exp} = 2.6030(24) \times 
10^{-8}$ s, for the decay $\pi^+ \to \mu^+ \nu_\mu$.

Now, to lowest order in the derivative expansion, the
matrix element of eq.~\pref{pionmatrixelement} can be
directly evaluated as a function of the parameter $F$ using
the second of  eq.~\pref{effcurrents}. Comparing these 
results permits the inference
\eq
\label{thevalueforf}
F = \fpi = 92 \, \MeV.
\eeq
With this constant in hand, we may now use the low-energy
effective lagrangian to predict the low-energy pion
self-interactions.

Before proceeding to these predictions, we first repeat
these steps for another  matrix element in order to infer
the value of the constant, $g$, which governs the size of
the pion-nucleon coupling. We once again 
consider the weak interaction, 
eq.~\pref{weakintunderlying}, but this time consider its
prediction for the decay rate of a free neutron into a
proton, an electron and an antineutrino: $n \to p e
\ol\nu_e$. In this case, the most general Poincar\'e-,
parity-, time-reversal- and isospin-invariant form for the
desired matrix element is:
\bg
\label{neutronmatrixelement}
\bra{N(k,\sigma)} \vec \jmath_\ssi^\mu(x) 
\ket{N(l,\zeta)} &=& {i e^{iqx} \over 2
(2 \pi)^3} \; \; \ubr(k,\sigma) \Bigl[ F_1(q^2) 
\gamma^\mu + F_2(q^2)
\gamma^{\mu\nu} q_\nu \Bigr] \, \vec \tau 
\; u(l,\zeta) , \nn\\
\bra{N(k,\sigma)} \vec\jmath_\ssa^\mu(x) 
\ket{N(l,\zeta)} &=& {i e^{iqx} \over 2(2
\pi)^3} \; \; \ubr(k,\sigma) \Bigl[ G_1(q^2) 
\gamma^\mu \gamma_5 + G_2(q^2)
\gamma_5 q^\mu \Bigr] \, \vec \tau \; 
u(l,\zeta). \nn\\ &&
\nd
Here, $l^\mu$ and $k^\mu$ are the four-momenta of the
initial and final nucleons, and $q^\mu = (l - k)^\mu$ is
their difference. $\zeta$ and $\sigma$ similarly represent
the spins of the initial and final nucleons. $u(k,\sigma)$
is the Dirac spinor for a free particle having momentum
$k^\mu$, with $k^2 + m_\ssn^2 = 0$, and spin $\sigma$. (Our
normalization is: $\ubr(k,\sigma) u(k,\sigma') =
(m_\ssn/k^0) \; \delta_{\sigma \sigma'}$.) Finally, 
$\gamma^{\mu\nu}$ stands for the commutator $\hf \; 
[\gamma^\mu, \gamma^\nu]$.

The unknowns in this matrix element are the four
Lorentz-invariant functions, $F_1, F_2, G_1$ and $G_2$, of
the invariant momentum transfer, $q^2$. These functions are
not completely arbitrary, however, since they must encode
the fact that we are working in a limit where $G =
SU_\ssl(2) \times SU_\ssr(2)$ is taken to be a symmetry of
the QCD lagrangian. The implications of $G$-invariance are
easily extracted by demanding current conservation,
$\partial_\mu \vec \jmath^\mu_\ssi = \partial_\mu 
\vec\jmath^\mu = 0$, for all of the currents. Keeping in
mind that the nucleons have equal mass in the $G$-invariant
limit in which we are working, this implies no conditions on
the functions $F_1$ and $F_2$, but implies for the others:
\eq
\label{currentconservationconds}
2i m_\ssn \; G_1(q^2) = q^2 \; G_2(q^2) . \eeq

In the rest frame of the decaying neutron, the components
of the momentum transfer, $q^\mu$, are at most of order 1
MeV. Since this is much smaller than the typical
strong-interaction scale, 
$\L \sim 1$ GeV, which characterizes the matrix element,
for neutron decay it suffices to simplify 
eqs.~\pref{neutronmatrixelement} using $q^\mu \approx 0$.
In this approximation the neutron decay rate depends only
on the two unknown constants, $F_1(0)$ and $G_1(0)$.

\begin{figure}
\epsfbox[-70 670 650 670]{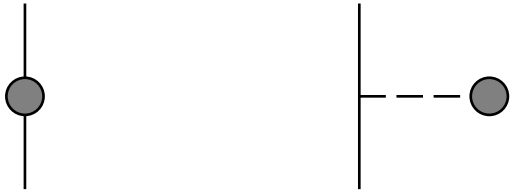}
\vspace{1in}
\caption{The Feynman graphs which give the dominant nucleon
matrix elements of the Noether currents in the low-energy
effective theory. Solid lines represent nucleons, and
dashed lines represent pions.}
\end{figure}

Since the constants $F_1(0)$ and $G_1(0)$ correspond to the
low-energy limit of these current matrix elements, they may
be related to the constants which appear in the dominant
terms of the low-energy effective lagrangian. This may be
done by using 
eqs.~\pref{effcurrents} to directly evaluate the matrix
elements of eqs.~\pref{neutronmatrixelement}. Doing so, we
find contributions from the two Feynman graphs of Fig.~2.1.
The first of these gives the direct matrix element of
eqs.~\pref{effcurrents}, and contributes to the form
factors $F_1$ and $G_1$. The second graph uses the $NN\pi$
interaction of the effective lagrangian,
eq.~\pref{pionnucleonderiv}, together with the vacuum-pion
matrix element of eq.~\pref{pionmatrixelement}. It
contributes only to the form factor $G_2$. Evaluating these
graphs, we find:
\eq
\label{formfactorpredictions}
F_1 = 1, \qquad G_1 = g, \qquad \hbox{and} 
\qquad G_2 = {2i g  m_\ssn \over q^2},
\eeq
from which we see $F_1(0) = 1$ and $G_1(0) = g$. The factor
$1/q^2$ in $G_2$ comes from the massless pion propagator in
the second of Figs.~2.1. Notice that this result for $G_2$
is precisely what is required to satisfy the
current-conservation condition of
eq.~\pref{currentconservationconds}.

The finding that $F_1(0) = 1$ states that this part of the
matrix element is not renormalized by the strong
interactions, since this value for $F_1(0)$ is the same as
would have been obtained if the matrix elements of
$\vec\jmath^\mu_\ssi$ were taken using the underlying quark
states rather than with the composite nucleon states.
$F_1(0)$ is the same for both quarks and nucleons because
$F_1(0)$ is the quantity which determines the matrix
elements in these states of the conserved isospin charges,
$\vec \Sci = \int 
d^3\bfr \; \vec\jmath^0_\ssi$. But these have matrix
elements which depend only on the $SU_\ssi(2)$
transformation properties of the states whose matrix
elements are taken. Since both quarks and nucleons are
isodoublets, and since inspection of 
eq.~\pref{noethercforqcd} shows that quarks have $F_1(0) =
1$, the same must be true for nucleons.

The same argument does not hold for the axial current
because this is a current for a symmetry which is
spontaneously broken. This turns out to imply that the
corresponding conserved charge is not well defined when
acting on particle states, and so $G_1(0)$ need not be
unity.

We finally arrive at the desired conclusion: the neutron
decay rate, which is completely determined by the constants
$F_1(0) = 1$ and $G_1(0) = g$, can be used to experimentally
infer the numerical value taken by the remaining constant,
$g$, of the effective lagrangian. The measured neutron mean
life --- which is $\tau_{\rm exp} = 887(2)$ s --- then
implies $g = 1.26$.

Having determined from experiment the values taken by $F$
and $g$, we are now in a position to use the effective
pion-nucleon lagrangian to predict the low-energy
properties of pion-pion and pion-nucleon interactions.

\subsection{The Goldberger-Treiman Relation}

Historically the trilinear $N-N-\pi$ interaction has been
written with no derivatives, as a Yukawa coupling: 
\eq
\label{yukawaform}
\Scl_{\ssn\ssn\pi} = i g_{\ssn\ssn\pi} \; 
(\Nbr \, \gamma_5 \vec\tau \, N)  \cdot
\vec\pi,
\eeq
with the constant $g_{\ssn\ssn\pi}$ found from
phenomenological studies to be close to 14. But the value
of this constant can be predicted in terms of the constant
$g$, and this prediction serves as the first test of the
low-energy pion-nucleon lagrangian.

The prediction starts with the trilinear $N-N-\pi$
interaction of
eq.~\pref{pionnucleonderiv}:
\eq
\label{trilinearnnpi}
\Scl_{\ssn\ssn\pi} = - \; {ig \over 2 \fpi} \, 
\left(\Nbr \gamma^\mu \gamma_5
\vec \tau N \right) \cdot \partial_\mu \vec \pi , \eeq
and performs an integration by parts to move the derivative
to the nucleon fields. One then uses the lowest-order
equations of motion for $N$: \ie\ $(\dslsh + m_\ssn)\, N =
0$, to simplify the result. One obtains a result of the
form of eq.~\pref{yukawaform}, but with  
\eq
\label{predictionforgpin}
g_{\ssn\ssn\pi} = {g m_\ssn \over \fpi}.
\eeq
Using the experimental values: $g = 1.26$, $m_\ssn = 940$
MeV and $\fpi = 92$ MeV gives the prediction
$g_{\ssn\ssn\pi} = 12.8$, which agrees well with the
phenomenologically inferred value. This prediction,
eq.~\pref{predictionforgpin}, is known as the {\em
Goldberger-Treiman relation}.

We turn now to one last dangling issue which remains to be
addressed before we can compute low-energy pion-pion and
pion-nucleon scattering.

\section{Explicit Symmetry-Breaking}

Notice that the effective lagrangian,
eqs.~\pref{pionderivterms} and \pref{pionnucleonderiv}, has
very definite implications for the masses of the pions and
nucleons. It states that the pion multiplet must be exactly
massless, and that the nucleon masses must be equal. Since
these predictions rely only on the assumption of unbroken
$G$ invariance, and since $G$-invariance only holds for QCD
in the limit that $m_u$ and $m_d$ vanish, corrections to the
pion and nucleon mass predictions can only be inferred by
including the effects of the symmetry-breaking quark mass
terms for the low energy effective theory. We do so, in
this section, to lowest order in the light-quark masses.

The quark mass terms in the QCD lagrangian are proportional
to $\qbr \; M \Pl \; q + \hc$, where $M = \pmatrix{m_u & 0
\cr 0 & m_d \cr}$ is the light-quark mass matrix. Under the
$G= SU_\ssl(2) \times SU_\ssr(2)$ symmetry, $q \to (g_\ssl
\Pl + g_\ssr \Pr) \; q$, this transforms into: 
\eq
\label{masstermtransfn}
\qbr \; M \Pl \; q  \to \qbr \; g_\ssr^\dagger 
M g_\ssl \Pl \; q + \hc.
\eeq
Although this is not invariant, it {\em would} have been
invariant if the mass matrix had been a field which had
also transformed under $G$ according to: $M \to g_\ssr M
g_\ssl^\dagger$.

We imagine the effective pion-nucleon theory having an
expansion in the light quark masses, $M$: $\Scl_{\rm eff} =
\Scl_0 + \Scl_1 + \cdots$, where the subscript indicates the
power of $M$ it contains. Each of these terms may be
separately expanded in powers of the derivatives, and of
the fields $\pi$ and $N$. The construction to this point
has given the lowest-derivative terms which can appear in
$\Scl_0$. Our goal now is to determine the most general
form which may be taken by $\Scl_1$, and which contains no
derivatives of any fields. This will give the dominant
symmetry-breaking contribution at low energies.

\subsection{Pions Only: Vacuum Alignment}

We start by focussing on the part of $\Scl_1$ which depends
only on the pion fields. The form taken by $\Scl_1$ may be
obtained from the following argument. We  require that
$\Scl_1$ be $G$-invariant, but only if we take $M \to
g_\ssr M g_\ssl^\dagger$ in addition to transforming the
fields $\pi$ in their usual way.

It is straightforward to construct one such a term
involving only the pion fields. The simplest construction
is to use the quantity $\Xi \equiv u_\ssl u_\ssr^\dagger 
= e^{i \vec \theta \cdot \vec\tau} = \cos\theta + i
\hat \theta \cdot \vec\tau \sin\theta$, defined in section 
(8.2.1), which transforms according to $\Xi \to \tw\Xi 
= g_\ssl \Xi g_\ssr^\dagger$. (Recall here that $\theta$
and $\hat\theta$ are defined by $\theta = \sqrt{\vec\theta 
\cdot \vec \theta}$ and $\hat\theta = \vec\theta/\theta$.) 

A possible lagrangian therefore is:
\bg
\label{vacalignterms}
\Scl_{1,\pi\pi} &=& - A \; \Re \Tr\, [M \; \Xi ] 
 -B \; \Im \Tr\, [M \; \Xi ] \\
&=& -A \, (m_u + m_d) \; \cos\theta 
-B\,  (m_u - m_d) \; \theta_3 \; {\sin\theta \over \theta} .
\nd
Clearly this generates a potential energy which is
a function of $\vec\theta$, as is possible because of the
explicit breaking of the $SU_\ssl(2) \times SU_\ssr(2)$
symmetry by the quark masses. As a result, all values
for $\vec\theta$ are not equally good descriptions of the vacuum,
and it is necessary to minimize the potential in order
to determine the vacuum value for $\vec\theta$. This 
choosing of the vacuum value for the pseudo-Goldstone
fields after the introduction of explicit symmetry-breaking 
is a process known as {\em vacuum alignment}. 

In the present instance the potential is minimized by
$\theta_1 = \theta_2 = 0$, and has the schematic form
$V(\theta) = - \Scl_1(\theta) = A \cos\theta + 
B \sin\theta = - |A| \cos(\theta
- \theta_0)$, for $\theta = \theta_3$ and
$A$ and $B$ (or, equivalently, $A$
and $\theta_0$) constants. This, once minimized (giving
$\theta_{\rm min} = \theta_0$) and expanded about 
the minimum (with $\theta = \theta_{\rm min} + 
\theta'$), the potential becomes $V(\theta') = 
- |A| \cos\theta'$, leaving
our lagrangian density of the form:
\bg
\label{pionmassterm}
\Scl_{1,\pi\pi} &=& {\Scm^3 \over 2} \; 
\Tr \, [ M \; (\Xi + \Xi^\dagger)] ,
\nn\\
&=& (m_u + m_d) \; \Scm^3 \; \cos \theta , \\  
&=& m_\pi^2 \; \left[\fpi^2 - \,
\hf \; \vec \pi \cdot \vec \pi - \nth{4! \,
\fpi^2} \; ( \vec \pi \cdot \vec \pi)^2 + 
O(\pi^6) \right] , \nn
\nd
with the constant $\Scm^3$ positive.
 
Eq.~\pref{pionmassterm} gives the required 
symmetry-breaking interaction, where
the $\vec\pi$'s are chosen so that the vacuum is
at $\vec \pi = 0$, and, in the last line, we have
also eliminated the arbitrary parameter, $\Scm$, which has
the dimensions of mass, in terms of the common mass,
$m_\pi$, we find for all three pions:
\eq
\label{pionmassprediction}
 m_\pi^2 =  (m_u + m_d) \; {\Scm^3 \over \fpi^2}.  
\eeq

There are several features here worth highlighting.
Firstly, notice that all of the pion self-interactions
necessarily preserve isospin to this order in the
derivative and quark-mass expansions. This implies, among
other things, degenerate masses for all three pions. This
preservation of isospin does {\em not} rely on the
isospin-breaking difference, $m_u - m_d$, being small in
comparison with $m_u$ or $m_d$. Rather, it relies only
on $m_u$ and $m_d$ both being small compared to the
characteristic scale of QCD. We must look elsewhere for
an understanding of the observed mass difference between
the charged and neutral pions, such as to the
isospin-breaking electromagnetic interactions.

Secondly, the lagrangian of eq.~\pref{pionmassterm}
necessarily implies a quark-mass-dependent contribution to
the vacuum-energy density, $- \rho_\ssv
= \Scl_{1,\pi\pi}( \vec\pi =
0) = m_\pi^2 \fpi^2 = (m_u + m_d) \, \Scm^3$. This
contribution permits a physical interpretation for the
parameter $\Scm$, as follows. In the underlying theory the
derivative of the total vacuum energy, 
$\rho_\ssv$, with respect to any quark mass is given by:  
\eq
\label{qqbarvev}
{\partial \rho_\ssv \over \partial m_q} =
\bra{\Omega} \qbr \, q \ket{\Omega},
\eeq
where $\ket{\Omega}$ is the QCD ground state. We see, by
comparison with the pion scalar potential, 
eq.~\pref{pionmassterm}, that  
\eq
\label{qqbarvevval}
\bra{\Omega} \ubr \, u \ket{\Omega} = 
\bra{\Omega} \dbr \, d  \ket{\Omega} =
- \Scm^3 + \cdots,
\eeq
where the ellipses here denote the contributions due to
quantum effects in the low-energy pion-nucleon theory.
Evidently $\Scm \approx \left[m_\pi^2 \fpi^2/(m_u+m_d)
\right]^{1/3}$ gives the size of the expectation value
which is responsible for the spontaneous breaking of the
chiral $SU_\ssl(2) \times SU_\ssr(2)$ symmetry. Using
the values $m_\pi = 140$ MeV, $\fpi = 92$ MeV and
$4.5 \; \hbox{MeV} < m_u + m_d < 14 \; \hbox{MeV}$
then gives $230 \; \hbox{MeV} < \Scm < 330 
\; \hbox{MeV}$ for this scale.

Next, we remark that since none of the terms which appear
in $\Scl_{1,\pi\pi}$ depend on derivatives of $\vec\pi$, they
do not at all affect expressions \pref{effcurrents} for the
conserved Noether currents of the theory. We therefore need
not at all change the above analysis which determined the
experimental values for the constants $F$ and $g$ from pion
and nucleon weak decays.

Finally, we note that the mass term given in 
eq.~\pref{pionmassterm} is the only possible term (up to
normalization) which is linear in $M$ and depends only on 
$\theta$, and not on its derivatives. This uniqueness
follows from the impossibility of building a $G$-invariant
scalar potential. To see this, we write $\Scl_1 = \Tr \,( M
\; \Sco ) + \hc$, for $\Sco(\vec\theta)$ a two-by-two 
matrix function of the Goldstone boson fields. 
$\Sco$ must transform under $G$ according to 
$\Sco \to g_\ssl \Sco g_\ssr^\dagger$. $\Sco_1 = \Xi$
satisfies these conditions, but suppose $\Sco = \Sco_2$
were a second, independent solution. In this case, the
combination $V(\vec\theta) = \Tr \, [ \Sco_1 \Sco_2^\dagger
]$ or $V(\vec\theta) =\det\, [ \Sco_1 \Sco_2^\dagger ]$ 
would be a $G$-invariant scalar potential, as would any of the
eigenvalues of the matrix  $\Sco_1 \Sco_2^\dagger
$. Since we know from the previous chapter no such
potential is possible, it follows that an independent quantity, 
$\Sco_2$, also cannot exist.

\subsection{Including Nucleons}

We next consider the part of $\Scl_1$ which involves
precisely two factors of the nucleon field, and no
derivatives. That is: 
$\Scl_{1,N \pi} = - \Nbr \; f(\vec\theta, M) \; \Pl N +
\hc$, where the transformation laws: $\vec\theta \to
\tw{\vec \theta}$ and 
$N \to \tw N = h(\vec\theta,g) \, N$ imply that the
matrix-valued function, $f(\vec\theta,M)$, must satisfy:
$f(\tw{\vec\theta}, 
g_\ssr M g_\ssl^\dagger) = h \; f(\vec\theta,M) \;
h^\dagger$.

The solution, unique up to normalization, to this condition
is: 
$f = u_\ssr^\dagger M u_\ssl$, where $u_\ssl =
u_\ssr^\dagger = 
\exp\left[ {i \over 2} \; \vec \theta \cdot \vec \tau
\right]$. We therefore find: 
\bg
\label{nucleonmasssplitting}
\Scl_{1,N\pi} &=& - \lambda \; \Nbr \, 
\left[ e^{ {i \over 2} \vec \theta
\cdot \vec \tau} \, M \, e^{{i \over 2} 
\vec \theta \cdot \vec \tau} \right] N
+ \hc   \nn\\  &=& - \lambda \, \Nbr \, 
M \, N - i \lambda \, \left( {\sin
\theta \over 2
\theta} \right) \; \Nbr \Bigl\{ \vec \theta 
\cdot \vec \tau, M \Bigr\} \gamma_5
\, N \nn\\
&& \qquad \qquad \qquad + 
\lambda \left( { \sin^2 { \theta \over 2} \over
\theta^2 } \right) \; \Nbr \Bigl[ \vec 
\theta \cdot \vec \tau M \vec \theta
\cdot \vec \tau + \theta^2 M \Bigr] \; N  \\ 
&=& - \lambda \, \Nbr \, M \, N -
{i \lambda \over 2 \fpi} \, \Nbr \Bigl\{ \vec
\pi \cdot \vec \tau , M \Bigr\} \, N \nn \\ 
&& \qquad \qquad \qquad + {\lambda
\over 4 \fpi^2} \; \Nbr \Bigl[
\vec \pi \cdot \vec \tau M \vec \pi \cdot 
\vec \tau + \vec \pi \cdot \vec \pi
\, M \Bigr] \; N + \cdots .\nn
\nd
As usual, curly braces denote the anticommutator of the
corresponding matrices: $\{ A, B \} = AB + BA$.

Notice that besides providing nonderivative pion-nucleon
couplings, this term also splits the neutron and proton
masses by an amount: 
\eq
\label{npmassdiff}
\delta_\lambda m_\ssn = \lambda (m_d - m_u) .
\eeq
Even though they do not contribute to the 
pion mass splittings, the differing $u$ and $d$ quark 
masses do act to split the masses of the
nucleon isodoublet. Now, $m_u$ and $m_d$
may be determined by repeating the above analysis for the
masses of the lightest eight mesons, 
$\pi, K, \eta$, under the assumption that these are all
pseudo-Goldstone bosons for the symmetry group $SU_\ssl(3)
\times SU_\ssr(3)$ which is appropriate when the $s$ quark
is assumed to be light in addition to the $u$ and $d$
quarks. In principle, once this has been done, 
eq.~\pref{npmassdiff}
permits the constant $\lambda$ to be extracted from
the experimental difference, $m_n - m_p = 
1.293318(9)$ MeV. It is important in so doing to 
include also the contributions of the electromagnetic 
interactions to this mass difference, since these are
similar in size to eq.~\pref{npmassdiff}.\footnote{I 
thank John Donoghue
for reminding me of the importance of this electromagnetic
contribution.}

\section{Soft Pion Theorems}

We may now proceed to work out some of the implications of
the effective lagrangian for low-energy pion-pion
scattering. As usual, the first question must be to ask
which interactions need be considered in which Feynman
graphs in order to properly mimic the low-energy expansion
of the underlying QCD theory. To this end we use the
powercounting results of Chapter 1.

\subsection{Power Counting}

For simplicity we consider here only the case where there
are no nucleons in the low-energy theory, since only in this
case we can directly use the power-counting results
obtained in Chapter 1. As has already been emphasized, 
these results cannot be directly applied to nucleons, 
because they were derived under the
assumption of very light, relativistic fermions, and the
nucleons in the low-energy pion-nucleon theory are very
massive and nonrelativistic at the energies of interest.

\subsubsection{Power Counting in the Symmetric Limit}

We start by omitting all symmetry-breaking terms of the
pion lagrangian which are proportional to the quark masses.
These are considered in the next section. In this case the
pions are massless, and their Goldstone-boson lagrangian
has the form given in 
eq.~\pref{selfintforpions}: 
\eq
\label{chiPT}
\Scl_{\pi \pi} = - {F_\pi^2 } \;
\left[ \frac12 \; \hat{g}_{mn}(\vec\theta) 
\; \partial_\mu \theta^m
\partial^\mu \theta^n +
 {c \over \L_\chi^2} \; h_{mnpq}(\vec\theta) 
\; \partial_\mu \theta^m
\partial^\mu \theta^n \, \partial_\nu \theta^p 
\partial^\nu \theta^q + \cdots
\right], \eeq
where $c$ is a dimensionless number which is, in principle,
calculable from QCD. The ellipses represent an infinite
sequence of additional terms, including several others
which also have four derivatives, as does the displayed
term proportional to $c$.

Eq.~\pref{chiPT} has the form of eq.~\pref{leffpc}, with:  
$f = \sqrt{\fpi \L_\chi}$, $v = \fpi$, and $M = \L_\chi
\sim 4 
\pi \fpi \sim 1$ GeV. The powercounting estimate of 
eq.~\pref{aedwdimreg} for the scattering of pions becomes:
\eq
\label{ChPTpcapp}
\tilde\Sca_\sse(q) \sim \fpi^2 \L_\chi^2 
\; \left( {1 \over \fpi} \right)^E \;
\left( {\L_\chi \over 4 \pi \fpi} \right)^{2L} 
\; \left( {q \over \L_\chi}
\right)^{2 + 2L + \sum_{ik} (k - 2) 
V_{ik}} , \eeq
which is a famous result, due first to Weinberg.

For a given observable, $\Sca_\sse(q)$, the number, $E$, of
external particles is fixed. In this case it is only the
last two factors of eq.~\pref{ChPTpcapp} which
differentiate different types of contributions. We remark
that in practical applications for pion scattering, it
happens that  $\L_\chi \sim 4 \pi \fpi \sim 1$ GeV. As a
result, the second-last factor, $(\L_\chi / 4 \pi
\fpi)^{2L}$, turns out to be $O(1)$ for realistic pion
scattering. This means that it is only the last factor
which controls the importance of various interactions.

According to eq.~\pref{ChPTpcapp}, the contribution of
higher-derivative interactions is clearly only suppressed
by the ratio 
$q/\L_\chi$, which limits us to considering only low-energy
pion dynamics near threshhold. The dominant term in the
expansion in powers of $q/\L_\chi$ corresponds to choosing
the smallest possible value for the quantity $P = 2 + 2L +
\sum_{ik} (k - 2) V_{ik}$. It is noteworthy, when using
this expression, to remark that all of the interactions
have at least two derivatives (we temporarily ignore 
pion masses \etc),
and so $k \geq 2$. Furthermore, it is only the first term in
the derivative expansion which has $k=2$, and so the $k=2$
interaction is unique. 

As a result the lowest value possible for $P$ is $P=2$, and
this is only possible if $L=0$ and if $V_{ik} = 0$ for all
$k > 2$. This implies that the dominant contribution to
pion scattering is computed by using only the first term in
the effective lagrangian, eq.~\pref{chiPT}, and working only
to tree level with these interactions.

The next-to-leading terms in $q/\L_\chi$ then have $P = 4$,
which can arise in either one of two ways. ($i$) We can have
$L=1$ and $V_{ik} = 0$ for all $k > 2$; or ($ii$) we can
have $L=0$ and $V_{i4} = 1$ for some $i$, while $V_{i2}$
can take any value. This states that the subleading,
$O(q^4)$, contribution is obtained by either working to one
loop order using only the interactions of the first term of
the lagrangian of 
eq.~\pref{chiPT}, or using tree graphs having exactly one
vertex taken from the four-derivative interactions in the
lagrangian, as well as any number of interactions from the
first term in the lagrangian.

In this way it is clear how to compute any given order in
the expansion in powers of $q/\L_\chi$.

\subsubsection{Symmetry-Breaking Terms}

Before proceeding to calculations, we must also include one
other feature. We must track the appearance of the explicit
symmetry-breaking terms, of which we only keep those which
are proportional to a single power of the light-quark
masses, 
$m_q \sim m_u, m_d$. These vertices, which come from the
symmetry-breaking term $\Scl_1$, can be very simply
included into the power-counting results of Chapter 1 by
considering all of the non-derivative interactions to be
suppressed by a dimensionless coupling, $c_{k=0} \sim (m_q
/ \L_\chi) \sim 
(m_\pi^2 / \L_\chi^2)$.

With these points in mind, eq.~\pref{ChPTpcapp} becomes: 
\eq
\label{chptpcwfermions}
\tilde\Sca_{E}(q) \sim \L_\chi^2 
\fpi^2 \; \left( {1 \over \fpi} \right)^{E}
\; \left( {q \over \L} \right)^P \; 
\left( { m_q \over \L_\chi}
\right)^{\sum_{k=0} V_{i k}},
\eeq
where $P$ can be written in either of two equivalent ways:  
\bg
\label{equivwaysforpagain}
P &=& 4 - E + \sum_{ik} \left( k +i  
- 4 \right) V_{i k},\nn\\
&=& 2 + 2L + \sum_{ik} \left( k - 
2 \right) V_{i k}.  \nd

Using the first of these forms for $P$ we see that the
contribution to the powercounting estimate due to the
insertion of the symmetry-breaking terms with $k=0$ is:
\eq
\label{kzeroterms}
\prod_{k=0} \left( {q \over \L} 
\right)^{\left( i - 4 \right) V_{ik}} \; \left(
{ m_q \over \L_\chi} \right)^{ V_{ik}} .  \eeq
The dangerous interactions are clearly those for which $i <
4$. For example, the pion mass term, $\sim m_\pi^2 \pi^2$,
has $i = 2$, and so contributes the factor
\eq
\label{pionmassfactor}
\prod_{k=0} \left( { m_q \L_\chi 
\over q^2 } \right)^{V_{ik}}.
\eeq

Now, we are interested in applications for which the
external momenta, $q$, are of order several hundred MeV,
and so $q \sim m_\pi \sim \sqrt{m_q \L_\chi}$. For these
momenta the factor $(m_q \L_\chi / q^2) \sim (m_\pi^2 /
q^2) \sim O(1)$. It follows that it is {\it not} a good
approximation to perturb in the pion mass term, and so we
should include this term in the unperturbed lagrangian.
That is, we should include the pion mass explicitly into
the pion propagator so that $G_\pi(q) = -i/(q^2 + m_\pi^2 -
 i\eps)$. It is legitimate to perturb in all of the other
symmetry-breaking interactions of the scalar potential,
however, since for these $k = 0$ and $i \geq 4$.

\begin{figure}
\epsfbox[-70 670 650 670]{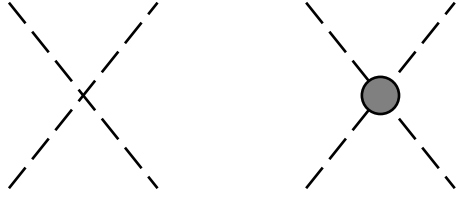}
\vspace{1in}
\caption{The Feynman graphs which give the dominant
contributions to pion-pion scattering in the low-energy
pion-nucleon theory. The first graph uses a vertex
involving two derivatives. The second involves the pion
mass, but no derivatives.}
\end{figure}

We now turn to specific interactions, starting with
pion-pion scattering, for which $E = 4$. In this case the
above powercounting shows that there are precisely two
dominant contributions. The first of these consists of the
tree graph of Fig.~2.2, using the four-point vertex from
the $G$-invariant term which involves two derivatives, 
eq.~\pref{pionderivterms}. The second contribution is also
obtained using the graph of Fig.~2.2, but this time takes
the four-point pion self-interaction from 
the symmetry-breaking scalar potential of 
eq.~\pref{pionmassterm}. Although the first term 
is unsuppressed by the light-quark masses, it gives a 
contribution which is down relative to the second term 
by two powers of external momenta, $q$. Both are  
therefore comparable in size for pions near threshhold,
$q^2 \sim m_\pi^2$. All other graphs are smaller than these
two by powers of either $m_q$ or $q$.

\subsection{Pion-Pion Scattering}

We now compute pion-pion scattering by evaluating the
graphs of Fig.~2.2 using the effective pion self-couplings
of 
eqs.~\pref{pionderivterms} and \ref{pionmassterm}. A
straightforward calculation gives the following $S$ matrix
element for the scattering $\pi_a \pi_b \to \pi_c \pi_d$:
\eq
\label{ampdef}
S(\pi_a \pi_b \to \pi_c \pi_d) = {i  
\delta^4 (p_a + p_b - p_c - p_d)
\over (2 \pi)^2 \sqrt{p_a^0 p_b^0 p_c^0 p_d^0}} \; 
\Sca_{ab,cd},
\eeq
with
\eq
\label{pionscatteringresult}
\Sca_{ab,cd} = { 1 \over \fpi^2} \; \Bigl[ 
\delta_{ab} \delta_{cd} \; (s -
m_\pi^2)  +  \delta_{ac} \delta_{bd} \; (t - m_\pi^2) + 
\delta_{ad}
\delta_{bc}
\; (u - m_\pi^2) \Bigr] ,
\eeq
where the Lorentz-invariant Mandelstam variables, $s = -
(p_a + p_b)^2$, $t = - (p_a - p_c)^2$ and $u = - (p_a -
p_d)^2$ are related by the identity: $s + t + u = 4
m_\pi^2$. In the CM frame $s$, $t$ and $u$ have simple
expressions in terms of the pion energy, $E$, and
three-momentum, $q$: $s = 4 E^2$, $t = -2E^2 + 2 q^2
\cos\vartheta$ and $u = -2 E^2 - 2 q^2 \cos\vartheta$. Here
$\vartheta$ denotes the scattering angle, also in the CM
frame.

Comparison with the data is made using channels having
definite angular momentum and isospin. If we decompose
$\Sca_{ab,cd}$ into combinations, $\Sca^{(\ssi)}$, having
definite initial isospin: 
\eq
\label{isodecomp}
\Sca_{ab,cd} = \Sca^{(0)} \; \nth{3} \, 
\delta_{ab} \delta_{cd}  +
\Sca^{(1)} \; \hf \, ( \delta_{ac} 
\delta_{bd} - \delta_{ad} \delta_{bc} )  +
\Sca^{(2)} \; \left[ \hf ( \delta_{ac} 
\delta_{bd} + \delta_{ad} \delta_{bc} )
-  \nth{3} \delta_{ab} \delta_{cd} \right], \eeq
then
\eq
\label{isoresults}
\Sca^{(0)} = {2s - m_\pi^2 \over \fpi^2}, 
\qquad  \Sca^{(1)} = {t - u \over
\fpi^2}, \qquad  \Sca^{(2)}_{cd} = - \; 
{s - 2 m_\pi^2 \over \fpi^2}.  \eeq

The next step is to resolve these amplitudes into partial
waves:
\eq
\label{pwavedefn}
\Sca^{(\ssi)}_\ell \equiv {1 \over 64 \pi} 
\int_{-1}^1 d\cos\vartheta \;
P_\ell(\cos \vartheta) \, \Sca^{(\ssi)} \eeq
where $P_\ell(\cos\vartheta)$, as usual, denote the
Legendre polynomials (so $P_0(x) = 1$ and $P_1(x) = x$).
Since all of the dependence on $\vartheta$ appears through
the variables $t$ and $u$, and since eqs.~\pref{isoresults}
give $\Sca^{(0)}$ and 
$\Sca^{(2)}$ as functions of $s$ only, it is clear that
only the partial wave $\ell = 0$ is predicted at lowest
order for the even isospin configurations. Also, since
$\Sca^{(1)}$ is strictly linear in $\cos\vartheta$, it only
involves the partial wave $\ell = 1$.

The actual comparison with the data is made by expanding
the (real part of) $\Sca^{(\ssi)}_\ell$ in powers of the
squared pion momentum: $q^2/m_\pi^2 = E^2/m_\pi^2 - 1 = (s
- 4 m_\pi^2)/4 m_\pi^2$. That is, writing
\eq
\label{energyexpansionforplwaves}
\Sca^{(\ssi)}_\ell = \left( {q^2 \over m_\pi^2} 
\right)^\ell \; \left(
a^\ssi_\ell + b^\ssi_\ell \; {q^2 \over m_\pi^2} 
+ \cdots \right) ,
\eeq
defines the pion scattering lengths, $a^\ssi_\ell$, and slopes, 
$b^\ssi_\ell$. Applying these definitions to
eqs.~\pref{isoresults} gives the predictions of the second
and third columns of Table 2.2. Column three gives the
numerical value corresponding to the analytic expression
which is given in column two. The predictions including the
next-order terms in the $q^2/\L_\chi^2$ expansion have also
been worked out, and are given in the fourth column of this
Table.\footnote{I have taken these values from the excellent
book {\it Dynamics of the Standard Model} 
by Donoghue, Golowich and Holstein (see bibliography).} 

\begin{table}
\begin{center}
\begin{tabular}{ccccc}
Parameter & \multicolumn{2}{c}{Leading Order} 
& Next Order & Experiment \\
&&&& \\
$a^0_0$ & ${7 m_\pi^2 / 32 \pi \fpi^2}$ 
& 0.16 & 0.20 & 0.26(5) \\
$b^0_0$ & ${m_\pi^2 / 4 \pi \fpi^2}$ 
& 0.18 & 0.26 & 0.25(3) \\
$a^1_1$ & ${m_\pi^2 / 24 \pi \fpi^2}$ 
& 0.030 & 0.036 & 0.038(2) \\
$a^2_0$ & $- \, {m_\pi^2 / 16 \pi \fpi^2}$ 
& -0.044 &  - 0.041 & - 0.028(12)\\
$b^2_0$ & $- \, {m_\pi^2 / 8 \pi \fpi^2}$ 
& -0.089 & -0.070 & -0.082(8) \\
\end{tabular}
\caption{Theory {\it vs} Experiment for Low-Energy Pion
Scattering} 
\end{center}
\end{table}

Comparison of these predictions with experiment is not
straightforward, since it is not feasible to directly
perform pion-pion scattering experiments. Instead, the
pion-pion scattering amplitudes at low energies are
inferred from their influence on the final state in other
processes, such as $K \to \pi \pi e \nu_e$ or 
$\pi N \to \pi \pi N$. The experimental results, as
obtained from kaon decays, for those quantities which are
predicted to be nonzero at lowest order are listed in the
right-hand-most column of Table 2.2. Data also exist for
other partial waves which are predicted to vanish at lowest
order, such as $I = 0, \ell = 2$, and these are found to be
in good agreement with the nonzero predictions which arise
at next-to-leading order in the low-energy expansion.

This example nicely illustrates the predictive power which
is possible with a low-energy effective lagrangian, even if
it is impossible to predict the values for the couplings of
this lagrangian in terms of an underlying theory. This
predictive power arises because many observables --- \eg\
the pion scattering lengths and slopes --- are all
parameterized in terms of a single constant --- the decay
constant, $\fpi$ --- which can be extracted directly from
experiment. We emphasize that this predictive power holds
regardless of the renormalizability of the effective
theory. Computing to higher orders involves the
introduction of more parameters, but predictions remain
possible provided that more observables are computed than
there are parameters to fix from experiment. The
information underlying these predictions comes from the
symmetries of the underlying theory, as well as the
restrictions due to the comparatively small number of
possible interactions which can appear at low orders of the
low-energy expansion.

\chapter{Magnons: Nonrelativistic Applications}

We now turn to a second application, this time to a
nonrelativistic system. Besides once again illustrating the
utility of the effective-lagrangian techniques, this example
shows how the analysis can be applied to more complicated
condensed-matter systems. It also illustrates how effective
lagrangians permit the separation of the generic predictions
which follow only from general properties such as the
symmetry-breaking patterns, from the details of the models
which may be used to establish these symmetry-breaking
patterns from the underlying physics.

We take as our application the macroscopic behaviour of
ferromagnets and antiferromagnets. These systems exhibit a
transition at low temperatures to a phase which is
characterized by a bulk order parameter, which we call
$\Bfs$ for the ferromagnet and $\Bfn$ for the
antiferromagnet, which transforms under rotations as a
vector. For ferromagnets this order parameter can be taken
to be the overall magnetization of the sample. Because this
order parameter spontaneously breaks the rotational
symmetry, Goldstone bosons must exist and so must appear in
any low-energy (or long-wavelength) description of these
systems. It is the low-energy interactions of these
Goldstone bosons which is described in this chapter.

The distinction between a ferromagnet and an
antiferromagnet requires more information concerning the
underlying material. As for most condensed-matter systems,
the underlying microscopic system consists of an enormous
number of electromagnetically interacting electrons and
atomic nuclei. One picture of what is going on in a
ferromagnet or antiferromagnet consists of imagining the
electrons being reasonably localized to their corresponding
atoms, with these atoms carrying a net magnetic moment due
to its having a net electronic spin. The electron in each
atom which carries the net spin interacts with its
counterparts on neighbouring atoms, resulting in (among
other things) an effective spin-spin interaction between
these atoms. This spin-spin interaction can come about due
to the exchange part of the Coulomb interaction, which
arises due to the antisymmetrization of the wavefunction
which is required because of the statistics of the
electrons. Under varying circumstances one might suppose
this spin-spin interaction to either favour the mutual
alignment of neighbouring spins, or their {\em
anti}alignment (where the spins line up to point in
opposite directions).

The behaviour of such mutually interacting electronic or
atomic spins may then be investigated by abstracting out
just this spin dynamics into a simplifying model. For
example, the interacting electrons can be replaced by a
system of spins which are localized to each of the lattice
sites which define the nuclear positions in the solid. The
mutual interactions of the atoms can be reduced to a
spin-spin coupling having a phenomenological sign and
magnitude, according to whether it is energetically
favourable for neighbouring spins to be aligned or
antialigned. Such models show that at low temperatures
macroscopic numbers of these spins tend to either align or
antialign, according to which of these takes less energy. A
system for which neighbouring spins tend to align, prefers
to acquire a net magnetization since the magnetic moment of
each atom adds to give a macroscopically large total. This
is a ferromagnet. If neighbouring spins prefer to
antialign, then the order of the ground state consists of
spins which alternate in their alignment, with every other
spin pointing in a fixed direction, and the others pointing
in the opposite direction. Such an arrangement is called
antiferromagnetic. These two alternative arrangements are
pictured in Fig.~3.1.

\begin{figure}
\epsfbox[30 740 650 740]{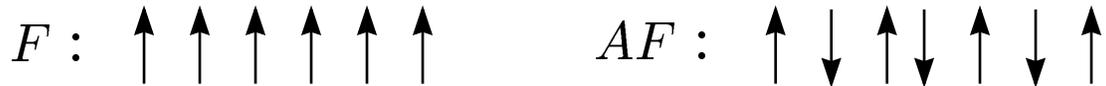}
\vspace{1in}
\caption{The Distinction Between Ferromagnets (F) and
Antiferromagnets (AF).}
\end{figure}

The statistical mechanics of such spin models can
successfully describe many features of real ferromagnets
and antiferromagnets. So long as calculations are based on
models, however, it is difficult to quantitatively assess
their accuracy. One of the purposes of the present chapter
is to show that some predictions for these systems are very
robust, since they do not rely on more than the qualitative
features of the models. The robust predictions are those
which can be formulated completely within the framework of
a low-energy effective theory, and which therefore rely
only on the spectrum and symmetries which dominate at low
energies. The accuracy of this kind of prediction {\em can}
be quantitatively assessed since this accuracy is controlled
by the domain of validity of the effective theory itself.
The role played in this kind of calculations by the details
of an underlying model is simply the prediction of the
quantum numbers and symmetries of the low-energy degrees of
freedom, and so the model need only get these qualitative
features right in order to accurately reproduce the proper
low-energy behaviour.

This line of reasoning, in which some quantitative
predictions can be justified as general low-energy features
of a given system can be of great practical importance. For
example, some very high-precision measurements are now
based on the macroscopic behaviour of complicated condensed
matter systems. Examples are the Josephson effect, or the
Integer Quantum Hall effect, both of which have been used
to fix the best measured value for the electromagnetic fine
structure constant, $\alpha$. These determinations are
accurate to within very small fractions of a percent. We
should only believe such a determination of $\alpha$ if we can
equally accurately justify the theoretical predictions of
the effects on which the determination is based --- a very
tall order if the prediction is to be based on a model of
the underlying system. Happily, such an accuracy is
possible, and is one of the fruits of an effective
lagrangian analysis of these systems.

We now turn to the application of these
effective-lagrangian techniques to the description of the
long-distance, low-energy behaviour of the ordered spin
systems.

\section{Antiferromagnetism: T Invariance}

We start with applications to the low-energy properties of
antiferromagnets. We do so because antiferromagnets
preserve a type of time-reversal symmetry, which makes the
analysis of its low energy behaviour fairly similar to what
would apply for relativistic systems.

For an antiferromagnet, the order parameter, $\Bfn$, can be
taken to be the staggered sum of the spins, $\bfs_i$, for
each lattice site, `$i$':
\eq
\label{aforderparam}
\Bfn = \sum_i (-)^i \; \bfs_i,
\eeq
where the sign, $(-)^i$, is positive for one sublattice for
which all spins are parallel in the ordered state, and is
negative for the other sublattice for which all spins are
antiparallel with those of the first lattice. We shall
refer to these sublattices in what follows as the `even'
and `odd' sublattice respectively. By taking such an
alternating sum we find the expectation $\Avg{ \Bfn} \neq
0$ in the system's ground state, $\ket{\Omega}$.

The action of time-reversal invariance, $T$, is to reverse
the sign of the spin of every site: $\bfs_i \to - \bfs_i$.
Although this transformation also reverses the order
parameter, $\Bfn \to 
- \Bfn$, it may be combined with another broken symmetry,
$S$, to obtain a transformation, $\tw T = TS$, which {\em
is} a symmetry of $\Bfn$. This other symmetry, $S$,
consists of a translation (or shift) of the whole lattice
by a single lattice site, taking the entire `even'
sublattice onto the `odd' sublattice, and vice versa. Since
both $S$ and $T$ act to reverse the direction of $\Bfn$,
they preserve $\Bfn$ when they are performed together.

We next turn to the construction of the general low-energy
lagrangian for the Goldstone bosons for the breaking of
rotation invariance --- called {\em magnons} --- for these
systems.

\subsection{The Nonlinear Realization}

The first step is to identify the symmetry breaking
pattern, $G \to H$. At first it is tempting to assume that
the role of $G$ should be played by the spacetime
symmetries, since these include rotations. This is not
correct, however, for several reasons. Firstly, the
spacetime symmetries of a lattice do not consist of the
full group of translations and rotations since these
symmetries are broken by the lattice itself. The unbroken
subgroup consists only of the group of lattice symmetries:
\ie\ those translations and rotations which take the
lattice to itself. There are indeed Goldstone bosons for
the spontaneous breaking of translational and rotational
symmetry down to this lattice group, but these are the
phonons and are not the focus of the present analysis.

In fact, the rotations of the spins on the lattice can be
taken to be an internal $SU(2)$, or $SO(3)$, symmetry,
rather than a spacetime symmetry. This is because the
action of rotations on the intrinsic spin of a particle
becomes an independent internal symmetry, separate from
spacetime rotations, in the limit that the particle
involved is nonrelativistic. This is because all of the
interactions which couple the orbital angular momentum with
the spin angular momentum vanish in the limit that the
particle mass tends to infinity. Since the spins of
interest for real systems are those for nonrelativistic
electrons or atoms, we may consider the broken symmetry
group to be an internal symmetry, $G = SU(2)$ (which equals
$G = SO(3)$, locally). In real life, the electron mass is
not infinite, so there are small `spin-orbit' effects which
really do break the internal spin symmetry. These introduce
small corrections to predictions based on this symmetry,
such as the exact gaplessness of the Goldstone mode. We
ignore any such symmetry-breaking effects in what follows.

The order parameter for the symmetry breaking is the
vector, 
$\Bfn$, itself, and so the group of unbroken
transformations is $H = U(1)$ (or, $SO(2)$), consisting of
rotations about the axis defined by $\Avg{\Bfn}$. The coset
space which is parameterized by the Goldstone bosons is
therefore the space $G/H = SU(2) 
/ U(1)$, or $SO(3)/SO(2)$. This last way of writing $G/H$
identifies it as a two-sphere, $S_2$, since this describes
the space swept out by the action of rotations on a vector,
$\Avg{\Bfn}$, of fixed length.

We now have two ways to proceed. We could, on the one hand,
follow the steps outlined in Chapter 1 to construct the
nonlinear realization of $G/H$ and its invariant
lagrangian. Instead we choose here to take a simpler route.
As discussed in Chapter 1, the most general possible
low-energy Goldstone boson lagrangian must necessarily take
the form of eq.~\pref{genlowdtermnr}:
\bg
\label{afmagnonlagrgenform}
\Scl_{\sss AF} &=& {F_t^2 \over 2} \;
\hat{g}_{\alpha\beta}(\theta) \;
\dot\theta^\alpha \dot\theta^\beta - \, 
{F_s^2 \over 2} \;
\hat{g}_{\alpha\beta}(\theta) \del
\theta^\alpha \cdot \del\theta^\beta \nn\\
&& \qquad \qquad \qquad + 
\hbox{(higher-derivative terms)}, \nd
where $\hat{g}_{\alpha\beta}(\theta)$ is an
$SO(3)$-invariant metric on the two-sphere. This form is
the most general consistent with the nonlinearly-realized
$SO(3)$ invariance, as well as with invariance with respect
to translations, rotations and the time-reversal-like
symmetry, $\tw T$, described above.\footnote{We use
translation and rotational invariance for simplicity, even
though these are too restrictive for real solids, for which
only the lattice symmetries should be imposed. For some
lattices, such as cubic ones, the implications of the
lattice group turn out to be the same as what is obtained
using rotation and translation invariance, at least for
those interactions involving the fewest derivatives which
are studied here.} The $\tw T$ invariance rules out
interactions having an odd number of time derivatives, such
as the term linear in time derivatives which was constructed
in Chapter 1.

The main point to be made is that the lagrangian given in 
eq.~\pref{afmagnonlagrgenform} is unique, a result which
follows from the uniqueness of the $SO(3)$-invariant metric
on the two-sphere. The uniqueness of this metric is a
consequence of the fact that the two broken generators of
$SO(3)/SO(2)$  form an irreducible representation of the
unbroken subgroup $SO(2)$ --- a condition which was shown
to imply a unique metric in Chapter 1. Since it is unique,
any representation of it is equally good and we choose here
to use the familiar polar coordinates, $(\theta,\phi)$, for
the two-sphere, in terms of which the invariant metric has
the usual expression: $ds^2 = d\theta^2 + \sin^2 \theta
d\phi^2$. With this choice the above lagrangian becomes:
\bg
\label{afmagnonpolcoords}
\Scl_{\sss AF} &=& {F_t^2 \over 2} 
\; \Bigl( \dot\theta^2 + \sin^2
\theta \; \dot\phi^2 \Bigr)  - \, 
{F_s^2 \over 2} \; \Bigl( \del\theta
\cdot \del \theta + \sin^2 \theta \; 
\del \phi \cdot \del \phi \Bigr) \nn\\
&& \qquad\qquad\qquad + 
\hbox{(higher-derivative terms)}. \nd

Alternatively, we can equally well parameterize $S_2$ using
a unit vector, $\vec n$, where $n_x = \sin\theta \cos\phi$,
$n_y = 
\sin\theta \sin\phi$ and $n_z = \cos\theta$, so $\vec n
\cdot 
\vec n = 1$. Then  
\bg
\label{afmagnonveccoords}
\Scl_{\sss AF} &=& {F_t^2 \over 2} 
\; \dot{\vec n} \cdot \dot{\vec n}  - \,
{F_s^2 \over 2} \; \del\vec n \cdot \del 
\vec n \nn\\ && \qquad\qquad\qquad +
\hbox{(higher-derivative terms)}.  \nd
This variable, $\vec n(\bfr,t)$, makes most clear the
physical interpretation of the Goldstone modes: they
describe long-wavelength variations in the direction of the
order parameter 
$\Avg{\Bfn}$. It has the drawback of hiding the
self-interactions which are implied by $\Scl_{\sss AF}$,
since the lagrangian of eq.~\pref{afmagnonveccoords} is
purely quadratic in $\vec n$. The self-interactions, which
are manifest in expression 
\pref{afmagnonpolcoords}, are nonetheless present, and are
hidden in the constraint $\vec n \cdot \vec n = 1$.

The nonlinear realization of $SO(3)$ transformations on
these variables is straightforward to work out, starting
with the transformation rule for $\vec s$: $\delta \vec n =
\vec \omega 
\times \vec n$, where the vector $\vec\omega$ represents
the three $SO(3)$ transformation parameters. This implies
the transformations:
\bg
\label{rotsinpolcoords}
\delta\theta &=& \omega_y \cos\phi - 
\omega_x \sin\phi  \nn\\
\delta\phi &=& \omega_z - \omega_x 
\cot\theta \cos\phi - \omega_y
\cot\theta \sin \phi.
\nd

With these transformation laws we may immediately write
down the first terms in a derivative expansion of the
Noether currents, 
$\vec\jmath^\mu = (\vec\rho, \vec\bfj)$, for the $SO(3)$
invariance in the low-energy effective theory. They may be
most compactly written:
\eq
\label{rotationcurrentsforafvec}
\vec\rho = F_t^2 \; (\dot{\vec n} 
\times \vec n) + \cdots \qquad \hbox{and}
\qquad \vec\bfj = - \, F_s^2 \; (\del 
\vec n \times \vec n) + \cdots ,
\eeq
where the dots are a reminder of the unwritten
higher-derivative contributions.

The other quantities which arise in the general nonlinear
realization may also be constructed in terms of these
variables. For example, the four independent components
of the covariant quantity (zweibein),
${e^\alpha}_\beta(\theta,\phi)$, are most easily constructed,
following the geometrical picture of Chapter 1, 
as the components of any two orthogonal vectors which are
tangent to the two sphere. These may be found 
by differentiating the unit vector, $\vec n$, because the 
identity $\vec n \cdot \vec n = 1$ implies 
$\vec n \cdot \delta \vec n = 0$, for any variation, 
$\delta \vec n$. Denoting these two vectors by 
$\vec e_\theta = \partial \vec n/\partial \theta$ and $\vec
e_\phi = \partial \vec n / \partial \phi$, we have in cartesian
components:
\bg
\label{zweibeinonsphere}
&& (\vec e_\theta)_x = \cos\theta 
\cos \phi, \qquad   (\vec e_\theta)_y =
\cos\theta \sin \phi,  \qquad  (\vec e_\theta)_z 
= - \sin\theta ; \nn\\ &&
(\vec e_\phi)_x = - \sin\theta \sin \phi, 
\qquad   (\vec e_\phi)_y = \sin\theta
\cos \phi,  \qquad   (\vec e_\phi)_z = 0 .
\nd
Clearly these vectors satisfy $\vec e_\theta \cdot \vec
e_\phi = \vec e_\theta \cdot \vec n = \vec e_\phi \cdot 
\vec n = 0$, and $\vec e_\theta \cdot \vec e_\theta = 1$, 
$\vec e_\phi \cdot \vec e_\phi = \sin^2 \theta$, so
$\vec e_\alpha \cdot \vec e_\beta = \hat g_{\alpha\beta}$,
as required. The two-by-two matrix, 
${e^\alpha}_\beta$, of components may be found by
expressing the two vectors, $\vec e_\beta$, defined by 
eqs.~\pref{zweibeinonsphere}, as linear combinations of any
two orthonormal basis vectors, ${\vec t}^\alpha$, which lie
tangent to the sphere: \ie\ ${e^\alpha}_\beta = {\vec
t}^\alpha \cdot \vec 
e_\beta$. Using the basis vectors $\vec e_\beta$ themselves
for this purpose leads to the result:
\eq
\label{zweibeinmatrix}
\pmatrix{{e^\theta}_\theta & 
{e^\theta}_\phi \cr {e^\phi}_\theta &
{e^\phi}_\phi \cr} = \pmatrix{ 1 & 0 \cr 0 & \sin\theta
\cr}.
\eeq

\subsection{Physical Applications}

Any physical question that could be asked of the low-energy
limit of the underlying theory can equally well be addressed
using the low-energy effective lagrangian. In particular,
the lagrangian just derived for the Goldstone modes for a
ferromagnet may be used to describe the response (at zero
or nonzero, but small, temperature) of the system to probes
which couple to the spin degrees of freedom.

In order to interpret the constants $F_s$ and $F_t$ it is
convenient to expand the field $\vec n$, or equivalently
$\theta$ and $\phi$, about its vacuum configuration, $\vec
n_0 = 
\Avg{\Bfn}$. We are free to perform an $SO(3)$ rotation to
choose the direction of $\vec n_0$ arbitrarily, and so we
use this freedom to ensure that $\vec n_0$ points up the
positive $x$-axis. This implies that $\theta$ and $\phi$
take the vacuum values, 
$\theta_0 = {\pi \over 2}$ and $\phi_0 = 0$. Writing the
canonically normalized fluctuation fields by $\theta ={ \pi
\over 2} + \vartheta/F_t$ and $\phi = \varphi/F_t$, the
lagrangian becomes:
\bg
\label{canonlagrforaf}
\Scl_{\sss AF} &=& \hf \; \Bigl( 
\dot\vartheta^2 - v^2 \;   \del\vartheta \cdot
\del \vartheta \Bigr) + \hf \, \cos^2 
\left({\vartheta \over F_t}
\right) \; \Bigl( \dot\varphi^2 - v^2 \; \del 
\varphi \cdot \del \varphi
\Bigr) \nn\\
&& \qquad \qquad \qquad + 
\hbox{(higher-derivative terms)}, \\
&=& \hf \; \Bigl( \dot\vartheta^2 - 
v^2 \; \del\vartheta \cdot \del \vartheta +
\dot\varphi^2 - v^2 \; \del \varphi 
\cdot \del \varphi  \Bigr) \nn\\
&& \qquad \qquad \qquad  - \, 
{\vartheta^2  \over 2 F_t^2} \; \Bigl(
\dot\varphi^2 - v^2 \; \del\varphi \cdot 
\del\varphi \Bigr) + \cdots .\nn
\nd
The constant, $v$, here represents the ratio $v = F_s/F_t$.
The ellipses denote terms that involve at least six powers
of the fields, or which involve more than two derivatives
with respect to either position or time.

The terms quadratic in the fields describe two real modes
which propagate according to the linear dispersion law:
$E(p) = v p$. These modes physically correspond to {\em
spin waves}: small, long-wavelength precessions of the
vector $\vec n$ about its vacuum value, $\vec n_0$. They
carry $\pm 1$ unit of the conserved $SO(2)$ spin in the
direction parallel to $\Avg{\Bfn}$. This gives the physical
interpretation of the parameter $v$ to be the velocity of
propagation of these modes. The condition that this
velocity must be smaller than the velocity of light is $v
\leq c = 1$ (in fundamental units), or, equivalently, $F_s
\leq F_t$. $F_t$ is similarly seen to govern the strength
of the interaction terms in eq.~\pref{canonlagrforaf}.

These modes and their interactions are amenable to
experimental study through their electromagnetic couplings.
Although magnons carry no electric charge, they do couple to
magnetic fields, $\Bfb$, due to the interactions of the
microscopic magnetic moments which participate in the
long-wavelength spin waves. This gives a coupling of the
magnetic field to the medium's spin density. This coupling
can be probed, for example, by scattering neutrons which
are also electrically neutral but which carry an intrinsic
magnetic moment which interacts with magnetic fields.

The interaction between magnons and electromagnetic fields
is therefore given by a term of the form: $\Scl_{\rm em} =
-\nu \; 
\vec{s} \cdot \Bfb$, where $\vec{s}$ is the system's spin
density. The lowest-dimension effective interaction between
magnons and electromagnetic fields is now obtained by
expressing the spin density in a derivative expansion,
using Noether's result, 
eq.~\pref{rotationcurrentsforafvec}, for $\vec{s}=
\vec\rho$. 
The result is:
\bg
\label{magnonemcoupling}
\Scl_{\rm em} &=& -\nu \; \vec{s} 
\cdot \Bfb, \nn\\ &=& -\nu F^2_t \Bigl[ B_x
\; (\dot\theta \, \sin \phi + \dot \phi \,
\sin\theta \cos\theta \cos\phi ) \nn\\    
&& \qquad \qquad
+ B_y \; (- \dot\theta \, \cos\phi + 
\dot\phi \, \sin\theta \cos\theta
\sin\phi)
- B_z \; \dot\phi \, \sin^2\theta) \Bigr] \nn\\  
&=& {\nu  F_t} \Bigl( B_y \;
\dot\vartheta + B_z \; \dot\varphi \Bigr) + \cdots ,
\nd
where $\nu$ is an effective coupling parameter having the
dimensions of magnetic moment (or: inverse mass, in
fundamental units).  Notice that the time derivative in
this interaction ensures invariance with respect to $\tw T$
transformations, under which $\vec s \to -\vec s$ and $\Bfb
\to - \Bfb$.

A nonrelativistic neutron couples to the magnetic field
with strength 
\eq
\label{neutronmoment}
\Scl = - \mu_\ssn \; n^\dagger \vec \sigma \, n 
\cdot \Bfb, \eeq
where $n(x)$ denotes the two-component neutron field, and 
$\vec\sigma$ denotes the Pauli matrices acting in the
two-component neutron spin space. The constant $\mu_\ssn$
is the neutron magnetic moment which is, in order of
magnitude, 
$\mu_\ssn \sim e/m_\ssn$. As usual $e$ is the
electromagnetic coupling constant (\ie\ the proton charge)
and $m_\ssn$ is the neutron (or nucleon) mass.

Using these interactions, eqs.~\pref{magnonemcoupling} and
\pref{neutronmoment}, the cross section per-unit-volume for
neutron scattering from the medium can be computed. For
slowly-moving neutrons, and under the assumption that only
the momentum, $\bfp'$, of the scattered neutron is measured
we find:
\eq
\label{crosssection}
{d \sigma \over V d^2\bfp'} = 
{\mu_\ssn^2 \nu^2 \over 4 \pi^3 v_\ssn}
\; V_{ij}(\bfp - \bfp') \; 
S_{ij}(E-E',\bfp - \bfp'). \eeq
Here $v_\ssn \ll 1$ is the speed of the incoming,
nonrelativistic, neutron, and $V$ is the volume of the
medium whose magnons are responsible for scattering the
neutrons. $E$ and $E'$ are the energies of the initial and
scattered neutrons, and $\bfp$ and 
$\bfp'$ are their momenta. The function, $V_{ij}(\bfq)$, is
the magnetic-moment interaction potential (in momentum
space) which arises from the electromagnetic interaction
between the neutron and magnon fields: 
\eq
\label{magmominteraction}
V_{ij}(\bfq) = \delta_{ij} - 
{ q_i q_j \over \bfq^2} . \eeq

The quantity $S_{ij}(\omega,\bfq)$ is the spin correlation
function, which contains all of the information about the
scattering medium which is relevant for analyzing the
neutron collision. It is defined by:
\eq
\label{defnofN}
S_{ij}(\omega, \bfq) = \int dt \, 
d^3\bfr \; \Avg{s_i(\bfr,t) \, s_j(0)} \;
e^{i \omega t - i \bfq \cdot \bfr}.
\eeq
The quantity $\Avg{s_i(\bfr,t) \, s_j(0)} =  \Tr[ \rho \; 
s_i(\bfr,t) \, s_j(0) ]$ appearing in here is the
expectation defined by the density matrix, $\rho$, which
characterizes the initial state of the medium with which
the neutron scatters. Notice that it is 
{\em not} a time-ordered product of operators which appears
in this expectation.

This correlation function may be explicitly computed at low
energies using the previously-derived effective lagrangian
which describes the low-energy magnon self-interactions.
The simplest case is that for which the medium is initially
in the no-magnon ground state, and where the energies
involved are low enough to neglect the magnon
self-interactions. In this case the spin density may be
well-approximated by the first terms of 
eq.~\pref{rotationcurrentsforafvec}. Then:
\eq
\label{resultforgroundstate}
S_{ij}(\omega,\bfq) = { \pi \omega^2 
F_t^2\over v |\bfq| } \; \delta_{ij} \;
\delta(\omega - v |\bfq|),
\eeq
where $v = F_s/F_t$ is the speed of magnon propagation.

We see that the cross section has sharp peaks when the
neutron energy and momentum transfers are related by the
magnon dispersion relation: $E - E' = v|\bfp - \bfp'|$.
This corresponds to inelastic scattering in which the
neutron transfers its energy and momentum to the medium by
creating a magnon. According to
eq.~\pref{resultforgroundstate} the resulting peaks in the
cross section are infinitely sharp, but in real systems
they have a finite width due to processes which cause the
produced magnons to scatter or decay. If the lifetime for
undisturbed magnon propagation, $\tau = 1/\Gamma$, is much
longer than the other interaction times of interest in the
neutron scattering, then the delta function in
eq.~\pref{resultforgroundstate} becomes replaced by the
lineshape:
\eq
\label{resultwithwidthincluded}
\delta(\omega - v |\bfq|) \to { \Gamma \over 2 \pi} 
\; {1 \over (\omega - v
|\bfq|)^2 + \nth{4} \; \Gamma^2}.
\eeq

Measurements of the positions and widths of these peaks as
functions of the scattered neutron energy and momentum can
be used to measure the magnon dispersion relation --- and
so the constant $v$ --- and its decay rate, $\Gamma$, for
the scattering medium. The predicted linear spectrum is
indeed found when neutrons are scattered from
antiferromagnets. The situation when neutrons scatter from
ferromagnets is different, as we shall now see.

\section{Ferromagnetism: T Breaking}

For ferromagnets, the order parameter is simply the total
magnetization, or the total spin, of the system. Since this
defines a vector in space, the spin symmetry, $G = SO(3)$,
is spontaneously broken to $H = SO(2)$ just as for an
antiferromagnet. The low energy behaviour of ferromagnets
and antiferromagnets are nevertheless quite different, and
this difference is due to the fact that time-reversal
symmetry is broken in a ferromagnet but not in an
antiferromagnet.

We denote by $\Bfs$ the order parameter for a ferromagnet,
which is related to the spins, $\bfs_i$, of the underlying
spin model by:  
\eq
\label{forderparam}
\Bfs = \sum_i \bfs_i.
\eeq
Ferromagnets are characterized by having ground states for
which there is a nonzero expectation for this quantity:
$\Avg{ \Bfs} \neq 0$.

The action of time-reversal invariance, $T$, is to reverse
the sign of the spin of every site, $\bfs_i \to - \bfs_i$,
and so it does the same for the order parameter, $\Bfs \to
- \Bfs$. The difference with the antiferromagnetic case
arises because for a ferromagnet it is not possible to find
another broken symmetry which combines with time reversal to
preserve $\Bfs$. The low-energy effective theory can
therefore contain $T$-violating terms, and this changes the
properties of its Goldstone bosons in an important way.

\subsection{The Nonlinear Realization}

Since the symmetry-breaking pattern for both ferromagnets
and antiferromagnets is $SO(3) \to SO(2)$, the nonlinear
realization of this symmetry on the Goldstone bosons is
identical for these two systems. We therefore use the same
polar coordinates in this case, $\theta$ and $\phi$, as in
the previous sections. As before it is convenient to use
these to define a unit vector, denoted by $\vec s$, with
components $s_x = \sin\theta \cos\phi$, $s_y = \sin\theta
\sin\phi$ and $s_z = \cos\theta$, so that $\vec s \cdot
\vec s = 1$. The field, $\vec s(\bfr,t)$, again describes
long-wavelength oscillations in the direction of
$\Avg{\Bfs}$.

The action of $SO(3)$ on these variables is once more given
by eq.~\pref{rotsinpolcoords}, and the term in the effective
lagrangian which involves the fewest spatial derivatives is
again determined to be:
\eq
\label{fspatialterms}
\Scl_{\ssf,s} =  - \, {F_s^2 \over 2} 
\; \Bigl( \del\theta \cdot \del \theta +
\sin^2 \theta \; \del \phi \cdot \del \phi \Bigr) .
\eeq

The new features appear once the term with the fewest time
derivatives is constructed. As is discussed in some detail
in Chapter 1, this involves only a single time derivative
because of the broken time-reversal symmetry. It has the
form given by 
eq.~\pref{tvterm}:
\eq
\label{tvtermagain}
\Scl_{\ssf,t} = - A_\alpha(\theta) \; 
\dot \theta^\alpha , \eeq
where the coefficient function, $A_\alpha(\theta)$, may be
considered to be a gauge field defined on the coset space
$G/H$. In Chapter 1 it was determined that the condition
that this term be $G$ invariant is that $A_\alpha$ must
only be $G$-invariant up to a gauge transformation, in the
sense that:
\eq
\label{inveqagain}
\Lie{\xi} A_\alpha \equiv \xi^\beta 
\partial_\beta A_\alpha + A_\beta
\partial_\alpha \xi^\beta = \partial_\alpha 
\Omega_\xi , \eeq
for each generator $\delta\theta^\alpha = \xi^\alpha$ of
$G$ on $G/H$, where $\Omega_\xi(\theta)$ are a collection
of scalar functions on $G/H$. This last condition is
equivalent to the invariance of the field strength for
$A_\alpha$: $\Lie{\xi} 
F_{\alpha\beta} = 0$. Our problem is to explicitly
construct such a gauge potential for the example of
interest, $G/H = SO(3) / SO(2) \equiv S_2$.

This construction is quite simple. Since our coset space is
two dimensional, it is always possible to write the field
strength in terms of a scalar field: $F_{\alpha\beta} =
\Scb(\theta) \; 
\eps_{\alpha\beta}$, where $\eps_{\alpha\beta}$ is the
antisymmetric tensor which is constructed using the coset's
$G$-invariant metric. The condition that $F_{\alpha\beta}$
be $G$ invariant is then equivalent to the invariance of
$\Scb$. That is: 
\eq
\label{invcondforb}
\Lie{\xi} \Scb \equiv \xi^\alpha \; \partial_\alpha \Scb =
0,
\eeq
which is only possible for all $G$ transformations if
$\Scb$ is a constant, independent of $\theta^\alpha$.

Our solution for $\Scl_{\ssf,t}$ for a ferromagnet
therefore simply boils down to the construction of a gauge
potential for which $F_{\alpha\beta} = \Scb \;
\eps_{\alpha\beta}$ on the two-sphere, $S_2 = SO(3)/SO(2)$.
But such a gauge potential is very familiar --- it is the
gauge potential for a magnetic monopole positioned at the
centre of the two-sphere. The result may therefore be
written (locally) as: $A_\alpha \; d\theta^\alpha = \Scb \;
\cos\theta \; 
d\phi$, and so the corresponding lagrangian is given by
\eq
\label{ftimeterms}
\Scl_{\ssf,t} = - \, \Scb \; \cos\theta \; \dot\phi, \eeq
where $\Scb$ is a constant. In terms of the vectors $\vec
s$, 
$\vec e_\theta = \partial \vec s / \partial \theta$ and
$\vec e_\phi = \partial \vec s / \partial \phi$ this may be
written:
\eq
\label{fancyftimeterms}
\Scl_{\ssf,t} = - \, \Scb \; \vec s 
\cdot \Bigl( \vec e_\theta \times \dot{\vec
e}_\theta \Bigr) .
\eeq

The complete Goldstone boson lagrangian containing the
fewest time and space derivatives is found by combining the
contributions of eqs.~\pref{fspatialterms} and
\pref{ftimeterms}, giving:
\eq
\label{ftotalterms}
\Scl_\ssf = - \, \Scb \; \cos\theta \; 
\dot\phi - \, {F_s^2 \over 2} \; \Bigl(
\del\theta \cdot \del \theta + \sin^2 
\theta \; \del \phi \cdot \del \phi
\Bigr) .
\eeq

It is instructive to compute the Noether currents for the
$SO(3)$ symmetry that is implied by this lagrangian
density. The conserved current density is the same as was
found for the antiferromagnet:
\eq
\label{conscurrentdens}
\vec\bfj = F_s^2 \; (\vec s \times 
\del \vec s) + \cdots . \eeq
In computing the corresponding expression for the charge
density, it is necessary to keep in mind that under these
transformations 
$\Scl_\ssf$ is not invariant, but instead transforms into a
total derivative:   
\eq
\label{totderivvariationofl}
\delta \Scl_\ssf = - \, {d \Omega \over dt} 
= - \, { \Scb \over \sin \theta} \;
\Bigl( \omega_x \,\cos\phi + \omega_y 
\, \sin\phi \Bigr). \eeq
Using this in the general expression, eq.~\pref{defofj},
for the Noether current gives the conserved charge density: 
\eq
\label{conschargedensityforfm}
\vec \rho = \Scb \; \vec s + \cdots .
\eeq
The ellipses in this equation, and in
eq.~\pref{conscurrentdens}, represent more complicated
terms which are suppressed by additional derivatives. As is
easily verified, the classical equations of motion for the
lagrangian, \pref{ftotalterms}, are equivalent to the
conservation condition for this current:
\eq
\label{landaulifshitzeq}
\dot{\vec s} + k \; \Bigl( \vec s \times 
\del^2 \vec s \Bigr) = 0.
\eeq
This equation has long been known to describe
long-wavelength spin waves in ferromagnets, and is called
the {\em Landau-Lifshitz} equation. The constant, $k$, here
is given in terms of $F_s$ and $\Scb$ by 
\eq
\label{lleqconst}
k = {F_s^2 \over \Scb}.
\eeq

Equation \pref{conschargedensityforfm} brings out a feature
of time-reversal breaking systems which is qualitatively
different from those which preserve time reversal. It
states that it is the conserved charge density itself,
$\vec \rho$, which acquires a vacuum expectation value and
breaks the $SO(3)$ symmetry:
\eq
\label{rhosvev}
\Avg{ \vec \rho} = \Scb \; \Avg{ \vec s} 
= \Scb \; \vec s_0 \neq 0.
\eeq
Clearly the breaking of time reversal (and lorentz
invariance) are prerequisites for the acquisition of a
nonzero ground-state expectation value for $\vec \rho$,
which is the time component of a current.

\subsection{Physical Applications}

The propagation of small-amplitude, long-wavelength spin
waves is therefore seen to be completely determined by the
underlying pattern of spontaneous symmetry breaking: $SO(3)
\to SO(2)$ together with $T$ violation. Linearizing the
Landau Lifshitz equation, eq.~\pref{landaulifshitzeq},
shows the resulting propagating modes to have the quadratic
dispersion relation: 
\eq
\label{ferromagnetdispreln}
E(p) = k p^2.
\eeq
This dispersion relation, and the value of the constant
$k$, can be measured by neutron scattering, in a manner
that is similar to what was found for antiferromagnets. We
highlight here only the differences which arise from the
antiferromagnetic example.

The lowest-dimension effective interaction which couples
the field $\vec s$ to electromagnetic fields in the
ferromagnetic case is:  
\bg
\label{ferromagnonemcoupling}
\Scl_{\rm em} &=& - \mu \; \vec s 
\cdot \Bfb, \nn\\ &=& - \mu \Scb \; \Bigl(
B_x \; \sin\theta \cos\phi + B_y \; \sin\theta
\sin\phi + B_z \; \cos\theta  \Bigr) , \nn\\ 
&=& - \mu \Scb \; B_x  - \mu \Scb
\; \Bigl(  B_y \; \delta \phi -
B_z \; \delta \theta \Bigr) + \cdots,
\nd
where $\mu$ is an effective coupling parameter and we have
taken the expectation value, $\Avg{\Bfs}$ to point in the
positive $x$ direction, so $\theta = {\pi \over 2} + \delta
\theta$ and $\phi = \delta \phi$. The constant term,
independent of $\delta\theta$ and $\delta\phi$, in the
final line of 
eqs.~\pref{ferromagnonemcoupling} gives the interaction
energy between the magnetic field and the expectation
value, 
$\Avg{\Bfs}$. This permits the physical interpretation of
the constant $\mu$ as the magnetic-moment density of the
material.

This interaction between $\vec s$ and $\Bfb$ breaks $T$
invariance, and has the following puzzling feature. It does
{\em not} involve any derivatives of the Goldstone boson
fields, 
$\theta$ and $\phi$, in apparent contradiction with the
general results of Chapter 1. In fact, the absence of
derivatives in 
eq.~\pref{ferromagnonemcoupling} is very much like the
absence of derivatives in the pion mass term. This is
because the coupling between $\vec s$ and $\Bfb$ relates
the internal spin $SO(3)$ symmetry to ordinary rotations in
space, and so destroys the freedom to consider both as
separate symmetries. But because rotation invariance is a
spacetime symmetry, the derivations of Chapter 1 do not
directly apply, since these assumed the action of internal
symmetries from the outset in the transformation rules of
the fields.

Once more taking the neutron coupling to the magnetic field
as in eq.~\pref{neutronmoment}, we may compute the cross
section for inelastic neutron scattering. For slowly-moving
neutrons, and under the assumption that only the momentum,
$\bfp'$, of the scattered neutron is measured we find:
\eq
\label{ferrocrosssection}
{d \sigma \over V d^2\bfp'} = 
{\mu_\ssn^2 \mu^2 \over 4 \pi^3 v_\ssn} \;
V_{ij}(\bfp - \bfp') \; S_{ij}(E-E',\bfp - \bfp').  \eeq
The variables are the same as for the antiferromagnetic
example: $v_\ssn$ and $\mu_\ssn$ are the speed and magnetic
moment of the slow incoming neutron, $V$ is the volume of
the medium, $E$ ($E'$) is the energy of the initial (final)
neutron, and $\bfp$ 
($\bfp'$) are the corresponding momenta. The
magnetic-moment interaction, $V_{ij}(\bfq)$, is as given in 
eq.~\pref{magmominteraction}.

The medium-dependent quantity, $S_{ij}(\omega,\bfq)$, once
more represents the spin correlation function, defined by 
eq.~\pref{defnofN}. As was the case for the ferromagnet
this is dominated by sharp peaks when the neutron scatters
to produce a magnon, and so has an energy and momentum
transfer related by the magnon dispersion relation. For a
ferromagnet this is: $E - E' = k (\bfp - \bfp')^2$.
Measurements of these peaks as functions of the scattered
neutron energy and momentum indeed verifies the quadratic
dispersion relation, and can be used to measure the
constant $k$.

A second consequence of the quadratic magnon dispersion
relation is the temperature dependence of the
magnetization, $M = |\Bfm|$, of a sample at very low
temperatures. Since the magnon field describes the
long-wavelength deviations of the net magnetization from
its ground state value, the net magnetization at very low
temperatures is simply proportional to the average magnon
occupation number. That is:
\eq
\label{lowtmagnetization}
M(0) - M(T)  \propto M(0) \; \int 
d^3\bfp \; n\left( {E\over T} \right),
\eeq
where $n(E/T) = (\exp[E/T] - 1)^{-1}$ is the Bose-Einstein
distribution. The temperature dependence of this result can
be determined by changing integration variables from $p$ to
the dimensionless quantity $x = E/T$. If $E(p) \propto
p^z$, for some power, $z$, then:
\eq
\label{scalinglaws}
p^2 \, dp = p^2 \, {dp \over dE} \; 
dE \propto E^{2/z} \; E^{-(z-1)/z} \; dE
\propto T^{3/z}.
\eeq
For $z = 2$ this predicts $[M(0) - M(T)]/M(0) \propto
T^{3/2}$, in agreement with low-temperature observations.

\chapter{$SO(5)$-Invariance and Superconductors}

The techniques described herein have recently proven useful
to analyze the consequences of a remarkable  proposal for
the existence of an $SO(5)$ invariance  amongst the
cuprates which exhibit high-temperature superconductivity.
This chapter presents a bare-bones outline of this
proposal, together with a brief summary of the
Goldstone-boson properties which emerge.

\section{$SO(5)$ Symmetry}

Although a proper presentation of the arguments  for ---
and against, since the subject remains controversial ---
the $SO(5)$ proposal is beyond the scope of this review, 
the form of the proposed symmetry itself is easy
to state. The starting point is the following experimental
fact: by performing small adjustments to any of the
high-\Tc\ cuprates, it is possible to convert them from
superconductors into antiferromagnets. This adjustment
is typically accomplished in practice
by altering the `doping', which means that
atoms having different valences are randomly substituted 
into a portion of the unit cells of the material of interest
For example the element Sr might be substituted
for the element La in some fraction, $x$, of the unit cells. 
Physically, this substitution
has the effect of changing the number of charge carriers 
in the band from which  the
superconducting electrons are taken.

\begin{figure}
\epsfbox[-40 700 650 700]{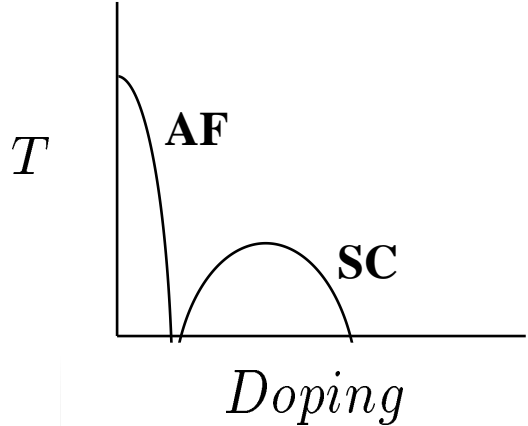}
\vspace{2.4in}
\caption{A typical Temperature--{\it vs}--Doping phase
diagram for a high-\Tc\ system.}
\end{figure}

This basic association of superconductivity and 
antiferromagnetism suggests a fundamental  connection
between the two. Zhang's proposal is that --- at least over
part of the theory's parameter  space --- these two phases
are related by an approximate  $SO(5)$ symmetry. The action
of the symmetry is simplest to state for the order
parameters for the two phases.

As we have seen, in \S3, the order parameter for the
antiferromagnetic (AF) phase is simply the direction in
space, 
$\vec n$, into which the alternating aligned spins point. A
nonzero value of this order parameter spontaneously breaks
the $SO(3) 
\simeq SU(2)$ symmetry of rotations amongst electron spins: 
\eq
\label{sothreedef}
\pmatrix{n_1 \cr n_2 \cr n_3 \cr} 
\to \Sco_3  \; \pmatrix{n_1 \cr n_2 \cr n_3
\cr} ,
\eeq
where $\Sco_n$ denotes an $n$-by-$n$ real, orthogonal
matrix.

For the superconducting (SC) phase, on the other hand, the
order parameter, $\psi$, is a quantity which carries the
quantum numbers of a pair of electrons (or holes).
Specifically, it has electric charge $q =  \pm2 e$, if $e$
is the proton charge, and is usually taken to have no
spin.  (The sign of this charge depends on whether the
charge carriers are electrons or holes.) A nonzero value
for $\psi$ signals the breaking of the symmetry of
electromagnetic phase rotations:  $\psi \to e^{iq \omega}
\; \psi$. This symmetry forms the group $SO(2) \simeq
U(1)$, as is easily seen by grouping the real and imaginary
parts of $\psi$ into a two-component vector:
\eq
\label{sotwodef}
\pmatrix{\psi_\ssr \cr \psi_\ssi \cr} \to 
\Sco_2  \; \pmatrix{\psi_\ssr \cr
\psi_\ssi \cr} ,  \eeq
where $\psi = \psi_\ssr + i \, \psi_\ssi$.

The $SO(5)$ proposal is that the system is approximately
invariant under five-by-five rotations of all five 
components of the order parameter,
\eq
\label{sofivedef}
\pmatrix{\psi_\ssr \cr \psi_\ssi \cr n_1 
\cr n_2 \cr n_3 \cr} \to \Sco_5
\; \pmatrix{\psi_\ssr \cr \psi_\ssi \cr n_1 
\cr n_2 \cr n_3 \cr} .
\eeq
This symmetry contains electron-spin rotations, 
eq.~\pref{sothreedef}, and electromagnetic phase
transformations, eq.~\pref{sotwodef}, as the
block-diagonal  $SO(2) \times SO(3)$ subgroup which acts
separately on the $\psi_k$ and the $n_a$.

As of this writing, there is a controversy over whether
such an approximate symmetry actually exists for  high-\Tc\
superconductors and, if so, to how large a  portion of the
phase diagram  it might apply. Regardless of how this
controversy ultimately becomes resolved,  two points on
which everyone must agree are:
\begin{enumerate}
\item
For any part of the phase diagram for which approximate
$SO(5)$ symmetry holds, and which lies within the ordered
(AF or SC) phases, there must be a total of four Goldstone
(or pseudo-Goldstone) bosons. This corresponds to one each
for the four $SO(5)$ generators which are broken by a
nonzero value for one of the $\psi_k$ or $n_a$. These
bosons include the usual Goldstone bosons for $SO(3)$ or
$SO(2)$ invariance (\eg, the magnons), plus some new
pseudo-Goldstone bosons which are consequences only of the
assumed $SO(5)$ symmetry.

\item
The low-energy properties of these Goldstone  and
pseudo-Goldstone bosons are completely dictated by the
assumed symmetry-breaking pattern, and  are independent of
the details of whatever microscopic electron dynamics gives
rise to the symmetry in the first place. These low-energy
properties may be efficiently described using an effective
lagrangian along the lines of those described in  Chapter 1.
\end{enumerate}
These two properties taken together make an unambiguous
detection of the $SO(5)$ pseudo-Goldstone bosons a
particularly attractive test of the $SO(5)$ proposal. Their
detection would be a `smoking gun' for the existence of an 
extended symmetry like $SO(5)$. Better yet, their
properties are unambiguously predicted theoretically,
without the  the usual complications which arise when
complicated electron dynamics is squeezed into a simple
theoretical model. We now construct the low-energy
effective lagrangian describing these pseudo-Goldstone
bosons, following the general techniques of the previous
sections.

\section{The Effective Lagrangian in the Symmetry Limit}

We start with the effective lagrangian in the (idealized)
limit where $SO(5)$ is not just approximate, but is instead
a {\it bona fide} symmetry of the system. In this case the
lagrangian symmetry is $G = SO(5)$, and this is
spontaneously broken (by the order parameters we are
considering) to the subgroup, $H = SO(4)$. We require in
this limit the nonlinear sigma model for the quotient space
$G/H = SO(5)/SO(4)$. As discussed in general in \S1, the
lowest terms in the derivative expansion of the Lagrangian
for this system are therefore completely determined up to a
small number of constants. (Precisely how many constants
depends on how much symmetry -- including crystallographic
symmetries -- the system has. As we describe shortly, more
possibilities also arise once explicit $SO(5)$-breaking
interactions are introduced.)

In this particular case the space $SO(5)/SO(4)$ is an old
friend: the four-sphere, $S_4$, which is defined as the
points swept out by an arbitrary five-dimensional vector,
$\vec{N}$, which has unit length: $\vec{N} \cdot \vec{N} =
\sum_{i=1}^5 N_i^2 \equiv 1$. In terms of such a field, the
invariant lagrangian obtained using the techniques of
previous sections is
\eq
\label{simpinvlagr}
\Scl_\inv = {f_t^2 \over 2} \;
\partial_t \vec{N} \cdot  \partial_t
\vec{N}  - {f_s^2 \over 2} \; \nabla
\vec{N} \cdot  \nabla  \vec{N}
\eeq
(For simplicity of this and later expressions, 
eq.~\pref{simpinvlagr}\ assumes rotational invariance,
which is 
{\it not} appropriate for real cuprates. For real systems,
the electrons believed responsible for superconductivity
and antiferromagnetism move preferentially along planes
made up of copper and oxygen atoms. The low-energy
lagrangian for such systems is better written either in two
space dimensions (for Goldstone bosons confined to the
planes), or in three dimensions with separate coefficients,
$f_a$ for each spatial direction, 
$\nabla_a$. These complications are ignored here, but are
discussed in more detail in the original references. As is
also discussed in these referencse, in specific dimensions
it is sometimes also possible to write more invariants than
are considered here, such as those which depend on the
completely antisymmetric tensor, $\epsilon_{ijk..}$.)

A convenient parameterization for the four-sphere, and so
of our Goldstone bosons, is given by polar coordinates:
\eq
\label{explangles}
\vec{N} = \pmatrix{\nq \cr \ns \cr}, 
\qquad \hbox{where}
\qquad
\nq = \cos\theta  \pmatrix{\cos\phi 
\cr \sin\phi \cr}, \qquad
\ns = \sin\theta \pmatrix{ \sin\alpha 
\cos\beta \cr  \sin\alpha \sin\beta \cr
\cos\alpha\cr} .
\eeq
(As usual, care is required to properly handle those points
where these coordinates are singular).

Using the standard expression for the round metric on $S_4$
then gives the unique Goldstone-boson effective lagrangian
for a rotational- and time-reversal-invariant system in the
$SO(5)$ symmetry limit:
\bg
\label{kintermanglesinv}
\Scl &=& {f^2_t \over 2} \Bigl[
(\partial_t \theta)^2 + \cos^2\theta
(\partial_t \phi)^2 + \sin^2\theta 
\Bigl( (\partial_t \alpha)^2
+ \sin^2\alpha (\partial_t \beta)^2
\Bigr) \Bigr] \nn\\
&& \qquad \qquad - \;  {f^2_s \over 2} \;\Bigl[
(\nabla \theta)^2 + \cos^2\theta (\nabla \phi)^2
+ \sin^2\theta \Bigl( (\nabla \alpha)^2
+ \sin^2\alpha (\nabla \beta)^2
\Bigr) \Bigr]  \, .\nn
\nd

\section{Symmetry-Breaking Terms}

Next consider how small $SO(5)$-breaking effects can change
the low-energy lagrangian. We may do so by following the
same steps as were taken in \S2\ to describe the
implications of quark masses on low-energy pion properties.
We do so here in two steps. We first classify the types of
violation of $SO(5)$ symmetry which can arise in real
systems. We then examine which consequences follow only
from $SO(2) \times SO(3)$ invariance, in order to be able
to disentangle these from predictions which are specific to
$SO(5)$. Finally we perturb in the various $SO(5)$
symmetry-breaking parameters to obtain the predictions of
approximate $SO(5)$ invariance. By contrasting what is
obtained with the result assuming only $SO(2) \times SO(3)$
invariance, the consequences of approximate $SO(5)$
invariance may be found.

\subsection{Kinds of Explicit Symmetry Breaking
(Qualitative)}

There are several kinds of $SO(5)$ symmetry breaking which
are worth distinguishing from one another. These are:

\begin{enumerate}

\item
{\it Electromagnetic Interactions:} One source of explicit
$SO(5)$ symmetry breaking is any couplings to applied
macroscopic electromagnetic fields. Electromagnetic gauge
invariance requires these to be incorporated into the
lagrangian using the usual precedure of minimal
substitution:
\eq
\label{minsubst}
\partial_t \psi \to (\partial_t - i \, q \; A_0) \psi,
\qquad \nabla \psi \to (\nabla - i\, q \; \Bfa) \psi.
\eeq
Such couplings necessarily break $SO(5)$ because they treat
the electrically-charged components of $\vec N$ differently
than the electrically-neutral ones. Although such couplings
do not pose any complications of principle, for simplicity's
sake we imagine no macroscopic electromagnetic fields to be
applied in what follows.

\item
{\it Doping:}
One of the physical variables on which the phase diagram of
the cuprates crucially depends is the doping. Since changes 
in the doping correspond to changes in the density of charge
carriers amongst the electrons which are relevant 
for both the antiferromagnetism and the superconductivity,
it may be described within the effective theory by using a 
chemical potential, $\mu$, coupled to electric charge. 
In this way adjustments in $\mu$ may be chosen to
ensure that the system has any given experimental charge
density. Mathematically a chemical potential is introduced 
by replacing the system Hamiltonian, $H$, with the 
quantity $H - \mu Q$, where $Q$ is the electric
charge. Within the Lagrangian formulation in which
we are working, this amounts to simply making 
the replacement $A_0 \to A_0 + \mu$ in the electrostatic
scalar potential, $A_0$.

\item
{\it Intrinsic Breaking:} The third, and final, category of
symmetry-breaking consists of everything apart from the
previous two. It is known that $SO(5)$ is not an exact
symmetry, even with no chemical potential, and in the
absence of any applied electromagnetic fields. It is the
interactions (involving the fewest derivatives) in this
last class of symmetry-breaking terms which we now wish to
classify.

\end{enumerate}

\subsection{General $SO(2) \times SO(3)$-Invariant
Interactions}

If $SO(5)$ were not a symmetry at all, then there would be
no guarantee that the low-energy spectrum should contain
particles described by all of the fields $\alpha$, $\beta$,
$\theta$ and 
$\phi$. For the purposes of later comparison, it is
nevertheless useful to ask what kinds of low-energy
interactions among such states are permitted by $SO(2)
\times SO(3)$ invariance.

The most general such Lagrangian involving these four
states may be written in terms of the fields $\ns$ and
$\nq$, where these fields satisfy the constraint $\ns\cdot
\ns + \nq \cdot \nq \equiv 1$, to the extent that we are
interested in only those modes which would be Goldstone or
pseudo-Goldstone modes in the $SO(5)$ limit. The most
general such result, which involves at most two
derivatives, supplements the invariant expression, 
eqs.~\pref{simpinvlagr} or \pref{kintermanglesinv}, with
the following terms:
\bg
\label{noninvform}
\Scl_\sb &=& -V + f^2_t \Bigl[ A 
\, \partial_t {\nq} \cdot  \partial_t {\nq} +
B \,\partial_t {\ns} \cdot
\partial_t {\ns} + C\,  (\nq \cdot \partial_t
\nq)^2 \Bigr] \nn\\
&&\qquad\qquad - f_s^2 \Bigl[ D \nabla_a {\nq}
\cdot \nabla_a {\nq} + E \nabla_a {\ns}
\cdot  \nabla_a {\ns} + F (\nq \cdot
\nabla_a \nq)^2 \Bigr] \nn
\nd
where $f_t$ and $f_s$ are the constants appearing in the
invariant lagrangian. The quantities $V, A, B, C, D, E$ and
$F$ are potentially arbitrary functions of the unique $SO(2)
\times SO(3)$ invariant which involves no derivatives: $\nq
\cdot \nq$. (Recall 
$\ns \cdot \ns$ is not independent due to the constraint
$\ns\cdot 
\ns + \nq \cdot \nq = 1$). 

In terms of polar coordinates, and inserting a chemical
potential as just described, the total effective lagrangian
becomes:
\bg
\label{kintermangles}
\Scl &=& - V+  {f^2_t \over 2}
\Bigl[ \Bigl(1 + 2A \sin^2\theta +
2 B \cos^2\theta + 2C\sin^2\theta
\cos^2\theta \Bigr) (\partial_t \theta)^2
\nn\\
&&\qquad\qquad + (1 + 2A ) \cos^2\theta
(\partial_t \phi + q\mu )^2  + (1+2B)\sin^2\theta
\Bigl( (\partial_t \alpha)^2 + \sin^2\alpha
(\partial_t \beta)^2 \Bigr) \Bigr] \nn\\
&& \qquad \qquad - {f^2_s \over 2} \Bigl[
\Bigl(1 + 2D  \sin^2\theta + 
2 E \cos^2\theta +
2F\sin^2\theta\cos^2\theta \Bigr)
(\nabla \theta)^2 \cr
&&\qquad\qquad + (1 + 2D ) 
\cos^2\theta (\nabla \phi)^2 + (1+2E) \sin^2\theta
\Bigl( (\nabla \alpha)^2 +
\sin^2\alpha (\nabla \beta)^2 \Bigr)
\Bigr] + \cdots,\nn
\nd
where all coefficient functions, $V$, $A$,...\etc, are now
to be regarded as functions of $\cos^2\theta$.

\subsection{Kinds of Explicit Symmetry Breaking
(Quantitative)}

Eq.~\pref{kintermangles} does not yet use any information
concerning the nature or size of the explicit symmetry
breaking (apart from the inclusion of $\mu$). This we must
now do if we are to quantify the predictions of approximate
$SO(5)$ invariance. We do so by making an assumption as to
how the symmetry-breaking  terms transform under $SO(5)$.

In \S2\ we saw, for pions, that the quark masses were
responsible for explicitly breaking the would-be chiral
symmetry of the underlying microscopic theory ($QCD$).
Although the same reasoning can be applied to $SO(5)$
breaking due to electromagnetic interactions and chemical
potential dependence, incomplete understanding of the
dynamics of the microscopic theory so far precludes a
similar identification of the other symmetry-breaking
parameters within some underlying condensed-matter system.
For these, we instead are forced to make an assumption.

We therefore assume all $SO(5)$-breaking terms of the
effective lagrangian to be proportional to one of two
possible quantities:

\begin{enumerate}

\item
{\it Chemical Potential:}
Since we know how the chemical potential appears in the
lagrangian, we know in detail how it breaks $SO(5)$. It
does so by an amount which is proportional to the electric
charge. For the fields appearing in $\vec{N}$, this is
represented by the five-by-five electric charge matrix: $Q
= \hbox{diag}(q,q,0,0,0)$.

\item
{\it Intrinsic Symmetry Breaking:}
In the absence of more information, we make the simplest
assumption for the form taken in the effective lagrangian
by all other microscopic effects which explicitly break
$SO(5)$. Since these break $SO(5)$ to $SO(3)\times SO(2)$
we take them to be proportional to a five-by-five matrix,
$M$, where $M = \epsilon 
\,\diag{3,3,-2,-2,-2}$. Here $\epsilon \ll 1$ is a measure
of the quality of the approximation that $SO(5)$ is a
symmetry.

\end{enumerate}

With these choices the Lagrangian is then the most general
function of the fields $\vec{N} = {\nq \choose \ns}$, $\mu
Q$ and $M$, subject to the following $SO(5)$ transformation
property
\eq
\label{transrule}
\Scl(\Sco_5 \vec N, \Sco_5 \mu Q 
\Sco_5^\sst, \Sco_5 M \Sco_5^\sst)
= \Scl(\vec N,\mu Q,M) ,
\eeq
where $\Sco_5$ is an $SO(5)$ transformation. The
implications of the approximate $SO(5)$ invariance may then
be extracted by expanding $\Scl$ in powers of the small
quantities $\epsilon$ and $\mu$. Since $M$ and $Q$ always
appear premultiplied by these small numbers, this expansion
restricts the kinds of symmetry breaking which can arise
order by order, which in turn constrains the possible
$\theta$-dependence of the coefficient functions in 
$\Scl$.

For example, a term in the scalar potential involving $2n$
powers of $\vec N$ must have the following form:
\eq
\label{potexpn}
V_{(n)} = \sum_{(k_1,l_1) \ne (0,0)}
\cdots \sum_{(k_n,l_n)\ne (0,0)}
C_{k_1l_1,\dots,k_n,l_n}\Bigl[ {\vec N} \cdot
(\eps M)^{k_1} (\mu Q)^{2l_1} 
{\vec N} \Bigr] \cdots
\Bigl[ {\vec N} \cdot (\eps M)^{k_n}
 (\mu Q)^{2l_n} {\vec N} \Bigr].
\eeq
Only even powers of $Q$ enter here due to its antisymmetry,
and the term $k_i = l_i = 0$ is excluded from the sums due
to the constraint ${\vec N}^\sst {\vec N} = 1$. Clearly,
expanding  
$\Scl$ to low order in the $SO(5)$-breaking parameters 
$\eps$ and $\mu$ necessarily also implies keeping only the
lowest powers of $\nq \cdot \nq = \cos^2\theta$ in $V$.

Similar conclusions may be obtained for the other 
coefficient functions in the Lagrangian of
eq.~\pref{noninvform}.  Working to
$O(\eps^2,\eps\mu^2,\mu^4)$ in $V$, and to $O(\eps,\mu^2)$
in the two-derivative terms  then gives:
\eq
\label{leadingpotterms}
V = V_0 + V_2 \,\cos^2\theta + \hf \,
V_4 \,\cos^4\theta,
\eeq
and
\bg
\label{leadingkinterms}
&&A = A_0 + A_2 \cos^2\theta,
\qquad B = B_0 + A_2 \cos^2\theta,
\qquad C = C_0, \nn\\
&&D= D_0 + D_2 \cos^2\theta, \qquad
E= E_0 + D_2 \cos^2\theta, \qquad
F= F_0 , \nn
\nd
for the coefficient functions in eq.~\pref{noninvform}.
Notice that the terms proportional to $\cos^2\theta$ in $A$
and $B$ are identical, as are the corresponding terms in $D$
and $E$. Expanding in powers of $\eps$ and $\mu$, the
constants  in 
eqs.~\pref{leadingpotterms}\ and \pref{leadingkinterms}\
start off linear in $\eps$ and $\mu^2$: $A_i = A_i^{10} 
\eps + 
A_i^{01}\mu^2 + \cdots$ \etc. The only exceptions to this
statement are: $B_0, E_0 \propto \eps$ (no $\mu^2$ term), 
$C_0, F_0 \propto \mu^2$ (no $\eps$ term), and $V_4 = 
V_4^{20} \eps^2 + V_4^{11} \eps \mu^2 + V_4^{02} \mu^4$.
Furthermore, since the $\mu^2 \; \nq\cdot \nq$ term in $V$
arises from substituting $\partial_t \to \partial_t -i \mu
Q$ in the kinetic term for $\nq$, we have: $V_2^{01} = - \,
\hf \, f_t^2  q^2$ to leading order. Higher powers of $\mu$
originate from terms in 
$\Scl$ which involve more than two derivatives.

\section{Pseudo-Goldstone Dispersion Relations}

We now turn to the calculation of the pseudo-Goldstone
boson dispersion relations. The scalar potential of  
eq.~ \pref{kintermangles}\ has three types of extrema:
\bg
\label{vacua}
&&(1) \qquad \theta_0 = 0 \quad\hbox{or}
\quad \pi ; \nn\\
&&(2) \qquad \theta_0 = {\pi \over 2}
\quad\hbox{or} \quad {3\pi \over 2} ; \nn\\
&&(3) \qquad \theta_0 \quad
\hbox{where $c = \cos\theta_0$
satisfies $V'(c^2) = 0$} . \
\nd
This leads to the four possible phases: ($i$) SC phase:
extremum (1) is a minimum, and (2) is a maximum; ($ii$) AF
phase: (2) is a minimum, and (1) is a maximum; ($iii$) MX
phase: both (1) and (2) are maxima, and (3) is a minimum;
or ($iv$) metastable phase: both (1) and (2) are minima,
and (3) is a maximum. We focus here purely on the AF and SC
phases.

\subsection{Superconducting Phase}

An expansion about the superconducting mimimum,
$\theta_0=0$, gives the dispersion relations in this phase
for the four bosons. Three of these --- $\theta$, $\alpha$
and $\beta$ --- form a spin triplet of pseudo-Goldstone
modes for which
\eq
\label{simpdispform}
E(k) = \Bigl[ c^2 \, k^2 + \Sce^2 \Bigr]^\hf,
\eeq
with the phase speed, $c^2_\alpha(SC)$, and gap,
$\Sce^2_\SC$, given to lowest order in $SO(5)$-breaking 
parameters by:
\bg
\label{SCpseudos}
c^2_\alpha(SC) &=& {f^2_s \over f^2_t} \;
\Bigl[ 1 + 2 \Bigl(E(1)-B(1)\Bigr) \Bigr]  \nn\\
 &=& {f^2_s \over f^2_t} \; \Bigl[ 1 + 
2 \Bigl(E_0-B_0\Bigr)
+ 2 \Bigl(D_2-A_2\Bigr) \Bigr], \nn\\
\Sce^2_\SC &=& { -2 V'(1)  \over f^2_t} \nn\\
&=& { -2 (V_2 + V_4)   \over f^2_t} .
\nd
In both of these results the first equation uses the
general effective theory, eq.~\pref{kintermangles}, while
the second equality incorporates the additional information
of 
eqs.~\pref{leadingpotterms}\  and \pref{leadingkinterms}.
An important part of the $SO(5)$ proposal is that these
states have been seen in neutron-scattering experiments in
the superconducting phase of the high-\Tc\ cuprates, even
quite far away from the antiferromagnetic regime.

The remaining field, $\phi$, would have been a {\it bona
fide} gapless Goldstone mode in the absence of
electromagnetic interactions. Its dispersion relation,
$E(k)$ is a more complicated function of $c^2 k^2$ and
$eq\mu$, whose form is not required here. The quantity $c$
which appears with $k$ throughout its dispersion relation
is given explicitly by
\bg
\label{SCGB}
c^2_\phi(SC) &=& {f^2_s \over f^2_t} \;
\Bigl[ 1 + 2 \Bigl( D(1)-A(1) \Bigr) \Bigr] \nn\\
&=& {f^2_s \over f^2_t} \; \Bigl[ 1 + 
2 \Bigl(D_0-A_0\Bigr) +
 2\Bigl(D_2 - A_2 \Bigr)\Bigr]  .
\nd

\subsection{Antiferromagnetic Phase}

Expanding about the AF minimum gives the usual two magnons,
as in \S3, satisfying dispersion relation of 
eq.~\pref{simpdispform} with:
\bg
\label{AFmagnons}
c^2_\GB(AF) &=& {f^2_s \over f^2_t} \;
\Bigl[ 1 + 2 \Bigl(E(0)-B(0)\Bigr) 
\Bigr] \nn\\
&=& {f^2_s \over f^2_t} \; 
\Bigl[ 1 + 2 \Bigl(E_0-B_0\Bigr) \Bigr] , \nn\\
\Sce^2_\GB(AF) &=& 0 .
\nd
The remaining two states group into an electrically-charged
pseudo-Goldstone state satisfying:
\eq
\label{compdispform}
E_\pm(k) = \Bigl[ c^2\, k^2 + 
\Sce^2 \Bigr]^\hf \pm  q\mu,
\eeq
with:
\bg
\label{AFpseudos}
c^2_\pGB(AF) &=& {f^2_s 
\over f^2_t} \;  \Bigl[ 1 + 2
\Bigl(D(0)-A(0)\Bigr)  \Bigr] \nn\\
&=& {f^2_s \over f^2_t} \; 
\Bigl[ 1 + 2 \Bigl(D_0-A_0\Bigr)  \Bigr],  \nn\\
\Sce^2_\AF = \Sce^2_\pGB(AF) 
&=& { 2 V'(0) \over f^2_t}  \nn\\
&=& { 2 V_2 \over f^2_t} .
\nd

These expressions imply a simple dependence of the gap on
the chemical potential:
\bg
\label{ourpredictions}
\Sce^2_\AF  &=&  m^2 - \kappa \mu^2, \nn\\
\Sce^2_\SC &=& -m^2 +
\kappa\mu^2 - \xi \mu^4 ,
\nd
where $m^2 := 2 V^{10}_2   \eps/f_t^2 + O(\eps^2)$ ,
$\kappa := -2 V_2^{01}/f_t^2  + O(\eps) = q^2 + O(\eps)$
and $\xi := 2 V_4^{02}/f_t^2  + O(\eps)$.  Within the AF
phase the pseudo-Goldstone boson gap is predicted to fall
linearly with $\mu^2$:
\eq
\label{AFlinearfall}
\Sce^2_\AF \approx \Sce^2_\AF(0) 
 [\mu^2_\AF - \mu^2] ,
\eeq
where $\mu_\AF$ represents the doping for which one leaves
the AF regime. Similarly $\Sce^2_\SC$ varies quadratically
with 
$\mu^2$.

Robust consequences of $SO(5)$ invariance are obtained from
expressions such as these by eliminating the free parameters
to obtain relations amongst observables. For example, if one
eliminates parameters  in favour of properties of the gap as
a function of $\mu$ we find:
\bg
\label{vepspreds}
\varepsilon^2_\AF(\mu) &=& 
{\varepsilon_\AF^2(0) \over \mu^2_\AF}
\, \Bigl[ \mu^2_\AF - \mu^2 \Bigr] ,\nn\\
\varepsilon^2_\SC(\mu) &=& 
{\varepsilon^2_\SC(\opt)
\over  \mu^4_\opt } \, (\mu^2 - 
\mu^2_{\SC-}) (2\mu^2_\opt - \mu^2) ,\nn\\
 {\varepsilon_\AF^2(0) \over \mu^2_\AF} &=& 2 \;
{\varepsilon^2_\SC(\opt) \over \mu^2_\opt } ,\nn\\
\mu^2_\AF &=& \mu^2_{\SC-} + O(\epsilon^2)  .
 \nd
Here $\mu_\opt$ denotes the chemical potential
corresponding to the maximum gap, $\Sce_\SC$.

Similarly, the phase velocities for all modes in both SC
and AF phases are equal to one another, and to
$f^2_t/f^2_s$, in the limit of strict $SO(5)$ invariance.
(The parameters $f_t$ and $f_s$ may be related to other
observables, such as the electric and magnetic screening
lengths.) It turns out that the $O(\eps)$ corrections to
this limit  are not arbitrary, but also satisfy some
model-independent relations, which follow by eliminating
parameters from the above expressions:
\eq
\label{speedrels}
c^2_\phi(SC) - c^2_\phi(AF) = 
c^2_\alpha(SC) - c^2_\alpha(AF) =
O(\eps).
\eeq

\section{Summary}

Approximate $SO(5)$ invariance clearly carries real
implications for the low-energy excitations of the system.
It predicts, in particular, the existence of a spin-triplet
pseudo-Goldstone state in the SC phase, and an
electrically-charged state in the AF phase. Furthermore,
$SO(5)$ invariance unambiguously relates the properties of
these states, like their gap and phase velocity, to one
another. Better yet, these properties are claimed to have
been measured  in the SC phase, since the spin-triplet
state is believed to have been observed, with a gap (at
optimal doping) of 41 meV. If true, this permits the
inference of the size of the rough order of the
$SO(5)$-breaking parameter $\epsilon$, and hence to
predictions for the properties of the hitherto undetected
boson in the AF phase.

It is extremely unlikely that an electrically-charged state
having a gap of only $\sim 40$ meV can exist deep within the
AF phase. Among other things it would make its presence felt
through the electromagnetic response of these systems in the
AF phase. At the very least, one can therefore conclude that
$SO(5)$ invariance cannot penetrate very far into the AF
part of the phase diagram, despite its appearance fairly
deep in the SC phase. Being based purely on
pseudo-Goldstone boson properties, this conclusion comes
independent of the details of how the underlying electrons
are interacting on more microscopic scales.

There is not yet a consensus as to how uncomfortable this
conclusion should make one feel about the remarkable
$SO(5)$ hypothesis. Either way, robust predictions based on
the low-energy consequences of symmetries are likely to play
a key role in forming any such final consensus.

\chapter{Bibliography}

This review is my personal presentation of the theory of
Goldstone bosons, and (apart from the $SO(5)$ section) were
first given as part of a lecture course for the Swiss {\it
Cours de Troisi\`eme Cycle}, in Lausanne and Neuch\^atel in
June 1994. I am indebted to the University of Neuch\^atel,
the University of Oslo, Seoul National University, and my
colleagues at McGill University for providing the forum in
which to work out the presentation of these ideas. I
thank Patrick Labelle and Oscar Hernandez for constructively criticizing them as they have evolved. Needless to say, any errors that remain are mine alone.

The classic early papers which founded this subject are

\begin{enumerate}

\item
S. Weinberg, {\it Physical Review Letters} {\bf 18} (1967)
188;
{\it Physical Review}  {\bf 166} (1968) 1568. 

\item
C.G. Callan, S. Coleman, J. Wess and B. Zumino, {\it
Physical Review} {\bf 177} (1969) 2247.

\end{enumerate}

My thinking has been  strongly influenced by graduate
courses given on this subject during the early 1980's in
the University of Texas at Austin, by Steven Weinberg.
This review is, in its presentation,  very similar
with  the logic of his textbook: {\sl The Quantum Theory of
Fields}, vols. I and II.

The following papers and reviews have also shaped how I
understand this subject:

\begin{enumerate}

\item
S. Weinberg, {\sl Phenomenological Lagrangians}, {\it
Physica} {\bf 96A} (1979) 327--340.

\item S. Weinberg, {\sl What is Quantum Field Theory, and
What Did We Think it is?}, Talk given at Conference on
Historical Examination and Philosophical Reflections on the
Foundations of Quantum Field Theory, Boston, MA, 1-3 Mar
1996; (hep-th/9702027).

\item
{\it The Quantum Theory of Fields, Volumes I,  II: Modern
Applications}, Steven Weinberg, Cambridge University Press,
1995 and 1996.

\item
J. Polchinski, {\sl  Effective Theory of the Fermi
Surface}, in the proceedings of the 1992 Theoretical
Advanced Study Institute, Boulder, Colorado,
(hep-th/9210046).

\item
W.E. Caswell and G.P. Lepage, {\it Physics Letters} {\bf
B167} (1986)  437; T. Kinoshita and G.P. Lepage, in {\it
Quantum Electrodynamics}, ed. by T. Kinoshita, (World
Scientific, Singapore, 1990), pp.  81--89.

\item
J. Gasser and H. Leutwyler, {\sl Chiral Perturbation Theory
to One Loop}, {\it Annals of Physics} (NY) {\bf 158} (1984)
142. 

\item
N. Isgur and M.B. Wise, {\sl Weak Decays of Heavy Mesons in
the Static Quark Approximation}, {\it Physics Letters} {\bf
B232} (1989) 113.

\item
S. Weinberg, {\sl Nuclear Forces from Chiral Lagrangians},
{\it Physics Letters} {\bf B251} (1990) 288-292.

\end{enumerate}

\section{Review Articles}

There have been a number of well-written review articles on
effective field theories, most of which concentrate on
chiral perturbation theory. Here is a partial list. I have
inevitably omitted several good ones, for which I give my
apologies now.

\begin{enumerate}

\item
G.S. Guralnik, C.R. Hagen and T.W.B. Kibble, in {\it
Advances in Particle Physics}, vol. 2, ed. by R.L. Cool and
R.E. Marshak (Wiley, NY, 1968).

\item
{\it Weak Interactions and Modern Particle Theory}, Howard
Georgi, Benjamin/Cummings, 1984.

\item
{\it Dynamics of the Standard Model}, John F. Donoghue,
Eugene Golowich and Barry R. Holstein, Cambridge University
Press, 1992.

\item
A. Manohar, {\sl Effective Field Theories},  in {\it
Schladming 1996, Perturbative and nonperturbative  aspects
of quantum field theory}, p.p. 311--362. (hep-ph/9606222).

\item
Mannque Rho, {\sl Effective Field Theory for Nuclei and
Dense Matter}. Submitted to Acta Phys.Pol.B,
(nucl-th/9806029).

\item
Antonio Pich, {\sl Effective Field Theory: Course} In the
proceedings of the Les Houches Summer School in Theoretical
Physics, Session 68: {\it Probing the Standard Model of
Particle Interactions}, Les Houches, France, 1997,
(hep-ph/9806303).

\item
D. Kaplan, {\sl Effective Field Theories}, in the
proceedings of the 7th Summer School in Nuclear Physics
Symmetries, Seattle, WA, 1995, (nucl-th/9506035).

\item
H. Georgi, {\sl Effective Field Theory}, {\it Annual review
of nuclear and particle science, vol. 43}, (1995), 209-252.

\item
H. Georgi, {\sl Heavy Quark Effective Field Theory}, in the
proceedings of the 1991 Theoretical Advanced Study
Institute, Boulder, Colorado, 1991. 

\end{enumerate}

\section{Particle Physics Data}

I have taken the experimental values for the section on
pions and nucleons from {\it The Review of Particle
Properties}, The European Physical Journal {\bf C3},
1 (1998).

\section{$SO(5)$ Invariance in High-$T_c$ Superconductors}

It is not the intention of these notes to provide more than
a superficial discussion of the $SO(5)$ proposal for the
high-$T_c$ cuprates, which was first made by Zhang, in

\begin{enumerate}
\item
S.C. Zhang, Science {\bf 275} (1997) 1089.
\end{enumerate}

\bigskip\noindent
This remarkable proposal immediately stimulated much
interest, with an associated literature. I do not attempt a
literature survey here, apart from listing three highlights
in which the `big picture' is addressed by several of the
field's major players. These are:

\begin{enumerate}

\item
 G. Baskaran, P.W. Anderson, {\sl On an $SO(5)$ Unification
Attempt for the Cuprates}, cond-mat/9706076.

\item
R.B. Laughlin, {\sl A Critique of Two Metals},
cond-mat/9709195.

\item
P.W. Anderson, G. Baskaran, {\sl A Critique of `A Critique
of Two Metals'}, cond-mat/9711197.

\end{enumerate}

\bigskip
The applications of the technique of nonlinear realizations
to identify the model-independent properties of the
resulting Goldstone and pseudo-Goldstone bosons, which is
followed in Chapter 4, is performed in:

\begin{enumerate}
\item
C.P. Burgess and C.A. L\"utken, {\it Physical Review} {\bf
B57} (1998) 8642 (cond-mat/9705216);  cond-mat/9611070.

\item
C.P. Burgess, J.M. Cline, C.A. L\"utken, cond-mat/9801303.

\end{enumerate}

\section{Condensed Matter Physics for Particle and Nuclear
Physicists}

There are several sources for condensed matter physics
which I have found useful. Some are

\begin{enumerate}

\item
R. Shankar, {\sl Effective Field Theory in Condensed Matter
Physics}, lecture given at Boston Colloquium for the
Philosophy of Science, Boston, Mass., 1996,
(cond-mat/9703210); {\sl Renormalization Group Approach to
Interacting Fermions}, {\it Reviews of Modern Physics},
(cond-mat/9307009).

\item
{\it Principles of Condensed Matter Physics}, P.M. Chaikin
and T.C. Lubensky, Cambridge University Press, 1995.

\item
{\it Principles of the Theory of Solids}, J.M. Ziman,
Cambridge University Press, 1972.

\item
{\it Introduction to Superconductivity, 2nd Edition},
Michael Tinkham, McGraw-Hill, 1996.

\item
A treatment of spin waves which is in the spirit of this
review has appeared in a preprint by J.M. Roman and J.
Soto, cond-mat/9709298; and by: Christoph Hofmann,
cond-mat/9805277.

\end{enumerate}

\end{document}

%% file: notemacros.tex

\def\ie{{\it i.e.\/}}
\def\eg{{\it e.g.\/}}
\def\etc{{\it etc.\/}}
\def\etal{{\it et.al.\/}}
\def\apriori{{\it a priori\/}}
\def\aposteriori{{\it a posteriori\/}}
\def\via{{\it via\/}}
\def\vs{{\it vs.\/}}
\def\cf{{\it c.f.\/}}
\def\adhoc{{\it ad hoc\/}}
\def\bll{$\bullet$}
\def\crc{$\circ$}
\def\expval{{\it vev}}


\def\anp#1#2#3{{\it Ann.\ Phys. (NY)} {\bf #1} (19#2) #3}
\def\arnps#1#2#3{{\it Ann.\ Rev.\ Nucl.\ Part.\ Sci.} {\bf #1}, (19#2) #3}
\def\cmp#1#2#3{{\it Comm.\ Math.\ Phys.} {\bf #1} (19#2) #3}
\def\ijmp#1#2#3{{\it Int.\ J.\ Mod.\ Phys.} {\bf A#1} (19#2) #3}
\def\jetp#1#2#3{{\it JETP Lett.} {\bf #1} (19#2) #3}
\def\jetpl#1#2#3#4#5#6{{\it Pis'ma Zh.\ Eksp.\ Teor.\ Fiz.} {\bf #1} (19#2) #3
[{\it JETP Lett.} {\bf #4} (19#5) #6]}
\def\jpb#1#2#3{{\it J.\ Phys.} {\bf B#1} (19#2) #3}
\def\mpla#1#2#3{{\it Mod.\ Phys.\ Lett.} {\bf A#1}, (19#2) #3}
\def\nci#1#2#3{{\it Nuovo Cimento} {\bf #1} (19#2) #3}
\def\npb#1#2#3{{\it Nucl.\ Phys.} {\bf B#1} (19#2) #3}
\def\plb#1#2#3{{\it Phys.\ Lett.} {\bf #1B} (19#2) #3}
\def\pla#1#2#3{{\it Phys.\ Lett.} {\bf #1A} (19#2) #3}
\def\prb#1#2#3{{\it Phys.\ Rev.} {\bf B#1} (19#2) #3}
\def\prc#1#2#3{{\it Phys.\ Rev.} {\bf C#1} (19#2) #3}
\def\prd#1#2#3{{\it Phys.\ Rev.} {\bf D#1} (19#2) #3}
\def\pr#1#2#3{{\it Phys.\ Rev.} {\bf #1} (19#2) #3}
\def\prep#1#2#3{{\it Phys.\ Rep.} {\bf C#1} (19#2) #3}
\def\prl#1#2#3{{\it Phys.\ Rev.\ Lett.} {\bf #1} (19#2) #3}
\def\rmp#1#2#3{{\it Rev.\ Mod.\ Phys.} {\bf #1} (19#2) #3}
\def\sjnp#1#2#3#4#5#6{{\it Yad.\ Fiz.} {\bf #1} (19#2) #3
[{\it Sov.\ J.\ Nucl.\ Phys.} {\bf #4} (19#5) #6]}
\def\zpc#1#2#3{{\it Zeit.\ Phys.} {\bf C#1} (19#2) #3}


\global\nulldelimiterspace = 0pt


\def\goto{\mathop{\rightarrow}}
\def\gotoo{\mathop{\longrightarrow}}
\def\mapstoo{\mathop{\longmapsto}}

\def\df{\mathrel{:=}}
\def\fd{\mathrel{=:}}


\def\frac#1#2{{{#1} \over {#2}}\,}  
\def\hf{{1\over 2}}
\def\nth#1{{1\over #1}}
\def\sfrac#1#2{{\scriptstyle {#1} \over {#2}}}  
\def\stack#1#2{\buildrel{#1}\over{#2}}
\def\dd#1#2{{{d #1} \over {d #2}}}  
\def\ppartial#1#2{{{\partial #1} \over {\partial #2}}}  
\def\del{\nabla}
\def\grad{\nabla}
\def\Square{{\vbox {\hrule height 0.6pt\hbox{\vrule width 0.6pt\hskip 3pt
        \vbox{\vskip 6pt}\hskip 3pt \vrule width 0.6pt}\hrule height 0.6pt}}}
\def\Dslsh{\hbox{/\kern-.6700em\it D}} 
\def\Scaslsh{\hbox{/\kern-.8200em${\cal A}$}} 
\def\dslsh{\hbox{/\kern-.5300em$\partial$}}
\def\pslsh{\hbox{/\kern-.5600em$p$}}
\def\sslsh{\hbox{/\kern-.5300em$s$}}
\def\epsslsh{\hbox{/\kern-.5100em$\epsilon$}}
\def\delslsh{\hbox{/\kern-.6300em$\nabla$}}
\def\lslsh{\hbox{/\kern-.4300em$l$}}
\def\elslsh{\hbox{/\kern-.4500em$\ell$}}
\def\kslsh{\hbox{/\kern-.5100em$k$}}
\def\qslsh{\hbox{/\kern-.5000em$q$}}
\def\transp#1{#1^{\sss T}} 
\def\slsh#1{\raise.15ex\hbox{$/$}\kern-.57em #1}
\def\Pl{\gamma_{\sss L}}
\def\Pr{\gamma_{\sss R}}
\def\pwr#1{\cdot 10^{#1}}
\def\Lie#1{\hbox{\pounds}_{#1}}



\def\supsub#1#2{\mathstrut^{#1}_{#2}}
\def\sub#1{\mathstrut_{#1}}
\def\sup#1{\mathstrut^{#1}}
\def\rsub#1{\mathstrut_{\rm #1}}
\def\rsup#1{\mathstrut^{\rm #1}}

\def\twi{\widetilde}
\def\mybar{\bar}

\def\roughly#1{\mathrel{\raise.3ex\hbox{$#1$\kern-.75em
   \lower1ex\hbox{$\sim$}}}}
\def\lsim{\roughly<}
\def\gsim{\roughly>}

\def\bv#1{{\bf #1}}
\def\scr#1{{\cal #1}}
\def\op#1{{\widehat #1}}
\def\tw#1{\tilde{#1}}
\def\ol#1{\overline{#1}}


\def\sss{\scriptscriptstyle}
\def\ss{\scriptstyle}


\def\bfa{{\bf a}}
\def\bfb{{\bf b}}
\def\bfc{{\bf c}}
\def\bfd{{\bf d}}
\def\bfe{{\bf e}}
\def\bff{{\bf f}}
\def\bfg{{\bf g}}
\def\bfh{{\bf h}}
\def\bfi{{\bf i}}
\def\bfj{{\bf j}}
\def\bfk{{\bf k}}
\def\bfl{{\bf l}}
\def\bfm{{\bf m}}
\def\bfn{{\bf n}}
\def\bfo{{\bf o}}
\def\bfp{{\bf p}}
\def\bfq{{\bf q}}
\def\bfr{{\bf r}}
\def\bfs{{\bf s}}
\def\bft{{\bf t}}
\def\bfu{{\bf u}}
\def\bfv{{\bf v}}
\def\bfw{{\bf w}}
\def\bfx{{\bf x}}
\def\bfy{{\bf y}}
\def\bfz{{\bf z}}


\def\Bfa{{\bf A}}
\def\Bfb{{\bf B}}
\def\Bfc{{\bf C}}
\def\Bfd{{\bf D}}
\def\Bfe{{\bf E}}
\def\Bff{{\bf F}}
\def\Bfg{{\bf G}}
\def\Bfh{{\bf H}}
\def\Bfi{{\bf I}}
\def\Bfj{{\bf J}}
\def\Bfk{{\bf K}}
\def\Bfl{{\bf L}}
\def\Bfm{{\bf M}}
\def\Bfn{{\bf N}}
\def\Bfo{{\bf O}}
\def\Bfp{{\bf P}}
\def\Bfq{{\bf Q}}
\def\Bfr{{\bf R}}
\def\Bfs{{\bf S}}
\def\Bft{{\bf T}}
\def\Bfu{{\bf U}}
\def\Bfv{{\bf V}}
\def\Bfw{{\bf W}}
\def\Bfx{{\bf X}}
\def\Bfy{{\bf Y}}
\def\Bfz{{\bf Z}}


\def\Sca{{\cal A}}
\def\Scb{{\cal B}}
\def\Scc{{\cal C}}
\def\Scd{{\cal D}}
\def\Sce{{\cal E}}
\def\Scf{{\cal F}}
\def\Scg{{\cal G}}
\def\Sch{{\cal H}}
\def\Sci{{\cal I}}
\def\Scj{{\cal J}}
\def\Sck{{\cal K}}
\def\Scl{{\cal L}}
\def\Scm{{\cal M}}
\def\Scn{{\cal N}}
\def\Sco{{\cal O}}
\def\Scp{{\cal P}}
\def\Scq{{\cal Q}}
\def\Scr{{\cal R}}
\def\Scs{{\cal S}}
\def\Sct{{\cal T}}
\def\Scu{{\cal U}}
\def\Scv{{\cal V}}
\def\Scw{{\cal W}}
\def\Scx{{\cal X}}
\def\Scy{{\cal Y}}
\def\Scz{{\cal Z}}


\def\ssa{{\sss A}}
\def\ssb{{\sss B}}
\def\ssc{{\sss C}}
\def\ssd{{\sss D}}
\def\sse{{\sss E}}
\def\ssf{{\sss F}}
\def\ssg{{\sss G}}
\def\ssh{{\sss H}}
\def\ssi{{\sss I}}
\def\ssj{{\sss J}}
\def\ssk{{\sss K}}
\def\ssl{{\sss L}}
\def\ssm{{\sss M}}
\def\ssn{{\sss N}}
\def\sso{{\sss O}}
\def\ssp{{\sss P}}
\def\ssq{{\sss Q}}
\def\ssr{{\sss R}}
\def\ssS{{\sss S}}
\def\sst{{\sss T}}
\def\ssu{{\sss U}}
\def\ssv{{\sss V}}
\def\ssw{{\sss W}}
\def\ssx{{\sss X}}
\def\ssy{{\sss Y}}
\def\ssz{{\sss Z}}


\def\Prob{\mathop{\rm Prob}}
\def\tr{\mathop{\rm tr}}
\def\Tr{\mathop{\rm Tr}}
\def\det{\mathop{\rm det}}
\def\Det{\mathop{\rm Det}}
\def\Log{\mathop{\rm Log}}
\def\Re{{\rm Re\;}}
\def\Im{{\rm Im\;}}
\def\diag#1{{\rm diag}\left( #1 \right)}


\def\bra#1{\langle #1 |}
\def\ket#1{| #1 \rangle}
\def\braket#1#2{\langle #1 | #2 \rangle}
\def\vev#1{\langle 0 | #1 | 0 \rangle}
\def\avg#1{\langle #1 \rangle}

\def\Bra#1{\left\langle #1 \right|}
\def\Ket#1{\left| #1 \right\rangle}
\def\Avg#1{\left\langle #1 \right\rangle}

\def\ddx#1#2{d^{#1}#2\,}
\def\ddp#1#2{\frac{d^{#1}#2}{(2\pi)^{#1}}\,}


\def\veps{\varepsilon}
\def\eps{\epsilon}
\def\L{\Lambda}
\def\G{\Gamma}

\def\vacbra{{\bra 0}}
\def\vac{{\ket 0}}

\def\rhs{right-hand side}
\def\lhs{left-hand side}

\def\hc{{\rm h.c.}}
\def\cc{{\rm c.c.}}

\def\pty{{\cal P}}
\def\trv{{\cal T}}
\def\ccj{{\cal C}}

\def\GF{G_\ssf}
\def\lqcd{\Lambda_{QCD}}
\def\msbar{{\ol{MS}}}
\def\dsbar{{\ol{DS}}}
\def\fpi{F_\pi}
\def\mpl{M_{\rm Pl}}

\def\sm{standard model}
\def\smh{standard-model}
\def\km{Kobayashi Maskawa}
\def\kmh{Kobayashi-Maskawa}
\def\edm{e.d.m.}


\def\eV{{\rm \ eV}}
\def\keV{{\rm \ keV}}
\def\MeV{{\rm \ MeV}}
\def\GeV{{\rm \ GeV}}
\def\TeV{{\rm \ TeV}}

\def\cm{{\rm \ cm}}
\def\sec{{\rm \ sec}}
\def\ecm{{\it e}{\hbox{\rm -cm}}}


\def\abr{\ol{a}}
\def\dbr{\ol{d}}
\def\hbr{\ol{h}}
\def\Nbr{\ol{N}}
\def\qbr{\ol{q}}
\def\sbr{\ol{s}}
\def\ubr{\ol{u}}
\def\vbr{\ol{v}}
\def\chibr{\ol{\chi}}
\def\psibr{\ol{\psi}}
\def\Scabr{\ol{\Sca}}


\def\twm{\tw{m}}
\def\twg{\tw{g}}
\def\twF{\twi{F}}


\def\st{\begin{equation}}
\def\stp{\end{equation}}
\def\eq{\begin{equation}}
\def\eeq{\end{equation}}
\def\bg{\begin{eqnarray}}
\def\nd{\end{eqnarray}}
\def\nn{\nonumber}

\def\pref#1{(\ref{#1})}

\renewcommand{\theequation}{\thechapter.\arabic{section}.\arabic{equation}}
\newcounter{problem}

\newenvironment{problems}{\begin{list}{[\arabic{chapter}.\arabic{problem}]}%
{\usecounter{problem}}}{\end{list}}

\newcommand{\subproblem}{\vspace{0.25in}\noindent}